\documentclass[tradiabstract]{aa}
\usepackage{amssymb,amsmath}
\usepackage{graphicx}
\usepackage[varg]{txfonts}
\usepackage{natbib}
\begin{document}

\title{Phase-space structure of protohalos: Vlasov versus Particle-Mesh}

\author{S. Colombi}

\institute{Sorbonne Universit\'e, CNRS, UMR7095, Institut d'Astrophysique de Paris, 98 bis boulevard Arago, F-75014 Paris, France}

\date{Submitted to {\it Astronomy} \& {\it Astrophysics}}

\abstract{The phase-space structure of primordial dark matter halos is revisited using cosmological simulations with three sine waves and Cold Dark Matter (CDM) initial conditions. The simulations are performed with the tessellation based Vlasov solver {\tt ColDICE} and a Particle-Mesh (PM) $N$-body code. The analyses include projected density, phase-space diagrams, radial density $\rho(r)$ and pseudo-phase space density, $Q(r)=\rho(r)/\sigma_v(r)^3$ with $\sigma_v$ the local velocity dispersion. Particular attention is paid to force and mass resolution. Because the phase-space sheet complexity, estimated in terms of total volume and simplices count, increases very quickly, {\tt ColDICE} can follow only the early violent relaxation phase of halo formation. During the latter, agreement between {\tt ColDICE} and PM simulations having one particle per cell or more is excellent and halos have a power-law density profile, $\rho(r) \propto r^{-\alpha}$, $\alpha \in [1.5,1.8]$. This slope, measured prior to any merger, is slightly larger than in the literature. The phase-space diagrams evidence complex but coherent patterns with clear signatures of self-similarity in the sine wave simulations, while the CDM halos are somewhat scribbly.  After additional mass resolution tests, the PM simulations are used to follow the next stages of evolution. The power-law progressively breaks down with a convergence of the density profile to the well known ``NFW''-like universal attractor, irrespectively of initial conditions, that is even in the three-sine wave simulations. This demonstrates again that mergers do not represent a necessary condition for convergence to the dynamical attractor. Not surprisingly, the measured pseudo phase-space density is a power-law $Q(r) \propto r^{-\alpha_Q}$, with $\alpha_{\rm Q}$ close to the prediction of secondary spherical infall model, $\alpha_{\rm Q} \simeq 1.875$. However this property is also verified during the early relaxation phase, which is non trivial.
}

\maketitle

\section{Introduction}
In our current understanding of large-scale structure formation, the main matter component of the Universe is Cold Dark Matter \citep[CDM, e.g.,][]{Peebles82,Peebles84,Blumenthal84}, which can be modelled as a self-gravitating collisionless fluid obeying Vlasov-Poisson equations:
\begin{eqnarray}
  \frac{\partial f}{\partial t} + \vec{u}.\nabla_{\vec{r}} f-\nabla_{\vec{r}}\phi . \nabla_{\vec{u}} f=0,\\
  \Delta_{\vec{r}} \phi= 4\pi\, G\, \rho=4\pi\, G \int f(\vec{r},\vec{u},t)\, {\rm d}^3 \vec{u},
\end{eqnarray}
where $f(\vec{r},\vec{u},t)$ is the phase-space density at physical position $\vec{r}$ and velocity $\vec{u}$, $\phi$ the gravitational potential and $G$ the gravitational constant. In the concordance model, dark matter is composed of very massive particles with very small initial velocity dispersion. This means that dark matter is in a very good approximation concentrated on a three dimensional phase-space sheet folding in six-dimensional phase-space. At early times, the phase-space density has the following form
\begin{equation}
  f(\vec{r},\vec{u},t=t_{\rm i})= \rho_{\rm i}(\vec{r})\, \delta_{\rm D}[\vec{u}-\vec{u}_{\rm i}(\vec{r})],
\end{equation}
where the initial density $\rho_{\rm i}$ and velocity $\vec{u}_{\rm i}$ are smooth functions of position $\vec{r}$. The scale length $\lambda$ below which initial density and velocity fluctuations are smooth depends on the dark matter particle mass. In this work, following the footsteps of previous numerical investigations \citep[e.g.,][]{Diemand05,Ishiyama10,Anderhalden13,Ishiyama14,Angulo17},  we shall consider standard  CDM model with neutralinos of mass $100$ GeV, which implies $\lambda$ of the order of a pc.

The phase-space sheet evolves under self-gravity and shell crossing occurs in various places of configuration space. In these regions, sheets, filaments and dark-matter halos form as the result of complex processes of multi-streaming dynamics. While perturbation theory can be used to predict the early stages of clustering \citep[e.g.,][]{Bernardeau02}, numerical simulations are required to follow accurately the dynamics of the phase-space sheet beyond shell crossing. Traditionally, dark matter is simulated with the $N$-body technique, in which dark matter elements are followed with a large ensemble of macro particles interacting with each other with a softened gravitational force. There exist many different $N$-body simulation codes, that differ mainly from each other through the way Poisson equation is solved \citep[e.g.,][for reviews]{Hockney88, Bertschinger98, Colombi01,Dolag08,Dehnen11,Vogelsberger20}.

Even though a consensus on the robustness of the results of $N$-body simulations is getting close, their numerical convergence is not yet fully proven in several situations. Indeed, $N$-body simulations are not free from biases, due to the discrete nature of the representation of the phase-space fluid with very massive macro-particles compared to the actual dark matter candidates. Discrete noise and close $N$-body encounters can drive the system away from the mean field limit embodied by the Vlasov equation \citep[e.g.,][but this list of references is far from comprehensive]{Aarseth88,Goodman93, Knebe00,Binney04,Diemand04,Joyce09,Colombi15b, beraldo17,beraldo19,Benhaiem18}, and this is particularly critical in the cold case \citep[][]{Melott97,Splinter98,Melott07,Wang07,Angulo13}. This is why it has been proposed recently to adopt a direct approach where the phase-space sheet is described in a smooth fashion with a fine tetrahedral tessellation \citep[][]{Hahn13,SC16,Hahn16}, an idea stemming directly from the waterbag approach \citep[e.g.,][]{DePackh62, Janin71,Cuperman71a,Cuperman71b,ColombiTouma08,ColombiTouma14}.\footnote{Note that neutrinos, not discussed here, compose a small fraction of dark matter. They have a large initial velocity dispersion which requires a different numerical approach from what is used for cold dark matter: in this case, direct Vlasov solvers rely on an Eulerian representation of the phases-space density on a six-dimensional mesh \citep[see, e.g.,][]{Yoshikawa13,Yoshikawa20}.}

In this article, we shall compare in details numerical results obtained from state of the art Vlasov simulations realised with the public code {\tt ColDICE} \citep[][]{SC16}\footnote{see {\tt www.vlasix.org} and\\ {\tt https://github.com/thierry-sousbie/dice}} to $N$-body simulations realised with a standard Particle-Mesh (PM) code \citep[e.g.,][]{Hockney88}. Particular attention will be paid to effects of force resolution and mass resolution. We shall confirm again that with the proper choice of parameters, basically typically more than one particle per softening length of the force, {\em the $N$-body approach remains perfectly reliable at the coarse level}. 

As the host of galaxies and clusters of galaxies, dark matter halos represent the main bricks of large-scale structure formation models and will be the main focus of this numerical investigation. Following the path of many previous works, this article will explore again their dynamical history, which, according to perturbation theory and cosmological simulations results, can be decomposed into the four following phases:
\begin{enumerate}
\item[(i)]  {\it The pre-collapse phase.} At the beginning, the phase-space sheet does not self-intersect in configuration space and its evolution can be followed analytically with a perturbative approach \citep[e.g.,][]{Bernardeau02}. For instance, linear Lagrangian perturbation theory, the Zel'dovich approximation \citep[][]{Zeldovich70}, provides a pretty good description of the evolution of the phase-space sheet, at least from the qualitative point of view. In the Zel'dovich solution, there are locally three orthogonal directions of motion: the first nonlinear structures to form are two-dimensional sheets orthogonal to one dimensional singularities building up along the main direction of motion. Higher-order Lagrangian perturbation theory \citep[e.g.,][]{Bouchet92,Buchert93,Buchert93b,Bouchet95,Rampf12,Zheligovsky14,Matsubara15} is needed for an accurate description of the pre-collapse motion \citep[e.g.,][]{Moutarde91}. At sufficiently high order, it has been shown, using the Vlasov simulations presented in the present work, to perform very well until shell-crossing \citep[][]{Saga18}.
   \vskip 0.2cm
\item[(ii)] {\it The early violent relaxation phase.} After shell-crossing, the system enters into a complex multi-stream violent relaxation phase. If collapse happens along another direction of motion, a filament forms. Then, if another shell crossing takes place along the third direction of motion, this creates the seed of a dark matter halo. The early stages of the dynamics thus build singular structures of various kinds according to the local properties of the displacement field \citep[][]{Arnold82,Hidding14,Feldbrugge18}. They correspond to foldings of the dark matter sheet in phase-space and quickly produce a very intricate spiral structure that we will study in details through phase-space slices and of which the complexity will be quantified in {\tt ColDICE} in terms of total volume and simplices count.

This early relaxation process is of monolithic nature, that is happens in the absence of merger. It takes place over short time scales and is very similar to the picture described in \citet[][]{LyndenBell67}, which explains the term of ``violent relaxation''.  During this phase, the dark matter ``protohalos'' build up a power-law density profile $\rho(r) \propto r^{-\alpha}$ of which the logarithmic slope $\alpha$ ranges in the interval $[1.3,1.7]$ according to various results in the literature \citep[][]{Moutarde91,Diemand05,Ishiyama10, Anderhalden13,Ishiyama14,Angulo17,Delos18a,Delos18b}. The exact value of the slope changes slightly according to the authors, although a consensus seems to emerge with $\alpha \simeq 1.5$. We shall examine this in detail again, by making sure that the measurements are performed during the {\em monolithic} phase of the evolution of halos extracted from CDM simulations with very small box size. Additionally, following the footsteps of \citet[][]{Nakamura85}, \citet[][]{Moutarde91} and \citet[][]{Moutarde95}, we shall study highly symmetric configurations with three sine waves initial conditions of various amplitudes. These investigations will be conducted with {\tt ColDICE} and the PM code, with thorough comparisons between both solvers. 
  \vskip 0.2cm
\item[(iii)]  {\it The convergence to a universal profile.} The initial power-law profile does not last for long because dark matter halos are the object of perturbations, in particular successive mergers with other halos. These mergers modify the profile that converges rapidly to a dynamical attractor, the so-called NFW profile \citep[][]{NFW96,NFW97}, where the radial density profile of dark matters halos has a form close to
  \begin{equation}
    \rho(r) \propto \frac{1}{(r/r_{\rm s})\,(1+r/r_{\rm s})^2}, 
      \label{eq:NFWpr}
    \end{equation}
with $r_{\rm s}$ a scale radius.  
The shape of dark matter halos and the form given by equation (\ref{eq:NFWpr}) have been the object of numerous discussions and convergences analyses \citep[e.g.,][]{Moore98,Jing00,Power03,Mansfield20}. Recent investigations suggest for instance that the Einasto profile that will be used in the present work \citep[equation \ref{eq:einasto} below,][]{Einasto65} provides better fits of dark matter halos profiles \citep[e.g.,][]{Navarro04}, but other forms have been suggested \citep[e.g.,][]{Stadel09}.

Even though it is still debated, the nearly universal nature of the dynamical attractor seems now unquestionable and the fact that mergers contribute to its establishment has been highlighted in several works \citep[e.g.,][]{Syer98,Ishiyama14,Ogiya16,Angulo17}. However, other numerical investigations suggest that it is possible to reach the dynamical attractor even without merger, as the consequence of radial instabilities or other sources of noise \citep[e.g.,][]{Huss99,MacMillan06,Ogiya18}. We shall revisit these questions in the presence and in the absence of merger, by pushing further in time the simulations described in point (ii). However, due to its adaptive nature, the computational cost of {\tt ColDICE}  prevents it from reaching sufficiently advanced stages of this dynamical phase, which will be approached solely with the PM code, after careful mass resolution tests. 

Another interesting property of the dynamical attractor is that the measured pseudo phase-space density,
\begin{equation}
  Q(r) \equiv \frac{\rho(r)}{\sigma_v(r)^3},
  \label{eq:Qofr}
\end{equation}
where $\sigma_v(r)$ is the local velocity dispersion, follows a pure power-law behaviour $Q(r) \propto r^{-\alpha_Q}$ over a very large dynamical range, with a logarithmic slope very close to the prediction of the secondary spherical infall model of \citet[][]{Bertschinger85}, $\alpha_Q=1.875$ \citep[][]{Taylor01,Navarro10,Ludlow10}. We shall see that this property stands as well during the monolithic phase (ii), for which very few studies of the pseudo phase-space density exist \citep[see however][]{Ishiyama10}.
 \vskip 0.2cm
\item[(iv)] {\it The quasi-static evolution.} Once the dynamical attractor is attained, the evolution of the halo becomes mainly monolithic again and its profile changes very little with time. 
\end{enumerate}
Note that, if on one hand, the pre-collapse phase (i) of the evolution of the phase-space sheet is well described by perturbation theory, on the other hand, the early violent relaxation phase (ii) and the convergence to the universal dynamical attractor (iii) are still poorly understood from the theoretical point of view despite numerous investigations,  involving maximum of entropy methods \citep[e.g.,][]{LyndenBell67, Hjorth10,Pontzen13,Carron13}, Jeans' equation \citep[e.g.,][]{Taylor01,Dehnen05,Ogiya18}, post-collapse perturbation theory \citep[e.g.,][]{Colombi15a,Taruya17,Rampf19} as well as self-similar solutions and secondary infall model \citep[e.g.,][]{Fillmore84, Bertschinger85,Henriksen95,Sikivie1997,Zukin10a,Zukin10b,Alard13,Sugiura20}. It is not the goal in the present work to investigate in detail analytical models, but some links between our measurements and predictions from self-similarity and solutions of Jeans' equations will be discussed. 

This article is organized as follows. Details about {\tt ColDICE} and the PM code are provided in section \ref{sec:simulations}, as well as the simulations suite realised for this project and other ongoing works. Then, sections \ref{sec:visudens} and \ref{sec:phaspasec} are respectively dedicated to visual inspection of the three-dimensional projected density and phase-space diagrams. Section \ref{sec:comp} measures complexity in the Vlasov simulations through plots of simplices counts and phase-space sheet volume.
This is followed in section \ref{sec:profiles} by a detailed examination of radial density profiles as well as the pseudo phase-space density. Finally, section \ref{sec:conclusion} summarises the main results and discusses a few prospects. 
\section{The simulations}
\label{sec:simulations}
This section is divided in three parts. The important features of the Vlasov solver {\tt ColDICE} are first summarised in \S~\ref{sec:coldicedes}. Full technical detail of the implementation of this massively parallel algorithm can be found in \citet[][]{SC16}. Then, the PM code written for this project is described in \S~\ref{sec:PMcode}. Finally, the simulation suite used in this work and other investigations \citep[][]{Saga18,Saga20,Sobolevski20} is described in \S~\ref{sec:simusuite}. The parameters of these numerical experiments are listed in Table~\ref{tab:tabsim}. 
\begin{table*}[htp!]
\begin{tabular}{lcccl}
\hline
  Designation          &      Initial conditions             &  $n_{\rm g}$ &    $n_{\rm s}$ or $n_{\rm p}$      &  Type \\ \hline
  {\it Quasi 1D}               &                                               &                    &                                                    & \\
  VLA-Q1D-HR           &     $\vec{\epsilon}=(1/6,1/8)$       &  512            &  256                                           & Vlasov, $I=10^{-6}$ \\
  VLA-Q1D-MRa         &      ''                                       &  256            &  256                                           & Vlasov, $I=10^{-6}$ \\
  VLA-Q1D-MR           &    ''                                         &  256            &  128                                          & Vlasov, $I=10^{-6}$ \\
  VLA-Q1D-LR           &     ''                                         &  128             &  64                                            & Vlasov, $I=10^{-5}$ \\
   PM-Q1D-UHR           &     ''                                       &  1024            &  1024                                            & PM \\
  \hline
  {\it Anisotropic  1}         &                                               &                    &                                                    & \\
 VLA-ANI1-HR           &     $\vec{\epsilon}=(5/8,1/2)$       &  512           &  256                                        & Vlasov, $I=10^{-6}$, Shift \\ 
  PM-ANI1-HR           &     ''                                        &  512           &  512                                        & PM  \\ \hline
  {\it Anisotropic  2}         &                                               &                    &                                                    & \\
   VLA-ANI2-UHR           &     $\vec{\epsilon}=(3/4,1/2)$       &  1024            &  512                                         & Vlasov, $I=10^{-7}$ \\
  VLA-ANI2-FHR           &     ''                                        &  512            &  512                                           & Vlasov, $I=10^{-7}$ \\
  VLA-ANI2-HR           &    ''                                           &  512            &  256                                           & Vlasov, $I=10^{-6}$ \\
  VLA-ANI2-HRS           &   ''                                           &  512            &  256                                           & Vlasov, $I=10^{-6}$, Shift \\
  VLA-ANI2-MRa         &    ''                                           &  256            &  256                                           & Vlasov, $I=10^{-7}$ \\
  VLA-ANI2-MR           &     ''                                         &  256            &  128                                          & Vlasov, $I=10^{-6}$ \\
  VLA-ANI2-LRa           &     ''                                          &  128             &  256                                            & Vlasov, $I=10^{-7}$ \\
  VLA-ANI2-LR           &     ''                                          &  128             &  64                                            & Vlasov, $I=10^{-5}$ \\
  PM-ANI2-UHR           &     ''                                        &  1024           &  1024                                      & PM  \\ 
  PM-ANI2-HR           &     ''                                        &  512           &  512                                        & PM  \\ 
  PM-ANI2-MR          &     ''                                         &  256           &  512                                        & PM  \\ 
  PM-ANI2-LR          &     ''                                         &  128          &  512                                        & PM  \\ 
  PM-ANI2-HR-D8          &     ''                                        &  512          &  256                                        & PM  \\ 
  PM-ANI2-HR-D64        &     ''                                        &  512         &  128                                       & PM  \\  
  \hline
  {\it  Anisotropic  3}         &                                               &                    &                                                    & \\
 VLA-ANI3-HR           &     $\vec{\epsilon}=(7/8,1/2)$       &  512           &  256                                         & Vlasov, $I=10^{-6}$, Shift \\ 
   PM-ANI3-HR           &     ''                                        &  512           &  512                                        & PM  \\ \hline
  VLA-SYM-FHR           &     $\vec{\epsilon}=(1,1)$              &  512            &  512                                           & Vlasov, $I=10^{-7}$ \\
  VLA-SYM-HR           &    ''                                           &  512            &  256                                           & Vlasov, $I=10^{-6}$ \\
  VLA-SYM-HRS           &   ''                                           &  512            &  256                                           & Vlasov, $I=10^{-6}$, Shift \\
  VLA-SYM-MRa         &    ''                                           &  256            &  256                                           & Vlasov, $I=10^{-7}$ \\
  VLA-SYM-MR           &     ''                                         &  256            &  128                                          & Vlasov, $I=10^{-6}$ \\
  VLA-SYM-LR           &     ''                                          &  128             &  64                                            & Vlasov, $I=10^{-5}$ \\ 
  PM-SYM-UHR           &     ''                                        &  1024          &  1024                                       & PM  \\ 
  PM-SYM-HR           &     ''                                        &  512          &  512                                       & PM  \\ \hline
{\it CDM halos 1, 2}      &                                               &                    &                                                    & \\
  VLA-CDM12.5-HR &   CDM, $L=12.5$ pc$/h$      &  512            & 256                                         & Vlasov, $I=10^{-6}$ \\
  VLA-CDM12.5-MR &   ''                                         &  256            & 128                                         & Vlasov, $I=10^{-6}$ \\
  VLA-CDM12.5-LR &   ''                                          &  128            & 64                                        & Vlasov, $I=10^{-6}$ \\
  PM-CDM12.5-HR &   ''                                          &  512            & 512                                        & PM \\ \hline
 {\it CDM halos 3, 4, 5}        &                                               &                    &                                                    & \\
  VLA-CDM25-HR &   CDM, $L=25$ pc$/h$      &  512            & 256                                         & Vlasov, $I=10^{-6}$ \\
  VLA-CDM25-MR &   ''                                         &  256            & 128                                         & Vlasov, $I=10^{-6}$ \\
  VLA-CDM25-LR &   ''                                          &  128            & 64                                        & Vlasov, $I=10^{-6}$ \\
  PM-CDM25-HR &   ''                                          &  512            & 512                                        & PM \\ \hline
 \end{tabular}
 \caption[]{Details on the ensemble of simulations performed for this work. The first column corresponds to the designation of the run. The second column gives the type of initial conditions, namely the relative amplitudes $\epsilon_i$ of the initial sine waves or the size $L$ of the box for the CDM simulations.  The third column indicates the spatial resolution $n_{\rm g}$ of the grid used to solve Poisson equation.  The fourth one mentions the spatial resolution $n_{\rm s}$ of the mesh of vertices used to construct the initial tessellation for the Vlasov runs or the number of particles $n_{\rm p}^3$ for the PM runs. Finally, the fifth column specifies which kind of code was used, as well as the value of the parameter $I$ used to bound violation to conservation of Poincar\'e invariants in the case of the Vlasov runs (equations~\ref{eq:poinca} and \ref{eq:poincac}). It also mentions, when relevant, if a small shift was applied to vertices positions in initial conditions.}
\label{tab:tabsim}
\end{table*}  
\subsection{Brief description of {\tt ColDICE}}
\label{sec:coldicedes}
In {\tt ColDICE}, the phase-space sheet is tessellated with an ensemble of tetrahedra (also called simplices) of which the vertices (the 4 corners of the tetrahedra) are initially disposed on a regular mesh of size $n_{\rm s}$ corresponding to $6n_{\rm s}^3$ simplices. Initial comoving positions $\vec{x}_{i,j,k}$, $i,j,k \in [1,\cdots,n_{\rm s}]$ and peculiar velocities $\vec{v}_{i,j,k}$ of the vertices are given by
\begin{eqnarray}
  \vec{x}_{i,j,k}(t=0) &\equiv& \vec{q}_{i,j,k}=\frac{L}{n_{\rm s}}\, (i, j,k), \label{eq:inigrida} \\
  \vec{v}_{i,j,k}(t=0) &=& (0,0,0), \label{eq:inigridb}
\end{eqnarray}
with $L$ the size of simulation volume, which is a periodic cube. The unperturbed positions define Lagrangian coordinates $\vec{q}_{i,j,k}$ of each vertex. These phase-space coordinates are slightly perturbed according to linear Lagrangian perturbation theory \citep[][]{Zeldovich70},
\begin{eqnarray}
  \vec{x}(t=t_{\rm i})&=&\vec{q} + D_+(t_{\rm i})\, \vec{P}(\vec{q}), \label{eq:linlag1}\\
  \vec{v}(t=t_{\rm i})&=&a(t_{\rm i})\,\frac{{\rm d} D_+}{{\rm d}t_{\rm i}}\, \vec{P}(\vec{q}), \label{eq:linlag2}
\end{eqnarray}
where we have dropped the subscripts $(i,j,k)$, $a$ and $D_+$ are respectively the expansion factor and the linear growing mode normalised to unity at present time, $\vec{P}$ the linear displacement field.

Then, the tessellation evolves dynamically under self-gravity by solving standard Lagrangian equations of motion for its vertices, similarly as for particles in a $N$-body simulation, with simple second order predictor corrector with slowly varying time step.

To compute the gravitational force, the tessellation is interpolated on a regular mesh of spatial resolution $n_{\rm g}$. This interpolation is performed by calculating the exact intersection between each simplex of the tessellation and each voxel cell) of the mesh.\footnote{A voxel is the 3D alter-ego of a pixel in 2D.} It is performed at linear order, which means that it accounts for the gradient of the volume density of the phase-space sheet inside each simplex. Once the three-dimensional projected density field is obtained, Poisson equation is solved in Fourier space. Calculation of the force field is performed with a standard four point finite difference stencil to compute the gradient of the gravitational potential. Finally, the force is interpolated to each vertex of the tessellation with second order Triangular Shaped Cloud (TSC) interpolation \citep[][]{Hockney88}.

As a time variable, {\tt ColDICE} uses the ``superconformal'' time $\tau$ given by
\begin{equation}
  {\rm d}\tau \equiv H_0\, \frac{{\rm d}t}{a^2},
\end{equation}
with $H_0$ the Hubble constant \citep[e.g.,][]{Doroshkevich73,Martel98}.
Time step constraints combine the classical Courant-Friedrichs-Lewy (CFL) condition,
\begin{equation}
  \Delta \tau \leq C_{\rm CFL}\, \frac{L}{v_{\rm max}\,n_{\rm g}},
  \label{eq:CFL}
\end{equation}
with $C_{\rm CFL}=0.25$,
and the optional additional dynamical condition
\begin{equation}
  \Delta \tau \leq \frac{C_{\rm dyn}}{\sqrt{\frac{3}{2}\, \Omega_{\rm m}\, a\, \rho_{\rm max}}},
  \label{eq:Crho}
\end{equation}
with $C_{\rm dyn}=0.01$, where $\Omega_{\rm m}$ is the matter density parameter of the Universe and $\rho_{\rm max}$ the maximum of the projected density (normalised to unity) computed in the mesh used to solve Poisson equation. Furthermore, to avoid too large variations of the expansion factor at early times and have correct behaviour of the system in the (quasi-)linear regime, the additional condition $\Delta \tau \leq 0.1 a ({\rm d}a/{\rm d}\tau)^{-1}$ is enforced. For most simulations, the dynamical condition (\ref{eq:Crho}) was ignored, except for the Vlasov runs with $n_{\rm g}=512$ and $n_{\rm s}=512$ in Table~\ref{tab:tabsim}, as well as VLA-ANI2-UHR, VLA-ANI2-MRa and VLA-SIM-MRa. It was indeed found in practice that condition (\ref{eq:Crho})  did not bring significant improvements on the results during the period of time covered by the Vlasov runs. 

During evolution, the phase-space sheets gets more and more intricate with time. In order to follow all the details of its complexity, local refinement with the bisection method is implemented in {\tt ColDICE} using a local quadratic interpolation of the tessellation mesh.\footnote{In the bisection method, refinement is performed on edges $[a,b]$ of tetrahedra, which are split in two segments $[a,v]$ and $[v,b]$ through the creation of a new vertex $v$. Then all the incident tetrahedra $i$, composed of vertices $[a,b,c_i,d_i]$, are split in two tetrahedra, composed of vertices $[a,v,c_i,d_i]$ and $[b,v,c_i,d_i]$. To preserve accuracy of refinement, a quadratic representation of the phase-space sheet inside each simplex is used, with the help of additional tracers corresponding to mid-points of the edges of each tetrahedron in Lagrangian space (that is in the space of initial, unperturbed positions). These tracers are actually used as the candidate refinement vertices. New tracers are created each time refinement is performed, by exploiting the local quadratic representation of the sheet. This refinement procedure preserves the conforming nature of the tessellation, by matching up vertices, edges and faces at the intersection between two tetrahedra, without hanging nodes, that is without an isolated vertex belonging to one tetrahedron on an edge, a face or in the bulk of another tetrahedron.} This anisotropic refinement attempts to preserve in the best way possible the Hamiltonian nature of the motion, by bounding local Poincare invariants
\begin{equation}
  I_{\rm p}=\frac{1}{H_0\,L^2}\left| \oint \vec{v}.{\rm d}\vec{x} (s) \right|
  \label{eq:poinca}
\end{equation}
measured over faces of candidate tetrahedra obtained from refinement. Because the initial unperturbed tessellation (eqs.~\ref{eq:inigrida} and \ref{eq:inigridb}) has strict zero velocity, we should have at all times $I_{\rm p}=0$ for any closed contour inside the phase-space sheet. Refinement is locally performed so that $I_{\rm p}$ remains very small:
\begin{equation}
  I_{\rm p}({\rm faces}) \leq I.
  \label{eq:poincac}
\end{equation}
Various values of $I$ used in the {\tt ColDICE} simulations are listed in last column of Table~\ref{tab:tabsim}, and range in the ensemble $\{10^{-5},10^{-6},10^{-7}\}$. Note that the choice of $I$ should somewhat relate to spatial resolution  $n_{\rm g}$, since the latter controls softening of the force field, which sources curvature of the phase-space sheet. 
\subsection{Brief description of the PM code}
\label{sec:PMcode}
The particle-mesh code \citep[PM, e.g.,][]{Hockney88} written for this project is standard. Using shared memory parallelisation with {\tt OpenMP}, it follows a set of $n_{\rm p}^3$ particles in a mesh of resolution $n_{\rm g}$ to solve Poisson equation. The initial conditions are the same for the PM particles as for the vertices in {\tt ColDICE}: particles are set on a regular network according to equations (\ref{eq:inigrida}) and (\ref{eq:inigridb}), with a slight perturbation on initial positions and velocities according to equations (\ref{eq:linlag1}) and (\ref{eq:linlag2}). The implementation of equations of motion of the particles is also exactly analogous to what is done for {\tt ColDICE} vertices, with the same constraints on the time step, except that condition (\ref{eq:Crho}) was systematically enforced using $C_{\rm dyn}=0.1$.

The main difference between the PM code and {\tt ColDICE} lies in the way Poisson equation is solved: first, the three-dimensional density is estimated on the computational mesh by projecting the particles using a TSC interpolation. Second, at variance with {\tt ColDICE}, an apodization of the Green function $G(k)$ is performed with a Hanning filter in order to reduce small scale anisotropies,
\begin{equation}
  G(k)= \frac{1}{2\, k^2} \left[ 1+\cos(L\, k) \right],
  \label{eq:hanning}
\end{equation}
where $k$ is the wavenumber. Otherwise all the other steps of force field calculation are exactly the same as in {\tt ColDICE}, including TSC interpolation of the force on the particles, exactly as what is done in {\tt ColDICE} for the vertices.

The other difference is that we do not perform local Lagrangian refinement, that is the number of particles does not change with time. The PM code is therefore equivalent to evolving the initial vertices of {\tt ColDICE} and let them carry the mass instead of the simplices.

The TSC interpolation used to compute the density along with Hanning filtering (\ref{eq:hanning}) help to reduce effects of the discrete noise of the particles, in particular local anisotropies, but add significant softening of the force field in the PM code compared to {\tt ColDICE}. Calculation of the density on the computational mesh in {\tt ColDICE} indeed corresponds in practice to Nearest Grid Point (NGP) interpolation \citep[e.g.,][]{Hockney88} in terms of convolution, so one can expect a loss of effective force resolution of the PM code compared to {\tt ColDICE} when using the same value of $n_{\rm g}$, which will turn in practice to be of about a factor 1.7 in the subsequent analyses. This approximate factor is obtained by comparing the force field generated by a particle interpolated on the grid with NGP scheme to the one obtained with TSC interpolation (which brings a factor 1.1 of effective resolution loss) combined with Hanning filtering (which brings a factor 1.5 of effective resolution loss). Note that a more complete comparison between the $N$-body approach and {\tt ColDICE} could include PM simulations without Hanning filtering or/and with Cloud-in-Cell interpolation \citep[see, e.g.,][]{Hockney88} instead of TSC interpolation. We leave this for future work as we believe that such additional analyses are not necessary to prove the main points of this article.

\subsection{The simulation suite}
\label{sec:simusuite}
We consider two kinds of initial conditions, CDM and three sine waves, as detailed in the next two sections.
\subsubsection{The ``CDM'' simulations}
For the CDM simulations, which start from fluctuations originating from a smooth random Gaussian field, initial vertices/particle positions and velocities are computed with the public software {\tt MUSIC} \citep[][]{Hahn11} used at linear order. It is important to notice that, because a quadratic representation of the phase-space sheet is used, it is needed to define special tracer vertices in initial conditions. Hence, a mesh of $n_{\rm p}^3$ PM particles corresponds to $n_{\rm s}^3$ vertices of the actual tessellation, with $n_{\rm s}=n_{\rm p}/2$, while the remaining $n_{\rm p}^3-n_{\rm s}^3$ vertices are used as tracers.

The assumed cosmological parameters of the CDM runs are the following \citep[][]{Planck18}: total matter density $\Omega_{\rm m}=0.315$, cosmological constant density parameter $\Omega_{\rm L}=0.685$, baryon density parameter $\Omega_{\rm b}=0.0493$, Hubble constant $H_0 \equiv 100\,h=67.4$ km/s/Mpc, rms density fluctuations in a 8 Mpc$/h$ sphere linearly extrapolated to present time $\sigma_8=0.811$, power law index of the density perturbation spectrum after inflation $n_{\rm spec}=0.965$. We use the extension of {\tt MUSIC} of \citet[][]{Angulo17} to have a transfer function consistent with a neutralino of mass $100$ GeV and a decoupling temperature of $T=30$ MeV.

Two simulations box sizes are considered, $L=12.5$ and $25$ pc$/h$, such that initial density fluctuations are smooth over several spatial resolution scales $L/n_{\rm g}$ and $L/n_{\rm s}$ of the computational mesh and of the initial tessellation.  These box sizes are obviously unrealistically small since we are not using any resimulation technique to account for tidal forces coming from scales larger than $L$. This is why in the figures, ``CDM'' is put in ``quotes'', because the halos extracted from these simulations most probably have a non-representative merger history. However, for the purpose of the present work, the random nature of the field is enough to have a qualitative idea on how the phase-space pattern changes compared to the highly symmetric case represented by the three sine waves case described below.

To make sure that transients related to the fact that only linear Lagrangian theory was used to set up initial conditions do not contaminate the measurements, the simulations are started at very high redshift, $a_{\rm i}=10^{-5}$ for the simulations with $n_{\rm g}=512$ and $a_{\rm i}=10^{-4}$ for those with lower spatial resolution.

\begin{figure*}[htp!]
\centering
\includegraphics[width=17.5cm]{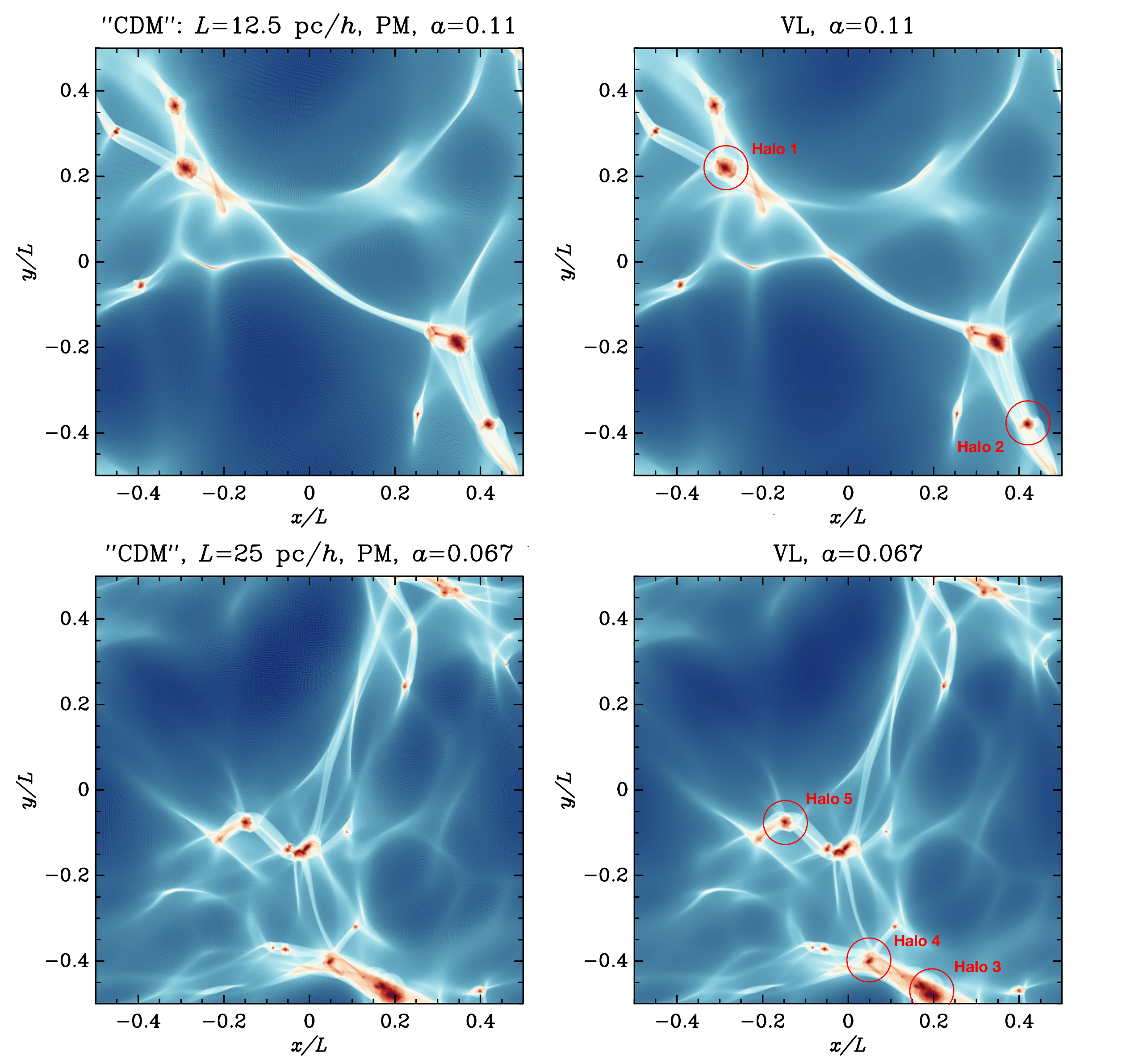}
\caption[]{Total projected density on $(x,y)$ plane of the ``CDM'' simulations: comparison of {\tt ColDICE} to PM. The color table, logarithmic, changes from dark blue to white and then dark red when the density increases. The left and right panels correspond respectively to PM and {\tt ColDICE} simulations, while top and bottom ones correspond respectively to a box size $L=12.5$ and $25$ pc$/h$. The simulations considered here are designated by PM-CDM12.5-HR (top left), PM-CDM25-HR (bottom left), VLA-CDM12.5-HR (top right) and VLA-CDM25-HR (bottom right) in Table~\ref{tab:tabsim}. The expansion factor indicated on each panel corresponds to the last snapshot of the Vlasov runs. Additionally, on the right panels, circles highlight the halos selected for detailed analyses.}
\label{fig:CDM_allbox}
\end{figure*}

Figure \ref{fig:CDM_allbox} shows the projected density over the whole simulation volume in the highest resolution Vlasov runs, with $n_{\rm g}=512$ and $n_{\rm s}=256$, and in the PM runs, with $n_{\rm g}=512$ and $n_{\rm p}=512$. At this level of detail, the differences between PM and {\tt ColDICE} are nearly invisible, but one can guess a faint signature of the particle pattern in underdense regions on left panels.

\begin{figure*}[htp!]
\centering
\includegraphics[width=18.3cm]{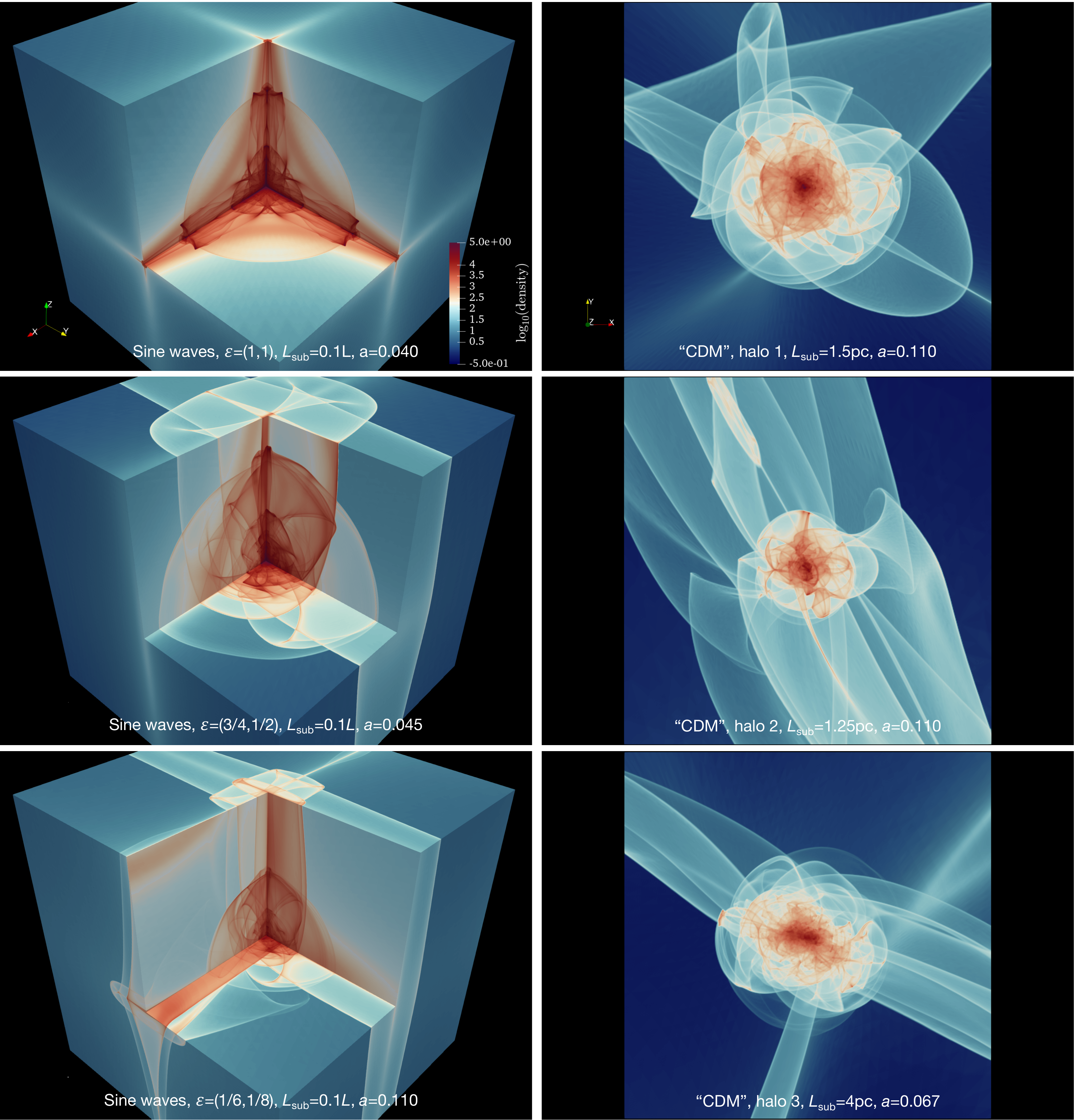}
\caption[]{Three dimensional spatial density $\rho(\vec{x})$ in typical configurations of protohalos studied in this work for the final snapshots of our highest force resolution Vlasov runs. The left panels correspond to initial conditions given by 3 crossed sine waves: from top to bottom,  VLA-SYM-HR with $\vec{\epsilon}=(1,1)$, VLA-ANI2-HR with $\vec{\epsilon}=(3/4,1/2)$, and VLA-Q1D-HR with $\vec{\epsilon}=(1/6,1/8)$. The right panels display protohalos extracted from our ``CDM'' runs: top two panels correspond respectively to halo 1 and halo 2, extracted from VLA-CDM12.5-HR; bottom panel corresponds to halo 3, extracted from VLA-CDM25-HR. The size of the subvolume on display as well as the expansion factor value are indicated on each panel. Note that the spatial resolution scale is $\varepsilon_{\rm r}=L/512 \simeq 0.002 L$ on left panels, which represents about $1/51$th of the subcube size; on two upper right panels, $\varepsilon_{\rm r}=L/512 \simeq 0.025$ pc$/h$, which corresponds respectively to about $1/61$th and $1/51$th of the subcube size on top right and middle right panels; on bottom right panel,  $\varepsilon_{\rm r}=L/512 \simeq 0.05$ pc$/h$, which corresponds to about $1/82$th of the subcube size.}
\label{fig:dens_3D_all}
\end{figure*}

Due to the smallness of the simulations volumes, only a few dark matter halo form. Five of them where selected for detailed analyses, as indicated on right panels of Fig.~\ref{fig:CDM_allbox}. The way these halos were extracted from the simulations consisted in simply identifying connected regions of density $\rho(\vec{x})$ larger than 400 in the computational volume sampled with a $512^3$ mesh, with $\rho(\vec{x})$ estimated as explained in Appendix~A.1. 
The centre of each halo was identified with the centre of mass of these regions. Projected density slices of three of these halos at the most evolved stages attained by the highest resolution {\tt ColDICE} runs are shown on right column of Fig.~\ref{fig:dens_3D_all}. 
\subsubsection{The three sine waves simulations}
\label{sec:3sines}
In the three sine waves case, the displacement field in equations (\ref{eq:linlag1}) and (\ref{eq:linlag2}) is given, for $q_{x,y,z} \in [0,L[$, by
\begin{eqnarray}
    \vec{P}=\frac{L}{2\pi} \left[ A_x \sin\left( \frac{2\pi}{L}\, q_x \right), A_y \sin\left( \frac{2\pi}{L}\, q_y \right),
    A_z \sin\left( \frac{2\pi}{L}\, q_z \right) \right],
\end{eqnarray}
where the vector $\vec{A}=(A_x,A_y,A_z)$ with $A_x\geq A_y\geq A_z \geq 0$ quantifies the linear amplitudes of the waves in each direction.

This rather symmetrical set-up is restrictive but remains quite generic. Near the center of the system, it indeed coincides up to quadratic order with the peak of a smooth random Gaussian field \citep[see, e.g.,][]{Bardeen86}. Following the evolution of these three sine wave configurations is thus expected to provide many insights on the dynamics, in particular during the early stages of the formation of dark matter halos, especially for those corresponding to high peaks of the initial field. The high level of symmetry also facilitates calculations of analytical predictions from perturbation theory \citep[][]{Saga18}. Of course, the three sine waves initial conditions remain unrealistic, since the only contribution to the external tidal field is given by the replica of the halo due to the periodic nature of the simulated box. Additionally, the system does not experience any merger in this set-up.
  
Again, the three sine waves simulations are all started at very high redshift with $a_{\rm i}= a(t=t_{\rm i})=0.0005$ to make the contamination by transients negligible, as actually proved by the very accurate comparisons with higher order perturbation theory predictions of \citet[][]{Saga18}. Hence, the important quantity is the relative amplitude of the waves, traced by the vector
\begin{equation}
  \vec{\epsilon}  \equiv \left( \frac{A_y}{A_x}, \frac{A_z}{A_x}\right).
\end{equation}
Five different values of $\vec{\epsilon}$ are considered, as listed in Table~\ref{tab:tabsim}, which defines three kinds of initial conditions: quasi one dimensional (Q1D) with $\vec{\epsilon}=(1/6,1/8)$, where one amplitude dominates over the two other ones, anisotropic (ANI1, ANI2, ANI3), where the amplitude of each wave is different but remains of the same order, and finally, axisymmetric (SYM) with $\vec{\epsilon}=(1,1)$. Information about the nomenclature used in the subsequent analyses is provided in Table~\ref{tab:regimes}, where 3 different regimes and the corresponding values of the expansion factor are introduced. {\em Early time} corresponds to the early violent relaxation phase (ii) described in the Introduction. {\em Mid time} corresponds to the intermediate step during which the system is progressively relaxing to the NFW dynamical attractor [point (iii) in Introduction], attained at what we call {\em late time}.
\begin{table*}[htp!]
\begin{tabular}{lcccc}
\hline
    Designation & Initial conditions: $\vec{\epsilon}$   & early time: $a$ & mid time: $a$ & late time: $a$ \\ \hline
    Q1D             &  $(1/6,1/8)$  &    0.110  &  0.260  & 0.450  \\
    ANI1            &  $(5/8,1/2)$  &    0.050  &   0.100  &  0.200  \\
    ANI2            &  $(3/4,1/2)$  &    0.040 and 0.045  &   0.095  & 0.185  \\
    ANI3            &  $(7/8,1/2)$  &    0.045  &   0.090   & 0.180  \\
   SYM               &  $(1,1)$         &    0.040  &   0.090   & 0.185  \\ \hline                                                 
\end{tabular}
\caption[]{Expansion factor values corresponding to early time, mid time and late time for the three sine waves simulations.}
\label{tab:regimes}
\end{table*}

Left panels of Fig.~\ref{fig:dens_3D_all} display three-dimensional views of the projected density in the central part of the computational volume for the most evolved stages of the highest force resolution Vlasov runs with $\epsilon=(1,1)$  (SYM), $(3/4,1/2)$ (ANI2) and $(1/6,1/8)$ (Q1D). They evidence the complex caustic pattern building up during the early violent relaxation phase, to be compared to the seemingly more intricate case of the CDM protohalos shown on right panels.

Full detail on all the three sine waves simulations is given in Table~\ref{tab:tabsim}. A large number of simulations was performed for extensive force resolution analyses by considering different values of $n_{\rm g}$ and mass resolution analyses by considering different values of $n_{\rm s}$ and $n_{\rm p}$. In addition, as discussed furthermore in sections \ref{sec:visudens} and \ref{sec:phaspasec} below, an asymmetry can appear during runtime in {\tt ColDICE} because of very small but cumulative rounding errors when projecting the tessellation on the computational mesh due to exact superposition between the tessellation and the mesh. To try to remedy this, a shift by half a voxel size is applied to the three sine waves initial conditions for some of the Vlasov runs (indicated as ``Shift'' on right column of Table~\ref{tab:tabsim}), so that vertices of the tessellation do not coincide exactly with edges of the mesh at the centre of the system.
\section{Visual inspection of the projected density}
\label{sec:visudens}
This section focuses on visual inspection of the projected density, $\rho(\vec{x})$. Its main objectives are three-fold and set up the way it is structured. First, through force resolution analyses performed in \S~\ref{sec:sparez}, we show that the caustic pattern of the protohalos remains robust when sufficiently far enough from the center of the system. Second, we want to confirm that the $N$-body approach still provides a good dynamical description of the system when particle density is large enough, despite the artificial pattern that might appear due to the discrete representation. This will be supported by detailed comparisons of the PM simulations to the {\tt ColDICE} simulations, along with a mass resolution analysis performed in \S~\ref{sec:massrez}, that includes, among other things, an analysis of the effect on the long run of changing the number of particles in the $N$-body simulations. Third, in order to be able to interpret the analyses performed in next sections, \S~\ref{sec:vistim} focuses on evolution with time of the projected density for three sine waves runs as well as CDM halos number 1 to 3.
\subsection{Visual inspection: force resolution}
\label{sec:sparez}
\begin{figure*}[htp!!]
\centering
\includegraphics[width=18.3cm]{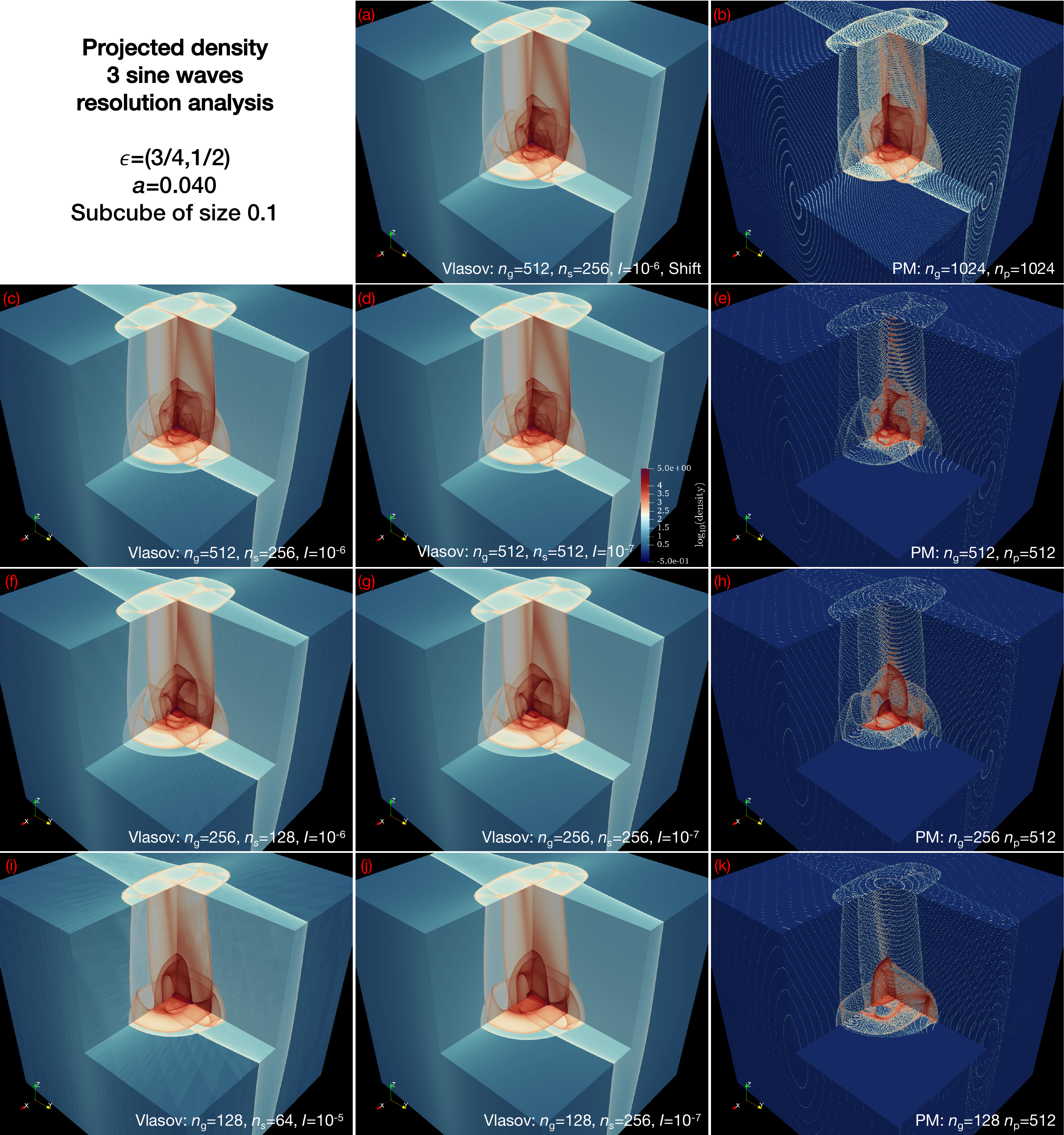}
\caption[]{Thorough force resolution analysis of the three-dimensional projected density $\rho(\vec{x})$ for a three sine waves initial setup with $\vec{\epsilon}=(3/4,1/2)$. A subcube of size $L/10$ is extracted from the snapshot corresponding to expansion factor $a=0.04$ and sampled with $512^3$ voxels. The nature of the run (Vlasov or PM) and its parameters are detailed on each panel, in particular the spatial resolution $n_{\rm g}$  of the grid used to solve Poisson equation. In addition, the resolution $n_{\rm s}$  of the mesh of vertices employed to construct the initial tessellation is indicated for the Vlasov runs and its analog $n_{\rm p}$ for the network of $n_{\rm p}^3$ particles in the PM simulations, along with the value of the refinement control parameter $I$ for Vlasov. The spatial resolution scale $\varepsilon_{\rm r}=L/n_{\rm g}$ represents  about $1/102$th of the subcube size on upper right panel, $1/51$th in middle top panel and second line of panels, and $1/26$th, then $1/13$th, in the two next lines of panels. For completeness, from top to bottom and left to right, the runs considered are designated respectively by VLA-ANI2-HR, VLA-ANI2-MR, VLA-ANI2-LR, VLA-ANI2-HRS, VLA-ANI2-FHR, VLA-ANI2-MRa, VLA-ANI2-LRa, PM-ANI2-UHR, PM-ANI2-HR, PM-ANI2-MR and PM-ANI2-LR in Table~\ref{tab:tabsim}.} 
\label{fig:dens_3D_rez}
\end{figure*}

\begin{figure*}[htp!]
\centering
\includegraphics[width=18.3cm]{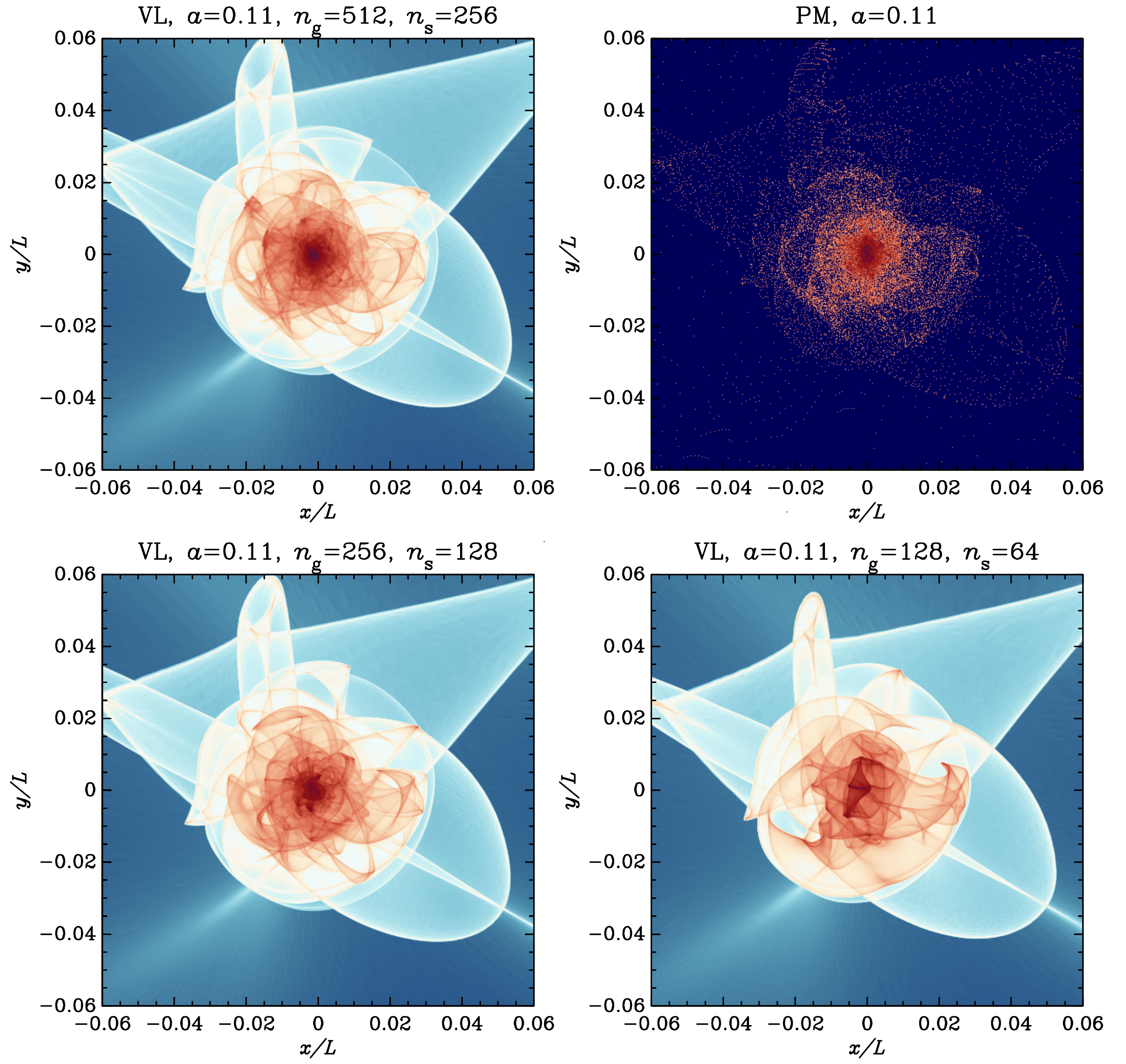}
\caption[]{Example of the effects of force resolution on one of the CDM halos. The three-dimensional projected density $\rho(\vec{x})$ is shown for halo 1 at expansion factor $a=0.11$. The top line of panels displays the results obtained from our high resolution Vlasov run VLA-CDM12.5-HR  (left) and the PM simulation PM-CDM12.5-HR (right), both with $n_{\rm g}=512$, that is a spatial resolution scale $\varepsilon_{\rm r}=L/n_{\rm g}$ of about $1/61$th the displayed slice size. Bottom left and bottom right panels correspond to Vlasov runs with $n_{\rm g}=256$ (VLA-CDM12.5-MR)  and $n_{\rm g}=128$ (VLA-CDM12.5-LR), respectively, hence a spatial resolution scale of about $1/31$th and $1/15$th of the displayed slice size.}
\label{fig:halo1_sl}
\end{figure*}

Figure \ref{fig:dens_3D_rez} shows, at early time, the three-dimensional density $\rho(\vec{x})$ for the three-sine wave simulations with $\vec{\epsilon}=(3/4,1/2)$. A subcube of size $L/10$  is considered, where $L$ is the simulation box size, and is sampled on $512^3$ voxels. Projection of the tessellation is performed in each voxel at linear order, but instead of the exact but complex ray-tracing procedure used in {\tt ColDICE}, we replace each tetrahedron by a dense regular network of particles which are then assigned to the voxels using Cloud-in-Cell interpolation \citep[e.g.,][]{Hockney88}, as detailed in Appendix~A.1. 
For the PM runs, a simpler NGP interpolation is performed on the voxels. 
Figure \ref{fig:halo1_sl} is analogous to Fig.~\ref{fig:dens_3D_rez}, but a zoom on halo 1 is considered. In this case, the subcube considered is of size $0.12 \, L=1.5$ pc/$h$. The color table is almost the same for both figures, spanning logarithmically the density from blue to red in the interval $\log_{\rm 10}\rho \in [-0.5, 5]$ and $[-1,4.5]$ respectively for the three sine waves and the CDM runs. The major differences visible in the color pattern between the Vlasov and the PM runs are obviously mainly due to discreteness effects: only in high density regions, such that there is a sufficient number of particles per sampling voxel, the PM colors become comparable to the Vlasov colors. 

When examining Figs.~\ref{fig:dens_3D_rez} and \ref{fig:halo1_sl}, the first important thing to notice is that PM simulations give very similar results to Vlasov runs if one put asides obvious discreteness effects related to representing the phase-space distribution function with particles. In other words, if we would reconstruct a {\tt ColDICE} like tessellation from the regular pattern used to build the Lagrangian, initial distribution of PM particles \citep[as first proposed by][]{Shandarin12,Abel12}, we would certainly obtain results very close to the Vlasov runs, with small differences due to effective force resolution. As predicted in \S~\ref{sec:PMcode}, because of Hanning filtering and TSC interpolation performed in the PM code, the results obtained for the $N$-body simulations with a given value of $n_{\rm g}\equiv n_{\rm g}^{\rm PM}$ actually lie between Vlasov runs with $n_{\rm g}\equiv n_{\rm g}^{\rm VL}=n_{\rm g}^{\rm PM}/2$ and those with $n_{\rm g}^{\rm VL}=n_{\rm g}^{\rm PM}$ with a best match obtained for $n_{\rm g}^{\rm VL}=n_{\rm g}^{\rm PM}/2$. This comparison shows that for $n_{\rm p} \ga n_{\rm g}$, gravitational dynamics is not significantly influenced by discrete sampling of the phase-space distribution function in the PM runs, at least during the early violent relaxation phase. 

Another important point concerns the caustic structure and force resolutions effects on its pattern. From examination of Figs.~\ref{fig:dens_3D_rez} and \ref{fig:halo1_sl}, we see that the caustic pattern loses complexity when force resolution is degraded, as expected, but is preserved at the coarse level. In particular, outer caustics keep their shape and their position if sufficiently far from the centre of the system, except for the lowest force resolution runs with $n_{\rm g}=128$. In this case, the caustic structures seem slightly less extended, which we can can associate to a less evolved dynamical state as a result of ``excessive'' force softening.

The fact that the caustic structure is mainly influenced by resolution effects near the centre of the system is natural. Indeed, persistent caustics are mainly kinematic objects and local self gravity affects weakly their dynamical evolution. The latter is mainly constrained by the global shape of the potential well in which the caustics evolve, sourced in large part by the singularity building up in the centre of the system. Note that what we call central part does not necessarily reduce to a point: it can also be a line or a surface, when considering the evolution of a structure such as a filament or a pancake.
\subsection{Visual inspection: mass resolution}
\label{sec:massrez}
Examining mass resolution in the Vlasov runs consists in analysing the effects of changing the resolution $n_{\rm s}$ of the mesh of vertices used to build the initial tessellation and combine this with an exploration of the space of values of the parameter $I$ controlling deviations from  local Poincar\'e invariants conservation (equations~\ref{eq:poinca} and \ref{eq:poincac}). Clearly, $I$ is the most important parameter since it controls local mass resolution during runtime by triggering refinement when necessary. However, the choice of $n_{\rm s}$ can influence dynamics in a subtle way. The scale length of fluctuations in the initial conditions need to be captured by the tessellation, which requires a sufficiently large value of $n_{\rm s}$. Indeed, at early time, when the phase-space sheet is nearly flat, tessellation resolution cannot be solely controlled by the refinement parameter $I$. In practice, it is wise to combine the choices of $I$ and $n_{\rm s}$ in a consistent way, depending on the level of accuracy required during runtime. In particular, increasing $n_{\rm s}$ should be associated with a decrease of the value of $I$. Similarly, taking a value of $n_{\rm s}$ very different from the parameter $n_{\rm g}$ controlling force resolution is possible but does not seem wise, for obvious reasons. 

Comparing first and second column of the group of 6 panels on left of Fig.~\ref{fig:dens_3D_rez} can give an indication of the effects of changing mass resolution of the tessellation. We notice, by comparing panels (c) [or (a)]  to (d), as well as (f) to (g), that differences induced by changes in control parameter $n_{\rm s}$ by a factor two and/or $I$ by a factor 10 are small. This is due to the fact that all the simulations considered in this paper already follow practical accuracy constraints derived in \citet{SC16}. Effects of mass resolution can however be distinguished in two bottom left panels, (i) and (j), for which variations in the choice of $n_{\rm s}$ and $I$ are the largest: the caustics are slightly shifted away from the centre on panel (j) which has a value of $n_{\rm s}$ four times larger and a value of $I$ hundred times smaller compared to panel (i).  This is an effect related to curvature: as a consequence of local convexity, the contours of the phase-space sheet get generally closer to the centre of the system when larger linear tetrahedra are used to sample it in order to compute projected density and solve Poisson equation. This effect can cumulate progressively with time and can have consequences on the dynamical properties of the system. Note that if we would use quadratic simplices to perform the projection during runtime, the difference between panels (i) and (j) would probably become undetectable at the visual inspection level. 

\begin{figure*}[htp!]
\centering
\includegraphics[width=13.5cm]{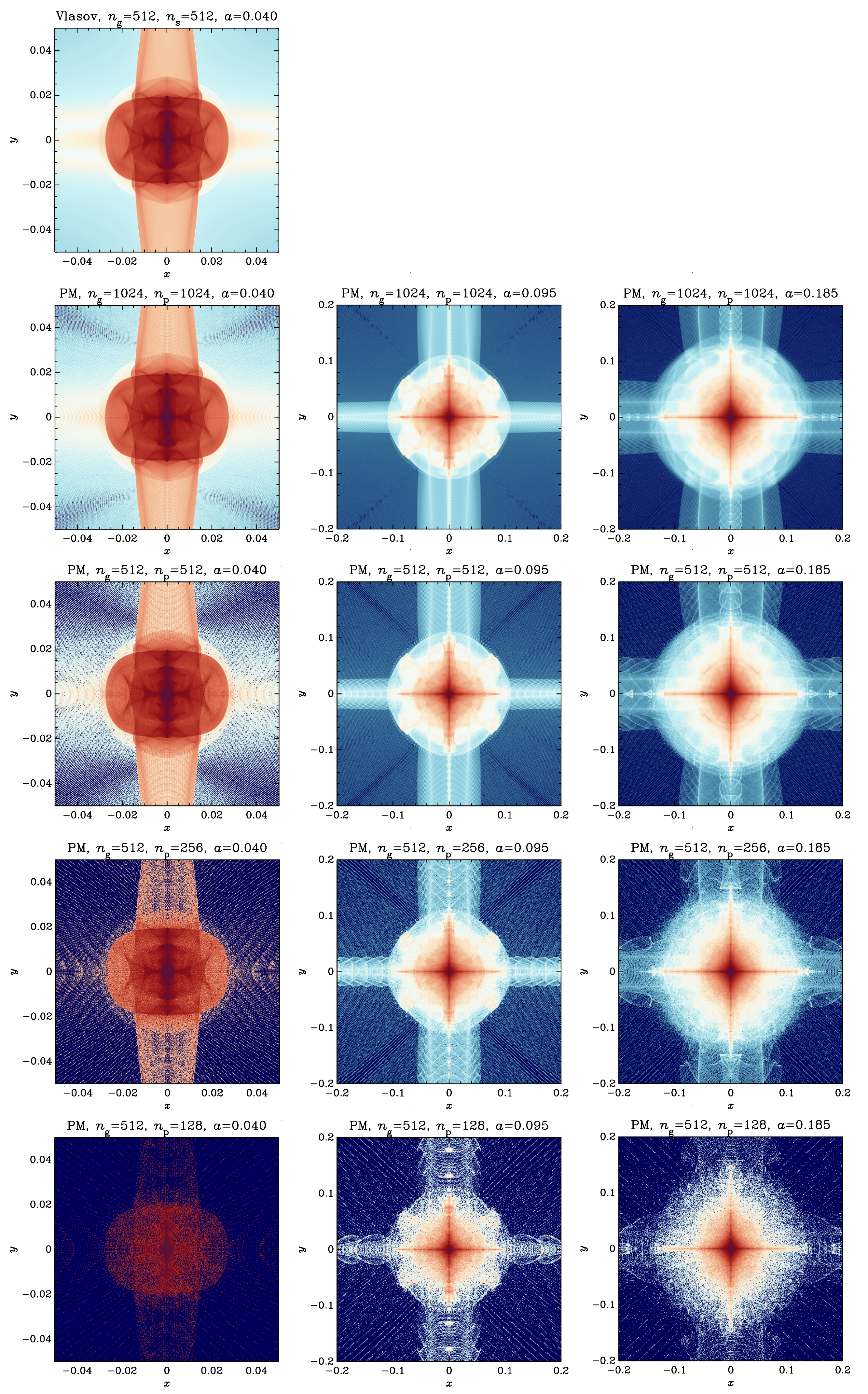}
\caption[]{Effect of mass resolution in the PM runs: total projected density on $(x,y)$ plane of the simulations of three crossed sine waves with $\vec{\epsilon}=(3/4,1/2)$. Except for top panel which stands for last snapshot of the highest resolution Vlasov run, each line of panels corresponds, for various expansion factors,  to a different number of particles, $n_{\rm p}^3$, with $n_{\rm p}=1024$, $512$, $256$ and $128$ when moving down. The spatial resolution is $n_{\rm g}=512$ for all the simulations, except for the first line of panels, which has $n_{\rm g}=1024$.  To help understanding furthermore the figures, remind that the logarithmic color table goes from dark blue to white, then to dark red. The images are computed from the projection on the $(x,y)$ plane of the density calculated on a $512^3$ mesh spanning a cube of size $L_{\rm sub}=L/10$ on left panels and $L_{\rm sub}=L/2.5$ otherwise, by using nearest grid point interpolation. The mass, hence the contribution of each particle, augments with dilution, explaining the change of colour towards dark red of individual particles in underdense regions when $n_{\rm p}$ decreases. From top to bottom, the simulations considered are designated by VLA-ANI2-FHR, PM-ANI2-UHR, PM-ANI2-HR, PM-ANI2-HR-D8 and PM-ANI2-HR-D64 in Table~\ref{tab:tabsim}.}
\label{fig:pdens_sinus_rez}
\end{figure*}

Effects of mass resolution on the PM runs are studied in details in Fig.~\ref{fig:pdens_sinus_rez}, again for the three sine waves initial conditions with  $\vec{\epsilon}=(3/4,1/2)$. The first column of panels corresponds to early time, where data from the Vlasov runs are available, as shown in top left panel. The second and third columns of panels stand for more evolved times, designated by mid time and late time in Table~\ref{tab:regimes}. Mass resolution decreases from top to bottom. Note that only the central part of the simulation is shown, with a zoom in the interval $(x,y,z) \in [-0.05,0.05]$ for left column of panels and $(x,y,z) \in [-0.2,0.2]$ for the two right columns of panels. At variance with previous figures, what is displayed here is the cumulated density over the $z$ line of sight extracted from the interval under consideration.

During early time, all the simulations seem to match each other closely, except for aliasing and discreteness effects due to the representation of the matter distribution with particles. Note the striking agreement between the highest resolution PM run, with $n_{\rm g}=n_{\rm p}=1024$, and {\tt ColDICE}, which has $n_{\rm g}=512$. This agreement deteriorates slightly when decreasing spatial resolution of the PM code to $n_{\rm g}=512$. Again, this is a consequence of force softening due to Hanning filtering and TSC interpolation, as extensively discussed in \S~\ref{sec:PMcode} and \ref{sec:sparez}. Except for this, there does not appear to be any artifact in the dynamics related to $N$-body relaxation, at least when $n_{\rm p} \geq 256$, while it is difficult to make any definitive conclusion in the  case $n_{\rm p}=128$ (bottom left panel).

Similarly, mid time stages of the evolution do not seem to be significantly affected dynamically by discreteness, at least inside the halo. In the filaments, however, we notice the appearance of clumps for $n_{\rm p}\leq 256$ that could be signatures of Jeans instability triggered by the discrete representation, although they could also be a mere signature of the pattern expected from a set of particles following regular orbits derived from the true potential: this has not been checked thoroughly and may require further investigation. Yet it seems obvious that these artificial structures will grow under gravitational instability.

The last column, corresponding to late time, highlights even more visual effects due to the discrete representation. It seems at this point unquestionable that dynamics is affected by $N$-body relaxation for $n_{\rm p} \leq 256$. This is also visible in the halo when examining carefully the highly concentrated cross building up during the course of dynamics: on lower right panels, it presents small fluctuations absent from from top right panels and of which the highly contrasted nature suggests they are sourced by some dynamical instability triggered by shot noise.

Hence, figure~\ref{fig:pdens_sinus_rez} suggests that, at the coarse level, the overal structure of the halo remains the same whatever the value of $n_{\rm p}$ we considered. This will be confirmed by measurements of radial density profiles in \S~\ref{sec:tirho}. Appearance of small artificial clumps due to $N$-body relaxation seems however unquestionable and is nothing fundamentally new \citep[see for instance nice illustrations of this effect in][]{Wang07,Hahn13}. Obviously, $N$-body relaxation is unavoidable, but is delayed when increasing $n_{\rm p}$. The practical condition $n_{\rm p} \ga n_{\rm g}$ suggested in previous works \cite[e.g.,][]{Melott97,Splinter98}  globally agrees well at the qualitative level with visual inspection of Fig.~\ref{fig:pdens_sinus_rez}. 
\subsection{Visual inspection: time evolution}
\label{sec:vistim}
\begin{figure*}[htp!]
\centering
\includegraphics[width=13.5cm]{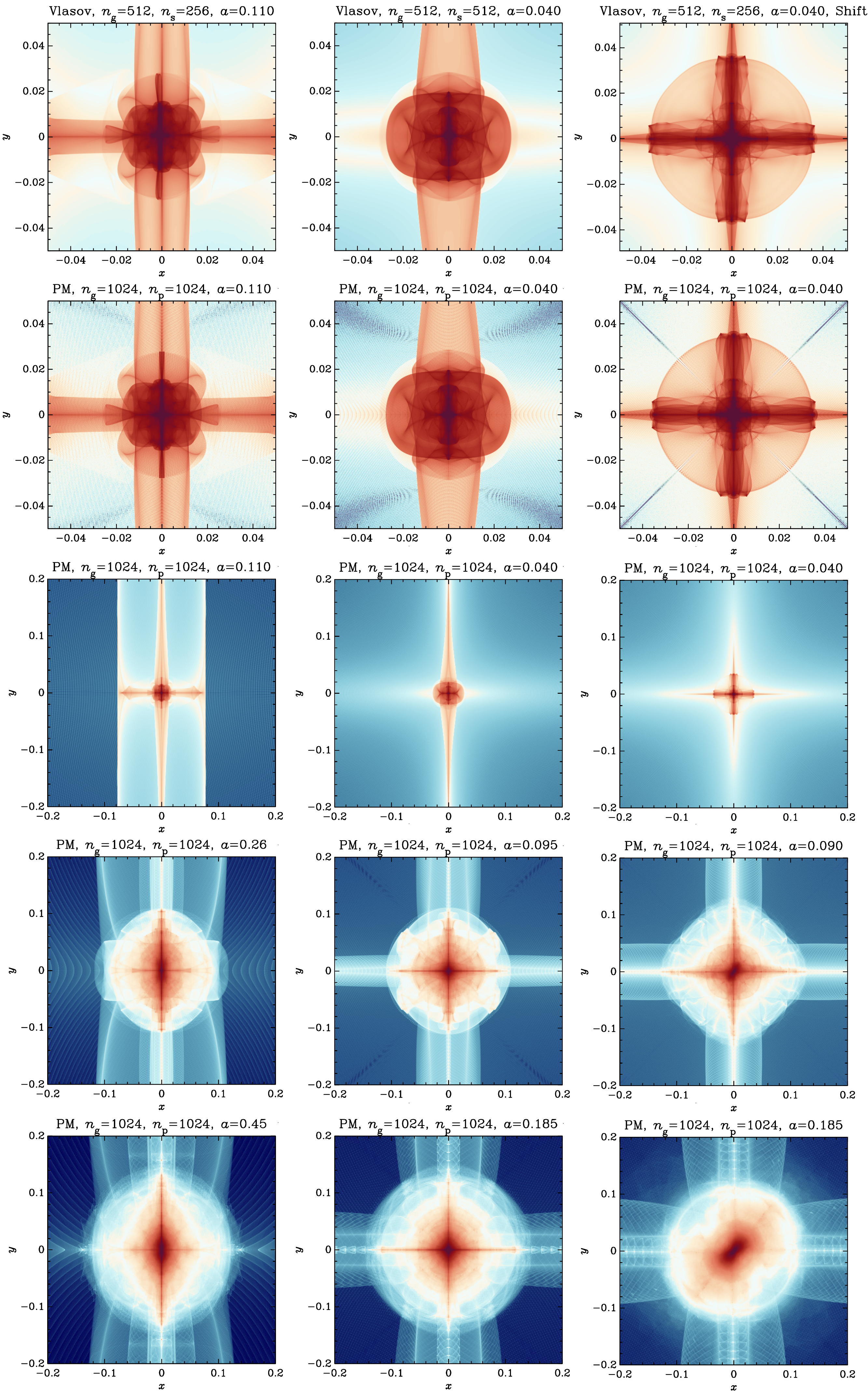}
\caption[]{Evolution of total projected density on $(x,y)$ plane for different configurations of the three sine waves simulations.  The left, middle and right columns of panels correspond respectively to the quasi one dimensional set-up, $\vec{\epsilon}=(1/6,1/8)$, the anisotropic set-up, $\vec{\epsilon}= (3/4,1/2)$, and the axisymmetric set-up, $\vec{\epsilon}= (1,1)$. Top line of panels provides the results obtained at early time from high resolution Vlasov runs with $n_{\rm g}=512$  (VLA-Q1D-HR, VLA-ANI2-FHR and VLA-SYM-HRS in Table~\ref{tab:tabsim}), to be compared to second line of panels, which corresponds to the highest resolution PM runs with $n_{\rm g}=n_{\rm p}=1024$ (PM-Q1D-UHR, PM-ANI2-UHR and PM-SYM-UHR). Third line of panels is the same as second line, but for a larger subcube. Last two lines of panels give the results obtained from the PM runs at mid and late time. Note that the region displayed is only a fraction of the full simulation size, namely $L_{\rm sub}=L/10$ for the 4 top panels and $L_{\rm sub}=L/2.5$ for the 6 bottom panels. Also, the density contributing to the projection comes only from the cubical sub-volume of size $L_{\rm sub}$.}
\label{fig:pdens_sinus}
\end{figure*}

Figure~\ref{fig:pdens_sinus} shows the evolution of the highest force resolution simulations with three initial sine waves of various relative initial amplitudes. 
When comparing first and second line of panels, we observe again the excellent agreement between the PM with $n_{\rm g}=1024$ and the Vlasov runs with twice smaller resolution. Note however the small asymmetry visible on upper left panel, which is due, as discussed in the end of \S~\ref{sec:3sines},  to very small computer rounding errors cumulating with time in {\tt ColDICE}. This effect can be significantly reduced by introducing a small shift in initial conditions, as performed in top right panel. 

Further evolution in time can be examined for the PM simulations in the three last lines of panels of Fig.~\ref{fig:pdens_sinus}. While early stages of the evolution of the system clearly reflect the nature of initial conditions, see for instance middle left panel which illustrates the quasi one-dimensional nature of the dynamics at large scales, in the centre of the system, all the simulations build up a roughly circular halo around a three-dimensional cross, of which the arms are more or less contrasted according to the strength of the initial waves. This cross is a particular feature related to the high level of symmetry of initial conditions and is obviously not present in the CDM halos discussed below.

In the axisymmetric case, $\vec{\epsilon}=(1,1)$,  the cross is perturbed, then destroyed at late time, most likely by radial orbit instabilities, but further detailed diagnostics of the dynamical properties of the flow will be needed to fully confirm this hypothesis. The high level of symmetry of initial conditions indeed makes the dynamics very radial and potentially prone to related instabilities especially in the axisymmetric configuration, which is analogous to the spherical case, that is known to be radially unstable when initial conditions are cold, as many works in the literature show \citep[e.g.,][and references therein]{Halle19}. This radial instability relates to force softening and particle number: for instance, in the $\vec{\epsilon}=(1,1)$ PM simulation with $n_{\rm p}=n_{\rm g}=512$, which is not shown here, this instability takes place later than for $n_{\rm g}=n_{\rm p}=1024$, with no visible symmetry breaking at mid time, unlike right panel of fourth line in Fig.~\ref{fig:pdens_sinus}. 

\begin{figure*}[htp!]
\centering
\includegraphics[width=17.5cm]{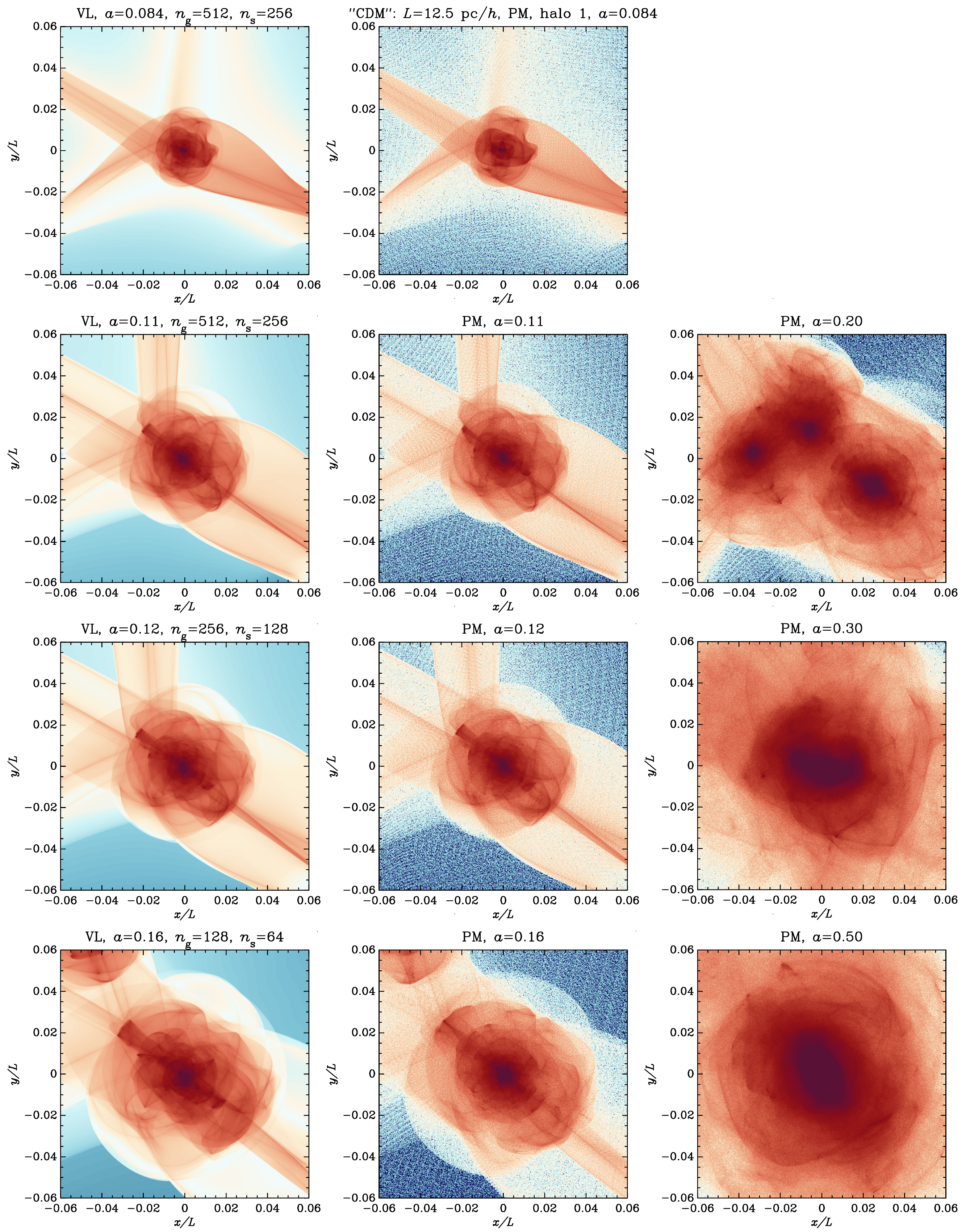}
\caption[]{Evolution of total projected density on $(x,y)$ plane for halo 1 of the ``CDM'' simulations with $L=12.5$ Mpc$/h$. The first column of panels corresponds, from top to bottom, to Vlasov runs with expansion factor values $a=0.084$, $0.11$, $0.12$ and $0.16$. The two top panels were generated using the highest resolution run with $n_{\rm g}=512$, VLA-CDM12.5-HR in Table~\ref{tab:tabsim}, while the two bottom ones used respectively VLA-CDM12.5-MR  with $n_{\rm g}=256$  and VLA-CDM12.5-LR with $n_{\rm g}=128$. The first column can be directly compared to the second one, which is analogous, but for the PM simulation, PM-CDM12.5-HR, with $n_{\rm g}=512$. Third column of panels corresponds to more advanced times in the PM simulation and highlights a multiple merger (halo 1 can be seen at the right of top panel of this column). Note that, similarly as in Fig.~\ref{fig:pdens_sinus}, the mass contributing to the projection comes only from the cubical subvolume displayed on each panel.}
\label{fig:pdens_halo1}
\end{figure*}

\begin{figure*}[htp!]
\centering
\includegraphics[width=17.5cm]{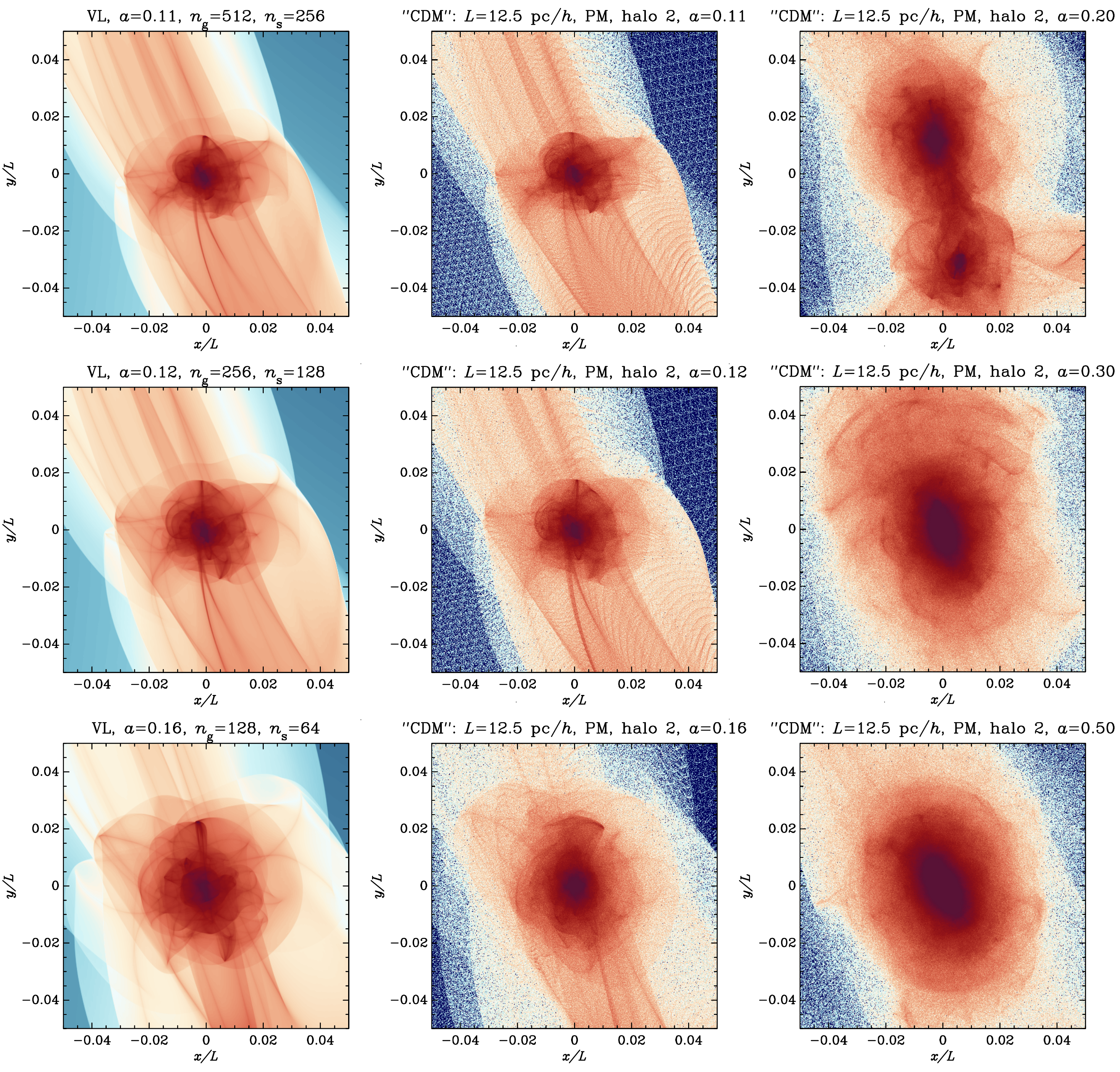}
\caption[]{Evolution of total projected density on $(x,y)$ plane for halo 2 of the ``CDM'' simulation with $L=12.5$ Mpc$/h$. This figure is exactly analogous to Fig.~\ref{fig:pdens_halo1} except that it does not have the two top panels corresponding to $a=0.084$, because this halo forms later than halo 1. }
\label{fig:pdens_halo2}
\end{figure*}

\begin{figure*}[htp!]
\centering
\includegraphics[width=17.5cm]{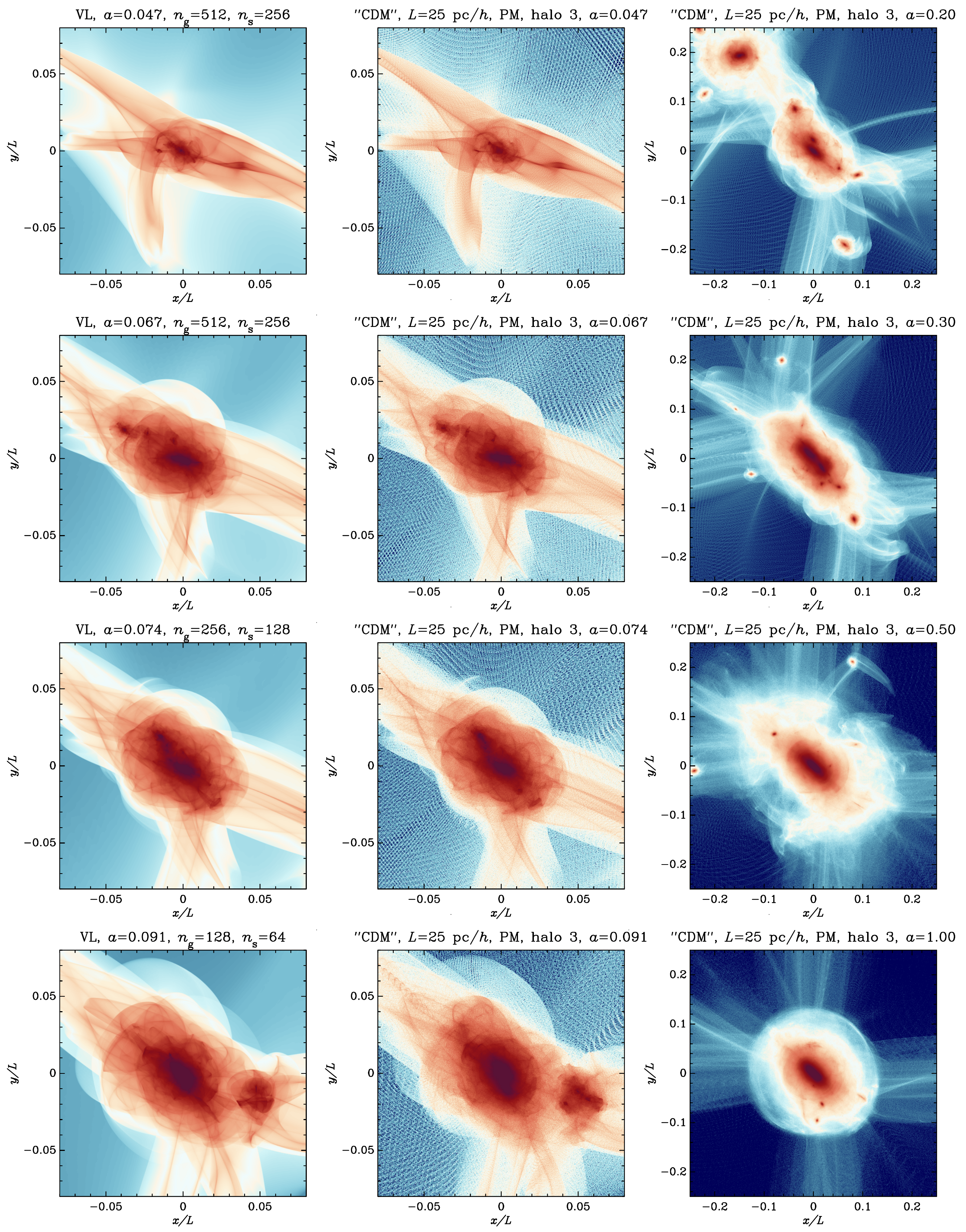}
\caption[]{Evolution of total projected density on $(x,y)$ plane for halo 3 of the ``CDM'' simulation with $=25$ Mpc$/h$. This figure is analogous to Fig.~\ref{fig:pdens_halo1} except that the times considered here are slightly different and that there is one extra panel on third column corresponding to present time, $a=1$. Note also that on third column of panels, the volume projected is larger, to have a better view of various structures at play.}
\label{fig:pdens_halo3}
\end{figure*}

Examining now figures~\ref{fig:pdens_halo1}, \ref{fig:pdens_halo2} and \ref{fig:pdens_halo3}, we turn to CDM simulations and inspect the evolution of halos number 1, 2 and 3. Comparison of first to second column of panels of these figures confirms the very good, if not spectacular, agreement between PM and Vlasov codes. Again, a careful examination of the figures shows that best visual match is obtained between PM runs with $n_{\rm g}=512$ and {\tt ColDICE} runs with $n_{\rm g}=256$.

The three halos considered in these figures are typical of what can be expected in the CDM scenario and remind of what was observed in other works \cite[e.g.,][]{Ishiyama10,Anderhalden13,Angulo17}. They form at the intersection of filaments of the cosmic web. Their early evolution is monolithic and similar to the three sine waves case. At later times, they are subject to successive mergers. This is illustrated by right columns of Figs.~\ref{fig:pdens_halo1}, \ref{fig:pdens_halo2} and especially by Fig.~\ref{fig:pdens_halo3}, which follows halo number 3 until present time, and is particularly rich of events. At the end of the simulation, halo number 3 has ``eaten'' almost all the matter available in the computing box and absorbed all the structures that formed at earlier times, in particular halos number 4 and 5 (not examined in detail here). Again, remind that these ``CDM'' simulations are totally unrealistic because of their very small box size, so their merger history is not representative. We shall discuss this more in detail when analysing radial density profiles in \S~\ref{sec:densprof}. One obvious consequence of the simulation volume smallness is that modes aligned with the sides of the box dominate large scale dynamics, hence the typical cross structure clearly visible on bottom right panel of Fig.~\ref{fig:pdens_halo3}. 
\section{Phase-space sections}
\label{sec:phaspasec}
This section corresponds to one of the truly innovative contributions of this article. For the first time, thanks to the finesse allowed by the tessellation technique, a detailed analysis of phase-space slices is performed in the {\tt ColDICE} simulations. Because the Vlasov code cannot follow the evolution of the system during many dynamical times, we consider only the early violent relaxation phase, but this will still provide us significant insights on the dynamics.

The objectives are once more three-fold, which sets up the way this section is organized. First, we aim in \S~\ref{sec:pss_tech} to test the robustness of the phase-space structure pattern with respect to force resolution. Second, in \S~\ref{sec:pss_PM}, through comparison of the results obtained with {\tt ColDICE} to PM simulations with large number of particles, we want to validate again the $N$-body approach when the particle density is high enough. Third,  in order to highlight specific patterns, e.g. related to self-similarity or to random perturbations, \S~\ref{sec:pss_inicdep} examines how the phase-space structure evolves with time and changes according to initial conditions.

\begin{figure*}[htp!]
\centering
\includegraphics[width=12.7cm]{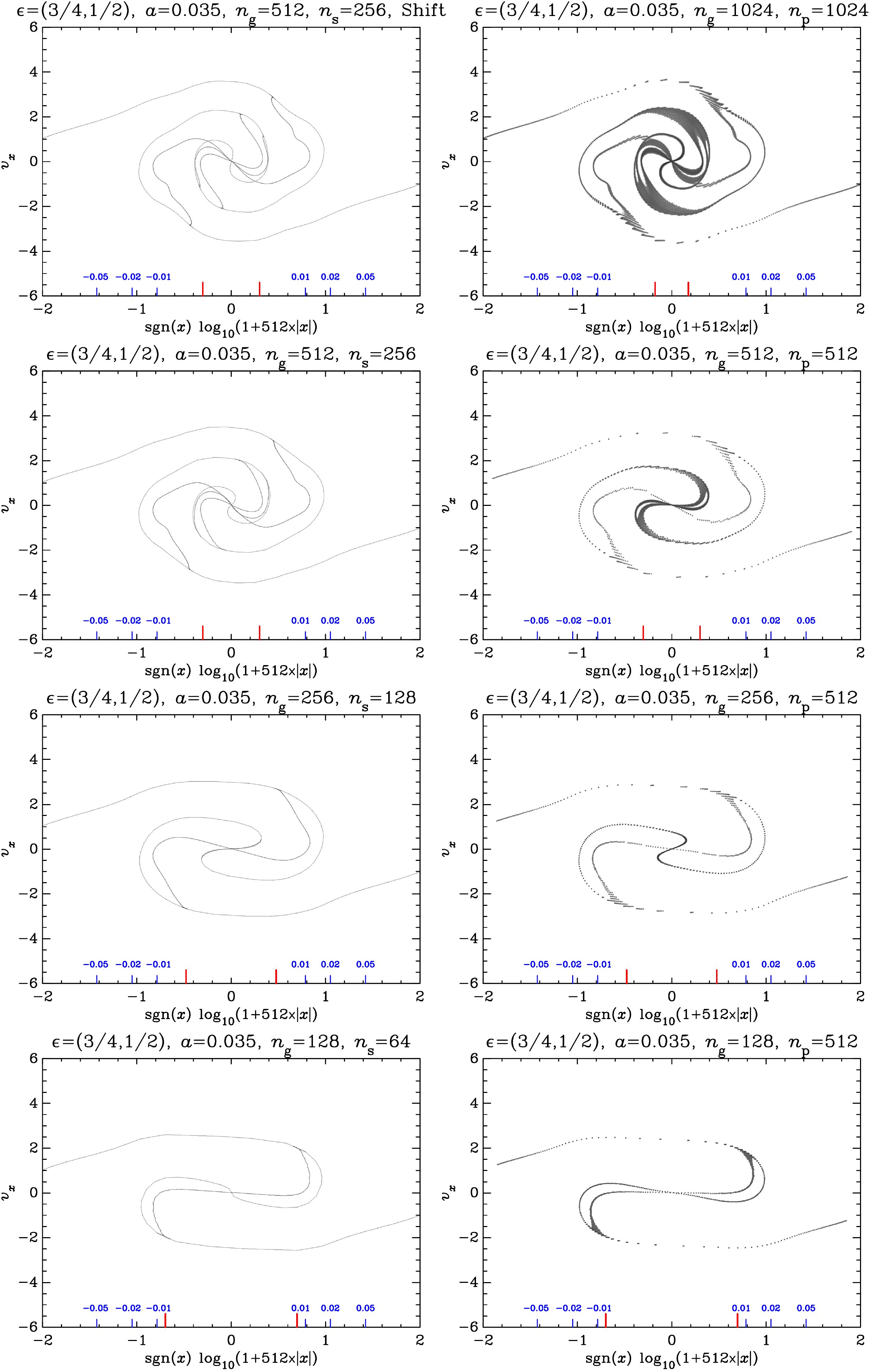}
\caption[]{Phase-space slice: force resolution analysis and comparison between Vlasov and PM for the three sine waves simulations with $\vec{\epsilon}=(3/4,1/2)$ and $a=0.035$. To have a better view of the fine structures of the system, the $x$ coordinate is represented in a logarithmic scale, ${\rm sgn}(x) \log_{10}(1+512 \times |x|)$. Some values of $x$ are indicated in blue inside each panel, while the two red vertical segments mark the force resolution scale, $L/n_{\rm g}$, which increases from top to bottom. {\it Vlasov runs are considered in left panels}. In this case, the intersection of the phase-space sheet with the hyperplane $y=z=0$ is calculated directly at linear order and represented in $(x,v_x)$ coordinates. The two top left panels consider Vlasov runs with $n_{\rm g}=512$ and $n_{\rm s}=256$. The only difference between both simulations is the initial shift of half a voxel size imprinted in initial conditions of the simulation considered in top left panel, VLA-ANI2-HRS in Table~\ref{tab:tabsim}, compared to the simulation in panel just below, VLA-ANI2-HR.  Note that our highest resolution run, VLA-ANI2-FHR (with $n_{\rm s}=512$ instead of 256), gives nearly identical results to VLA-ANI2-HR, and is not shown here. The two bottom left panels consider lower force resolution Vlasov runs with $n_{\rm g}=256$ (VLA-ANI2-MR)  and then $n_{\rm g}=128$ (VLA-ANI2-LR). {\it PM runs are examined in right panels}. In this case, we consider a very thin slice of particles with $(y,z) \in [-5\times 10^{-4}, 5 \times 10^{-4}]$ as tracers of the phase-space sheet. The top right panel shows the result obtained from our highest resolution PM run, PM-ANI2-UHR, with $n_{\rm g}=1024$ and $n_{\rm p}=1024$. The three next panels have all the same number of particles, $n_{\rm p}^3=512^3$, but decreasing spatial resolution, $n_{\rm g}=512$, 256 and 128 for PM-ANI2-HR, PM-ANI2-MR and PM-ANI2-LR, respectively.}
\label{fig:intersection_rez_a}
\end{figure*}

\begin{figure*}[htp!]
\centering
\includegraphics[width=12.7cm]{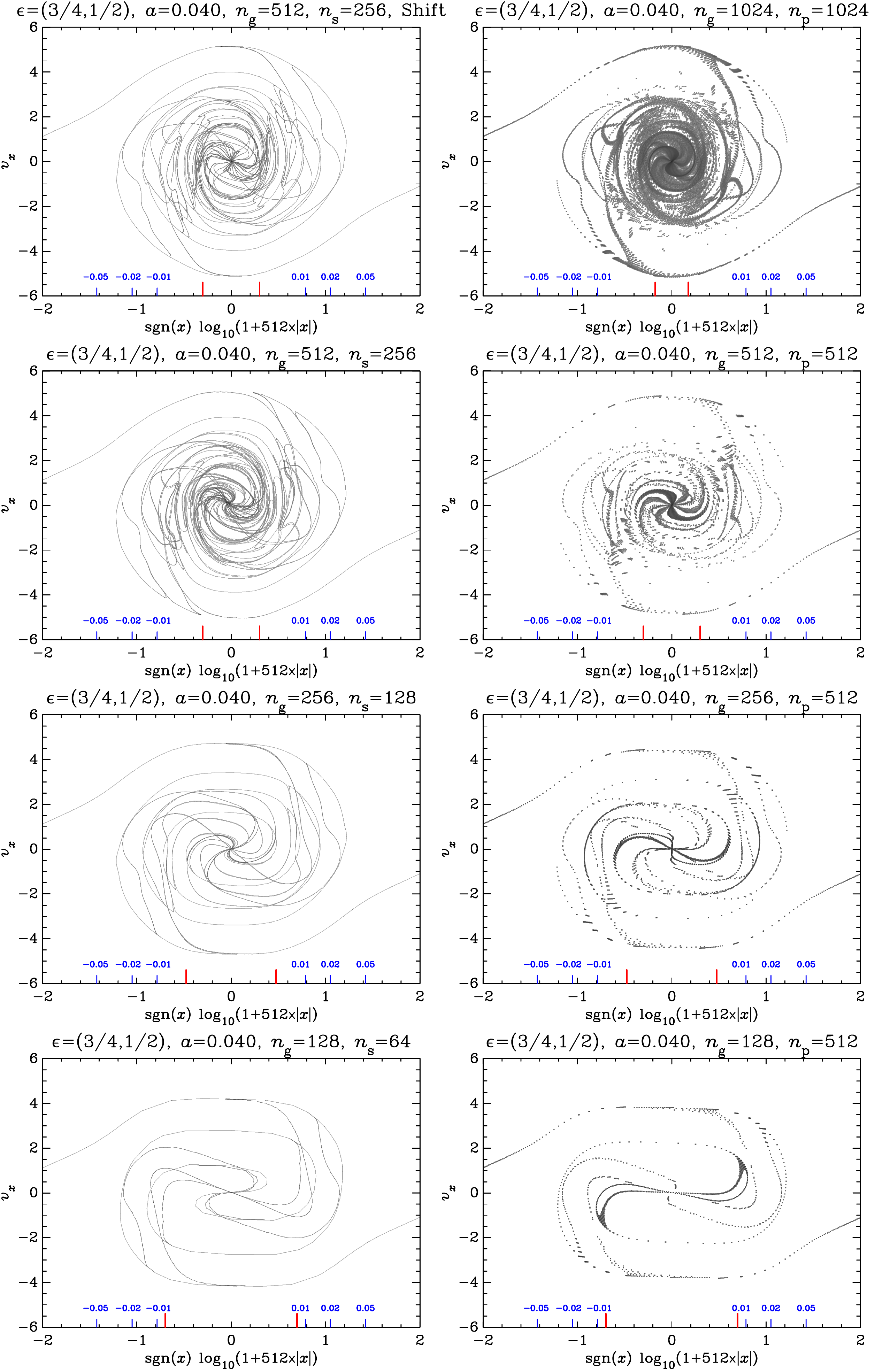}
\caption[]{Phase-space slice, continued: force resolution analysis and comparison between Vlasov and PM for the three sine waves simulations with $\vec{\epsilon}=(3/4,1/2)$ and $a=0.04$. This figure is exactly the same as Fig.~\ref{fig:intersection_rez_a}, but for a later time, $a=0.04$, which is the same expansion factor as in Fig.~\ref{fig:dens_3D_rez}.}
\label{fig:intersection_rez_b}
\end{figure*}

\begin{figure*}[htp!]
\centering
\includegraphics[width=12.7cm]{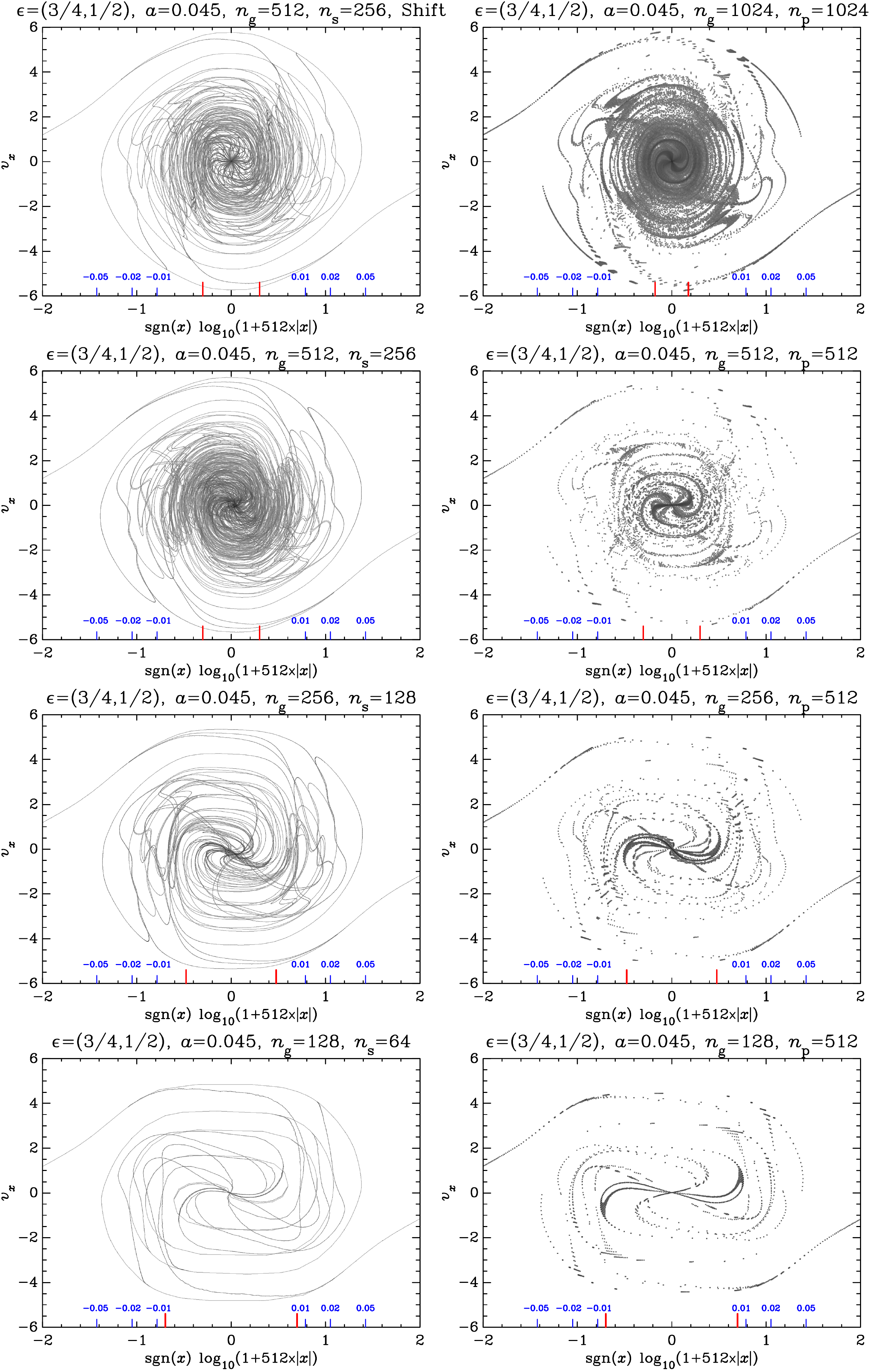}
\caption[]{Phase-space slice, continued: force resolution analysis and comparison between Vlasov and PM for the three sine waves simulations with $\vec{\epsilon}=(3/4,1/2)$ and $a=0.045$. This figure is exactly the same as Fig.~\ref{fig:intersection_rez_a}, but for the most evolved time available to the high force resolution Vlasov runs.}
\label{fig:intersection_rez_c}
\end{figure*}

\begin{figure*}[htp!]
\centering
\includegraphics[width=17cm]{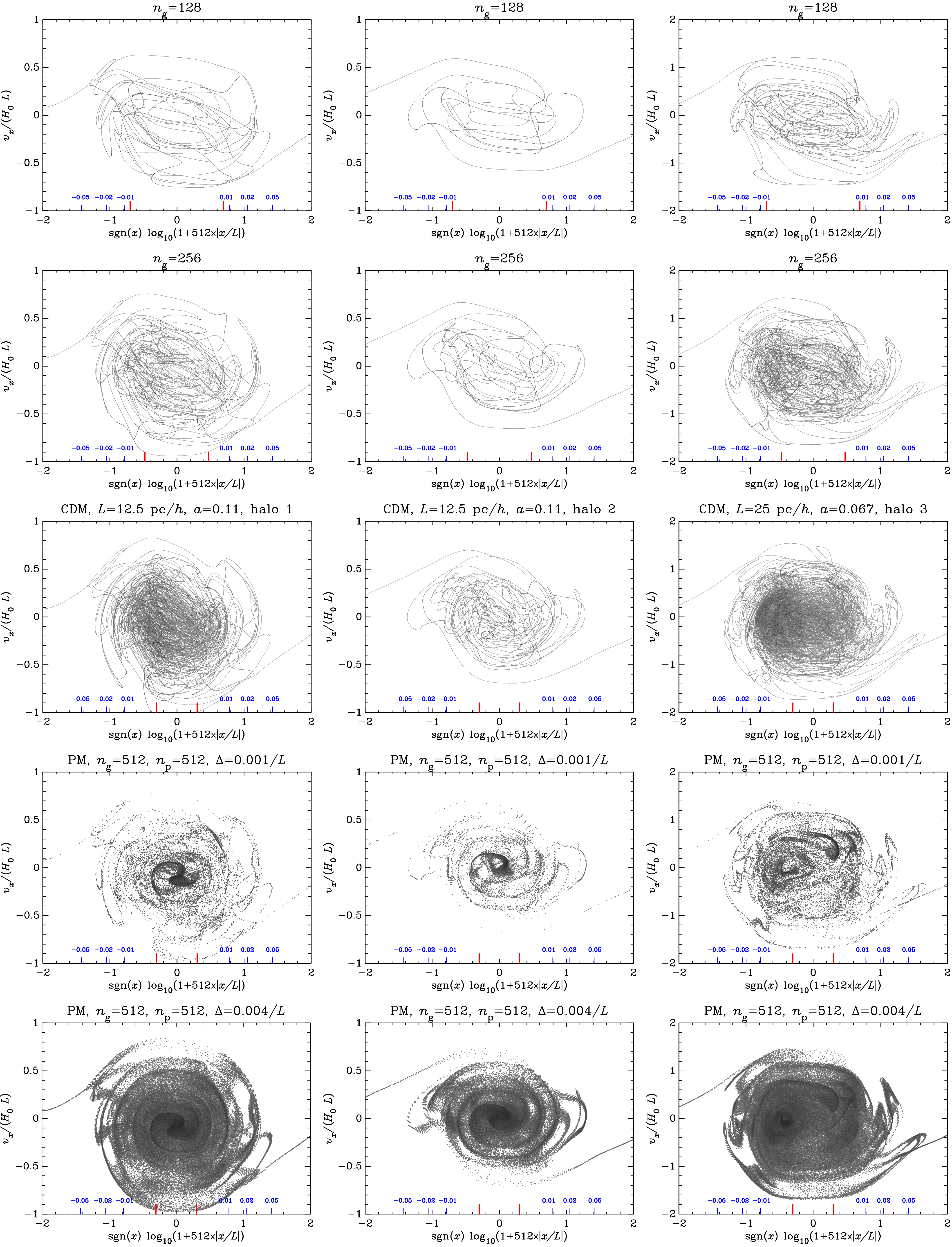}
\caption[]{Phase-space slice of ``CDM'' halos. Halos number 1, 2 and 3 are examined in  first, second and third column of panels, respectively. From top to bottom, Vlasov runs with increasing force resolution are considered, with $n_{\rm g}=128$ (VLA-CDM12.5-LR on left two panels and VLA-CDM25-LR on right one), 256 (VLA-CDM12.5-MR and VLA-CDM25-MR) and 512 (VLA-CDM12.5-HR and VLA-CDM25-HR), then PM simulations with $n_{\rm g}=512$ (PM-CDM12.5-HR and PM-CDM25-HR). In the latter case, a very thin slice of particles having $(y,z) \in [-\Delta/2, \Delta/2]$ is represented, with  $\Delta= 10^{-3}L$ and $4\times 10^{-3}L$ respectively in fourth and fifth line of panels.}
\label{fig:intersection_CDM_all}
\end{figure*}

\begin{figure*}[htp!]
\centering
\includegraphics[width=16.8cm]{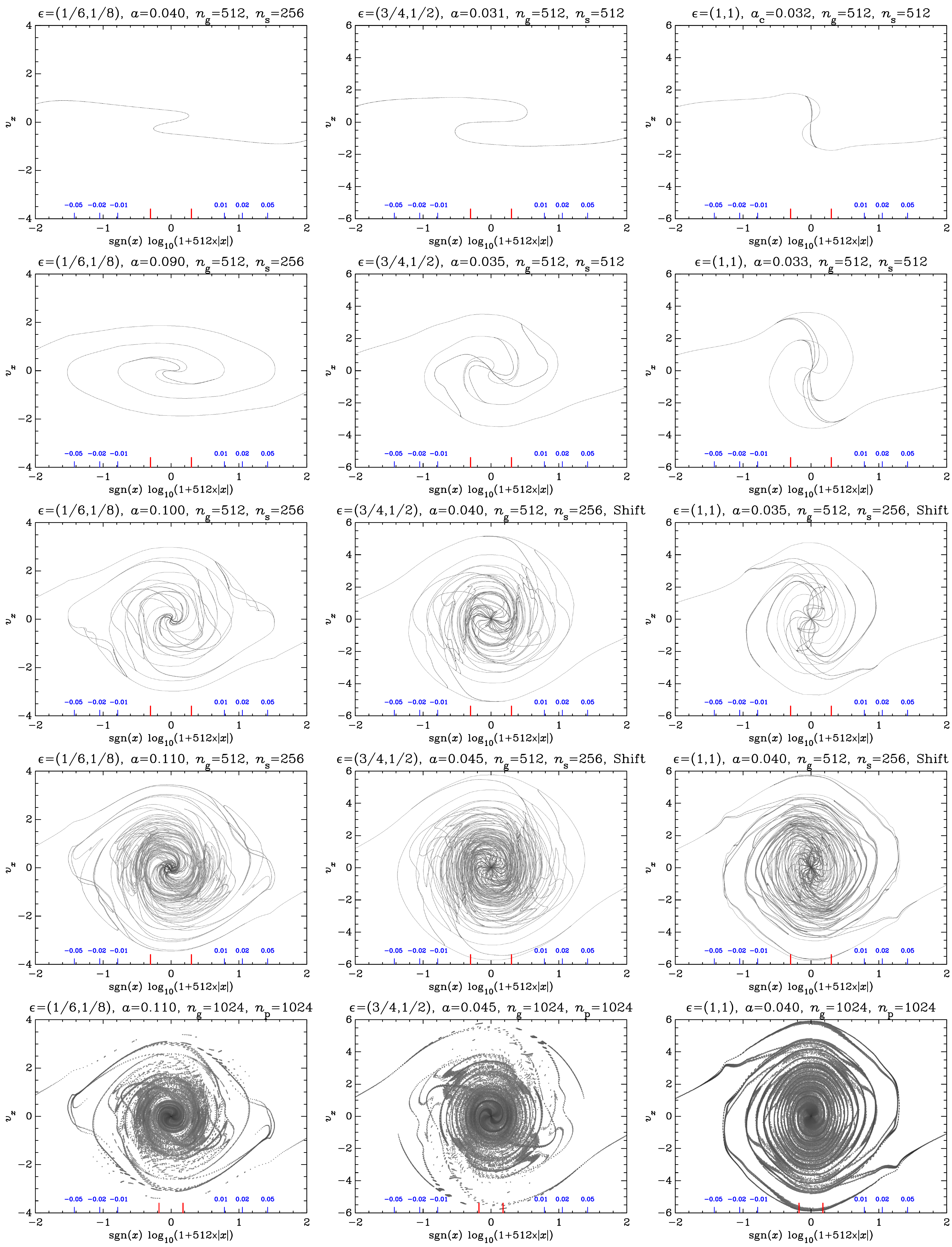}
\caption[]{Phase-space slice: early time evolution of three crossed sine waves simulations with various amplitudes. 
  Except for the last line of panels which concerns PM simulations, Vlasov runs with $\vec{\epsilon}=(1/6,1/8)$, $(3/4,1/2)$ and $(1,1)$ are considered in left, middle and right columns, respectively, with time augmenting from top to bottom, starting shortly after first shell-crossing. On the two top lines of panels, the highest resolutions Vlasov run with $n_{\rm g}=n_{\rm s}=512$  are under scrutiny, when available, namely VLA-ANI2-FHR and VLA-SYM-FHR in the middle and right columns, otherwise, VLA-Q1D-HR, VLA-ANI2-HRS and VLA-SYM-HRS are considered, respectively for left, middle and right column, with $N_{\rm g}=512$ and $n_{\rm s}=256$. To complete the figure, the last line of panels shows, at the same time as fourth line, the results obtained from our highest resolution PM runs, with $n_{\rm g}=n_{\rm p}=1024$, namely, from left to right, PM-Q1D-UHR, PM-ANI2-UHR and PM-SYM-UHR. In this case, we consider, like in Figs.~\ref{fig:intersection_rez_a} to \ref{fig:intersection_rez_c}, a very thin slice of particles with $(y,z) \in [-5\times 10^{-4}, 5 \times 10^{-4}]$ as tracers of the phase-space sheet.}
\label{fig:intersection_all}
\end{figure*}

To achieve these goals, we rely on detailed visual inspection of Figs.~\ref{fig:intersection_rez_a} to \ref{fig:intersection_CDM_all}. To be more specific, Figures~\ref{fig:intersection_rez_a}-\ref{fig:intersection_rez_c} display phase-space slices extracted from three sine waves simulations with $\vec{\epsilon}=(3/4,1/2)$. These three figures allow us to study thoroughly, at three different times, the effects of force resolution, which decreases from top to bottom, and to compare in detail {\tt ColDICE} (left panels) to the PM code (right panels).  To complete the analyses, CDM halos 1, 2 and 3, already shown in Figs.~\ref{fig:pdens_halo1}, \ref{fig:pdens_halo2} and \ref{fig:pdens_halo3}, are considered for {\tt ColDICE} at different force resolutions in the 3 top lines of panels of Fig.~\ref{fig:intersection_CDM_all} along with the PM results in the two bottom lines. Finally, figure~\ref{fig:intersection_all} examines high resolution sine waves simulations with different values of $\vec{\epsilon}$. Again, for comparison, the highest resolution PM results are shown in the last line of panels. 


\subsection{Phase-space sections: technical details}
\label{sec:pss_tech}
The phase-space slices displayed on each figure correspond, in the Vlasov code, to the intersection of the phase-space sheet with the hyperplane $y=z=0$ (with the origin of coordinate system centered on the halos). Remind that the intersection of a hypersurface of dimension $D=3$ with a hyperplane of dimension $D'=4$ in six-dimensional phase-space is expected to be, in the non trivial and non degenerate case, of dimension $D+D'-6=1$, that is it corresponds to a set of curves. Additionally, since the phase-space sheet is a connected periodic smooth hypersurface with no hole, this set of curves should also be fully connected, which means that in all the left panels of Figs.~\ref{fig:intersection_rez_a}, \ref{fig:intersection_rez_b} and \ref{fig:intersection_rez_c}, on top nine panels of Fig.~\ref{fig:intersection_CDM_all} and top twelve panels of Fig.~\ref{fig:intersection_all}, there should be no lose point in the curve pattern (except the two ends on each side of each panel), which is indeed the case after a detailed visual inspection.

Note that the intersection of the tessellation representing the phase-space sheet with the hyperplane $y=z=0$ is calculated at linear order, which means that the curves are actually sets of segments corresponding to the intersection of each simplex with the hyperplane. On can clearly guess the segmentation pattern in bottom left panel of Figs.~\ref{fig:intersection_rez_a}, \ref{fig:intersection_rez_b} and \ref{fig:intersection_rez_c}. This pattern can present some small oscillations that would disappear if a second order representation of the phase-space sheet (quadratic simplices) would be used, so these features are not artifacts related to dynamical instabilities. Another remark is that the figures only account for the intersections of which the count can cumulate but no weight is given to account for volume density of the phase-space sheet, which has to be taken into account when comparing the Vlasov phase-space slices to the PM ones, that are mass weighted.

As a final remark, top two left panels of Figs.~\ref{fig:intersection_rez_a}, \ref{fig:intersection_rez_b} and \ref{fig:intersection_rez_c} consider Vlasov simulations with exactly the same runtime parameters except that a small shift of half a voxel size was imprinted in initial conditions of the simulation considered in top left panel. As discussed in the end of \S~\ref{sec:3sines}, this procedure was introduced at some point to try to reduce the asymmetry that develops during time due to cumulative rounding errors in the Vlasov code. The effects of this asymmetry are indeed stronger without the shift. They are clearly visible in second left panel of Figs.~\ref{fig:intersection_rez_b}  and \ref{fig:intersection_rez_c}, but do not change significantly the phase-space pattern, except in the centre of the system. 
\subsection{Phase-space sections: force resolution}
\label{sec:sparezsli}
Figure~\ref{fig:intersection_rez_a} considers a moment at which the halo experienced only a few dynamical times, so that the phase-space structure of the system is not yet significantly intricate. As expected, this structure broadly follows a spiral pattern reminiscent of what is already well known in one dimension or in spherical symmetry \citep[e.g.,][]{Fillmore84,Alard13}, but is of course more complex. One can clearly guess on top left panel the multiple crossings the system experienced along each coordinate axis. These crossings relate to duplications of portions of the phase-space spiral. Here we find that the system experienced three crossings along $x$-axis, two along $y$-axis and one along $z$-axis. More specifically, duplication of the external arm of the spiral pattern reflects shell crossing along $y$-coordinate, while the scission in three of the central part of the spiral corresponds to one additional shell crossing along $y$-axis and one along $z$-axis. Obviously, these claims cannot be derived only from the examination of upper left panel of Fig.~\ref{fig:intersection_rez_a}: they result from combined analysis of the caustics pattern in Eulerian and Lagrangian spaces, that is not shown here to avoid multiplying the figures.  More discussions about the link between the Lagrangian pattern of the phase-space sheet and the internal dynamics of halos are deferred to a dedicated separate work \citep[][]{Sobolevski20}. Interestingly enough, from a dynamical point of view, the fact that only one crossing happened along $z$-axis suggests that the protohalo has just formed as a gravitationally bounded object.

When examining two bottom left panels of Fig.~\ref{fig:intersection_rez_a}, which correspond to lower force resolution simulations with $n_{\rm g}=256$ and $n_{\rm g}=128$, one can see only one duplication of each arm of the spiral, which means that collapse happened only along $x$ and $y$ directions, hence that the halo is not fully formed yet. Lowering force resolution indeed delays collapse time and halo formation time. To have accurate estimate of these times, it is required to resolve sufficiently the initial fluctuations. Softening of the force also obviously simplifies the structure of the phase-space sheet, which undergoes less foldings in the centre of the system. Yet, the outer part of the phase-space pattern remains approximately the same at the coarse level, even at later times (left panels of Figs.~\ref{fig:intersection_rez_b} and \ref{fig:intersection_rez_c}), which confirms the conclusions from visual inspection of the projected density in \S~\ref{sec:sparez} (see for instance Fig.~\ref{fig:dens_3D_rez}, which corresponds to the same time as Fig.~\ref{fig:intersection_rez_b}). Naturally, these conclusions also stand when examining Fig.~\ref{fig:intersection_CDM_all}, which considers CDM halos. In this case, the phase-space structure appears much less coherent than for the three sine waves simulations. When decreasing force resolution, this yarn ball pattern simplifies, especially in the centre of the system, but the outer parts remain mostly preserved.
\subsection{Phase-space sections: comparison with PM}
\label{sec:pss_PM}
Beside sparseness effects due to the discrete nature of the particle distribution, the match between PM  and {\tt ColDICE} is excellent, except that, as already extensively discussed in \S~\ref{sec:PMcode} and \ref{sec:sparez}, additional softening of the force due to the TSC interpolation and Hanning filtering in the PM code makes its effective force resolution nearly twice worse than in {\tt ColDICE}. This is very nicely illustrated by the phase-space diagrams, which can really allow an accurate comparison of the morphology of the phase-space sheet obtained in both codes, not only at the early stages of the evolution shown in Fig.~\ref{fig:intersection_rez_a}, but also at later times, as illustrated by Figs.~\ref{fig:intersection_rez_b}, \ref{fig:intersection_rez_c}, and independently of initial conditions (Fig.~\ref{fig:intersection_all}), even in the less coherent case of the CDM halos (Fig.~\ref{fig:intersection_CDM_all}).

Let us indeed repeat that if we would tessellate properly the particle distribution in Lagrangian space \citep[][]{Shandarin12,Abel12} and compute the intersection of this tessellation with the hyperplane $y=z=0$, the corresponding network of curves obtained from the PM particles would be very similar to that of the Vlasov code.  This means that the discrete nature of the representation of the phase-space density in the PM code does not significantly affect the gravitational force field, in agreement with what we concluded from visual inspection of the density in \S~\ref{sec:sparez}. Again, this result is not surprising, since, as advocated by earlier works \citep[e.g.,][]{Melott97,Splinter98} we consider, for phase-space diagram analyses, only $N$-body simulations with at least one particle per mesh element, $n_{\rm p} \geq n_{\rm g}$.

Note that the most evolved stages shown in Figs.~\ref{fig:intersection_rez_c}, \ref{fig:intersection_all} and \ref{fig:intersection_CDM_all} still correspond to rather early phases of the evolution of the halos. As discussed in \S~\ref{sec:sparez}, some instabilities due to particle shot noise in the PM code can develop at later times. Here, contrary to \S~\ref{sec:sparez}, more advanced times for the PM runs are not considered because the phase-space pattern becomes really intricate and indecipherable just by using directly a particle representation. A proper analysis would require the special tessellation technique on the Lagrangian particle distribution mentioned just above, which has not been used here because it was deemed unnecessary to prove the important points raised in this article. Another reason is that $N$-body particles are unable to trace at advanced times all the complexity of the phase-space sheet, hence reconstruction of the latter with the tessellation technique is expected to become inaccurate when there are too many foldings \citep[e.g.,][]{Hahn13}. 

\subsection{Phase-space sections: pattern analysis in various cases}
\label{sec:pss_inicdep}
Figure~\ref{fig:intersection_all} illustrates how phase space diagrams evolve with time in the high resolution sine waves simulations. Due to the highly symmetric nature of these systems, the phase-space structure remains coherent over time, taking the form of an intricate spiral. We however only see a section of it, so calling this complex pattern a spiral is actually an abuse of language. This ``spiral'' structure indeed experiences successive folds along the three coordinate axes, with orbital times related to the amplitude of the initial sine wave in each direction. As discussed in \S~\ref{sec:sparezsli} for $\vec{\epsilon}=(3/4,1/2)$, shell crossings along $y$ and $z$ axes translate into splits of the spiral arms. For instance, one can see, for the quasi one-dimensional case considered in left part of Fig.~\ref{fig:intersection_all}, a split of the central part of the spiral in second panel, which corresponds to shell-crossing along $y$-axis. On the right panels, since the system is axisymmetric, each shell crossing in one direction is associated to simultaneous crossings in the two orthogonal directions, hence, for each fold, each arm of the spiral is split in three parts.

Another feature of Fig.~\ref{fig:intersection_all} is the apparent self-similar pattern of the spiral structure when it builds up complexity. Naturally, this is true only outside the central region delimited by the two red vertical segments which mark spatial resolution of the computational mesh used to estimate the force field. Additionally, one has to take into account, in the quasi one-dimensional case (left panels), the asymmetry induced by very small but cumulative rounding errors in {\tt ColDICE}. Indeed we did not apply in this case any shift to initial conditions to remedy this defect, as discussed at the end of \S~\ref{sec:3sines}. Note furthermore that signatures of self-similarity are less easy to decipher for this value of $\vec{\epsilon}$, due to the large difference in dynamical times associated to each axis. Self-similarity will be discussed furthermore in \S~\ref{sec:profiles}.

Turning to more realistic configurations without imposed symmetries, examination of Fig.~\ref{fig:intersection_CDM_all} suggests that the phase-space structure is much less coherent in the CDM halos than in the sine waves simulations, which is not very surprising. Yet, this lack of coherence is not as strong as it seems and this is probably partly due to the choice of representation of the phase-space slices in the Vlasov simulations, since the phase-space sheet was not weighted according to its local volume density. When examining the PM results, which are mass weighted, we can clearly guess a clean and rather symmetric regular spiral pattern at the coarse level on two bottom left panels of Fig.~\ref{fig:intersection_CDM_all}. This is because the two halos corresponding to these plots are still in the monolithic phase which is very analogous to the three sine waves case. On the contrary, in the bottom right panel, that treats a composite halo, i.e. which already experienced some merger, the structure of the spiral is more intricate. Obviously, mergers contribute significantly to disorder in the phase-space structure.  
\section{Complexity}
\label{sec:comp}
This section corresponds to another innovative contribution of this article. We study, for the {\tt ColDICE} runs, what shall be referred to as complexity of the phase-space sheet, through the analysis of the simplices count and of the phase-space sheet volume as functions of time. This will allow us to estimate the degree of winding of the phase-space sheet, that we try shall to relate to self-similarity and, if relevant, to chaotic instabilities, depending on if it grows as a power-law of time or exponentially. Confirming earlier analyses by \citet[][]{SC16}, we shall see that the phase-space sheet intricacy always increases very quickly, which unfortunately makes the adaptive tessellation method impracticable on the long run, whatever the level of optimisation.

\begin{figure*}[htp!]
\centering
\includegraphics[width=18.5cm]{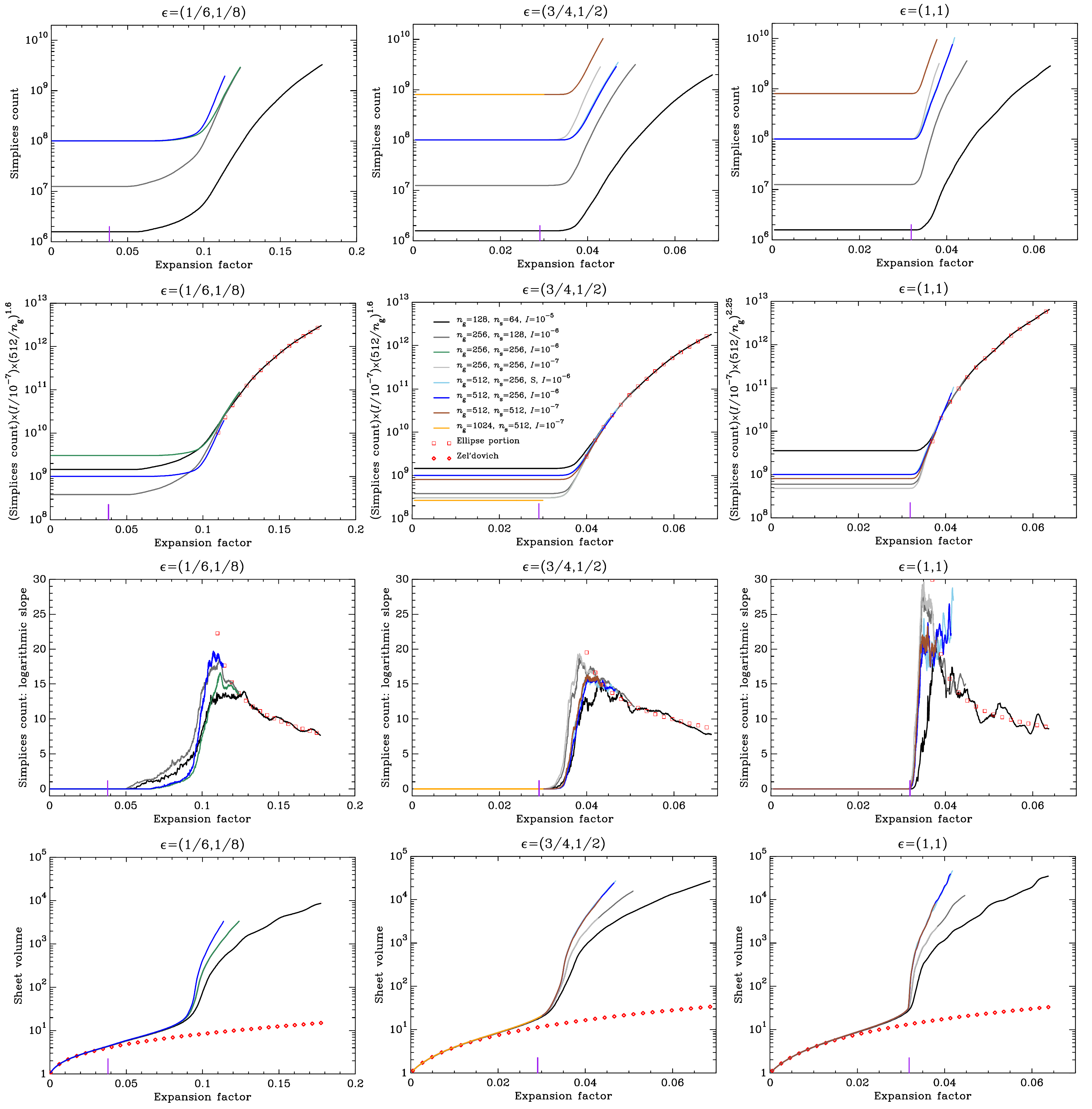}
\caption[]{Complexity of the phase-space sheet: simplices count and sheet volume in the simulations with three crossed sine waves initial conditions. Details on all the curves represented on each panel are provided in middle panel of second line. From left to right, different values of the initial relative amplitudes of the waves are considered, $\vec{\epsilon}=(1/6,1/8)$, $(3/4,1/2)$ and $(1,1)$, with collapse time indicated by the vertical purple segment.  The top line of panels displays the simplices counts as functions of expansion factor for all the Vlasov runs listed in Table~\ref{tab:tabsim} except VLA-ANI2-LRa. In the second line, the simplices counts are rescaled by the factor $(I/10^{-7})\times (512/n_{\rm g})^{\alpha}$, with $\alpha=1.6$ in two left panels and $\alpha=2.25$ in right panel, as discussed in the main text. A fit with an ellipse portion in linear-logarithm space is also shown with red symbols (equation \ref{eq:ellipsefit}). Next line of panels considers the logarithmic slope of the simplices count, to be compared again to the red squares, which correspond to the logarithmic slope derived from the ellipse portion. The last line shows the phase-space sheet volume as a function of expansion factor. Note that all the curves should superpose to each other: the differences relate to spatial resolution $n_{\rm g}$. The red losanges provide the prediction from Zel'dovich approximation.}
\label{fig:simplices_count}
\end{figure*}

\begin{figure*}[htp!]
\centering
\includegraphics[width=13cm]{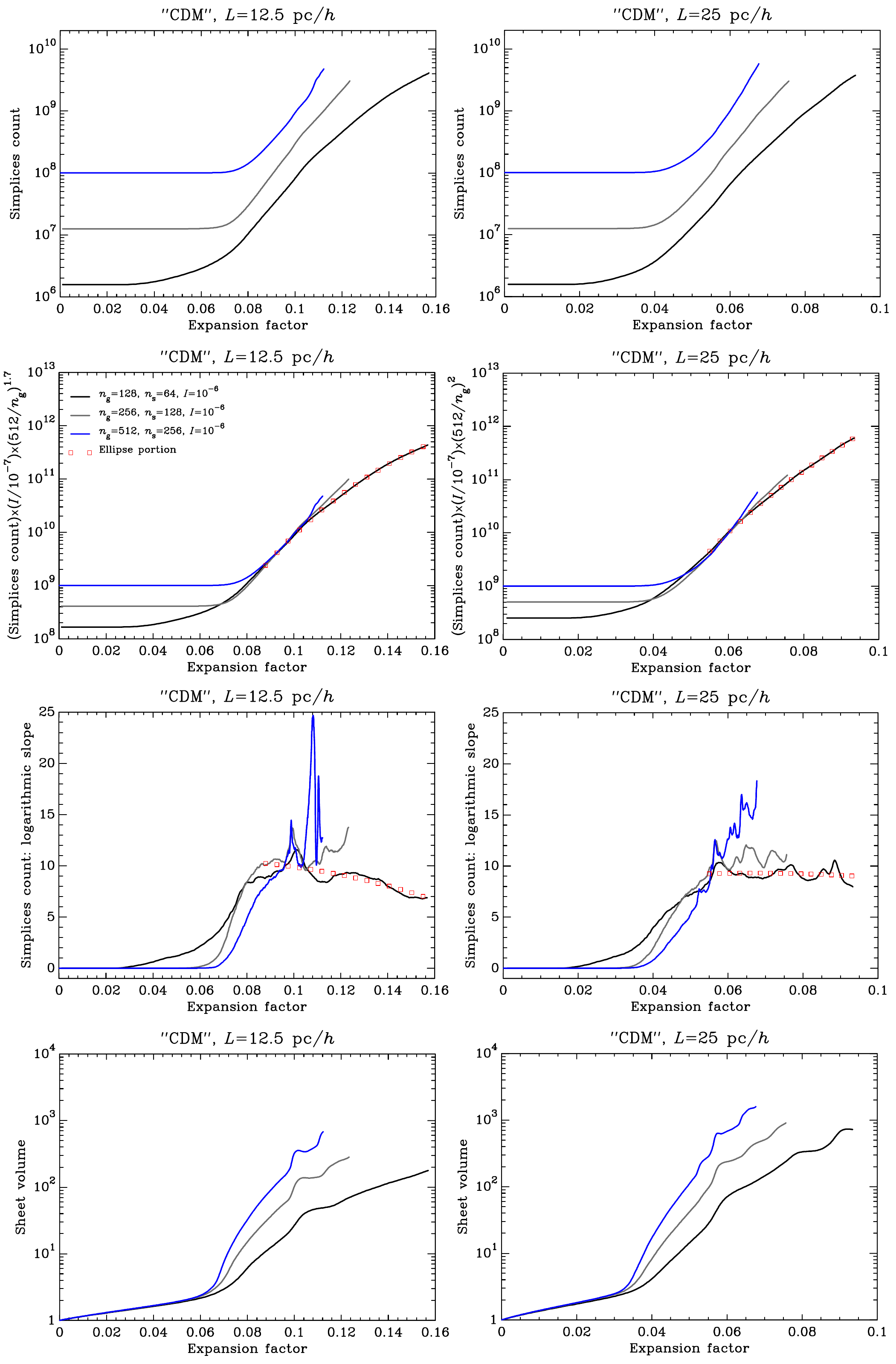}
\caption[]{Complexity of the phase-space sheet: simplices count and sheet volume in the ``CDM'' simulations. This figure is exactly analogous to Fig.~\ref{fig:simplices_count}, but left and right panels correspond to all the Vlasov CDM runs with box size $L=12.5$ Mpc$/h$ and 25 Mpc$/h$, respectively, as listed in Table~\ref{tab:tabsim}. Details on the curves are given in second left panel. Here, comparison with the Zel'dovich approximation in bottom panels is lacking but would provide qualitative results similar to what can be seen in bottom panels of Fig.~\ref{fig:simplices_count}.}
\label{fig:simplices_count_cdm}
\end{figure*}

Figures \ref{fig:simplices_count} and \ref{fig:simplices_count_cdm}
examine respectively the three sine waves simulations and the CDM runs. The top panels of these figures show the rough simplices counts as functions of expansion factor. As expected, during the early phases of the dynamics, the displacement field remains linear and the number of simplices $N_{\rm s}$ stays stable. At some point, $N_{\rm s}$ increases very quickly until it reaches a few billions at the end of the simulations (about 10 billions in the highest resolution runs). The way this happens is related to the formation and relaxation of dark-matter halos, which makes the phase-space sheet more and more intricate with time, as illustrated very well by the diagrams of previous section. The phase-space sheet volume shown in last line of panels of Fig.~\ref{fig:simplices_count} and \ref{fig:simplices_count_cdm} provides a precise quantitative estimate of the actual level of complexity of the sheet. It suddenly starts to increase after formation of the first halos.\footnote{Note that full formation of the halo requires, as described in Introduction, collapses along all of the three main axes of the dynamics. In the quasi one-dimensional case, this event takes place significantly later than first shell crossing, which is marked as the vertical purple segment on each panel of Fig.~\ref{fig:simplices_count}.}  This obviously triggers intense refinement of the tessellation. 

Simplices count is controlled by local refinement. The pattern and finesse of the latter depend on three factors: the dynamical state of the halos, force resolution traced by $n_{\rm g}$, and Poincar\'e constraint parameter $I$. Second line of panels of Figs.~\ref{fig:simplices_count} and \ref{fig:simplices_count_cdm} combine these three elements by considering the rescaled count $N_{\rm s,rescaled}$ defined as follows,
\begin{equation}
  N_{\rm s} \rightarrow N_{\rm s,rescaled}\equiv N_{\rm s} \times\left( \frac{I}{10^{-7}} \right)\times \left( \frac{512}{n_{\rm g}}
    \right)^\alpha, \label{eq:nsresca}
\end{equation}
where the parameter $\alpha$, which spans the interval $[1.6,2.25]$, corresponds roughly to the logarithmic slope of the radial density $\rho(r)$ in the early relaxation phase of the halos, as measured in \S~\ref{sec:profiles}. The factor proportional to $I$ stems from assuming that refinement is performed, in practice, in a locally isotropic fashion \citep[][]{SC16}. This is not imposed by the algorithm, which allows for anisotropic refinement, but it results from the dynamics.

The factor proportional to $n_{\rm g}^{-\alpha}$ in equation (\ref{eq:nsresca})  does not come from an analytic prediction. It is simply an educated guess relating to a supposed self-similar evolution of the phase-space sheet, of which we had clear hints in previous section for the three-sine waves simulations and that will be discussed furthermore in \S~\ref{sec:profiles}. 

After rescaling (\ref{eq:nsresca}), as expected, the curves on second line of panels of Fig.~\ref{fig:simplices_count} superpose to each other when refinement starts to dominate in terms of simplices count compared to the initial value, $N_{\rm s} \gg 6 n_{\rm s}^3$. Note however that the value of $\alpha$ was adjusted so that the superposition is visually optimal, but remains fixed in each panel of second line of Figs.~\ref{fig:simplices_count} and \ref{fig:simplices_count_cdm}, as shown in the caption of the ordinates, so this result still demonstrates to a large extent the numerical consistency of {\tt ColDICE} with respect to refinement in a self-similar framework. 

Turning now to the way $N_{\rm s}$ increases with time, we confirm the early findings of \citet[][]{SC16}, that this increase is very dramatic, which forces us to stop the runs only after a limited number of dynamical times. This highlights again the main weakness of the adaptive tessellation approach. Yet, in most cases, the increase is not exponential. This is best illustrated by the third line of panels of Figs.~\ref{fig:simplices_count} and \ref{fig:simplices_count_cdm} which display the logarithmic slope of $N_{\rm s}$ as a function of expansion factor. For the three sine waves case, this slope suddenly grows up to a peak around $10-30$, then slowly decreases with time but still with very high values, of the order of $7-9$ at the latest times considered in the figures. Note that the decrease is not obvious in the high resolution simulations of the axisymmetric case but this is inconclusive due to the limited time range available. 

To model the behaviour of $N_{\rm s}$ with expansion factor after the peak, an ellipse is adjusted to the black curves of second line of panels of the figures (red squares). These curves correspond to the lowest force resolution simulations, with $n_{\rm g}=128$,  for which the available time range is the largest. The portion of ellipse is given by the following function,
\begin{equation}
  \log_{10} N_{\rm s,rescaled}=w_0 -w_1 \sqrt{ 1-(a-w_2)^2/w_3^2},
  \label{eq:ellipsefit}
 \end{equation}
where the four parameters vector $\vec{w}=(w_0,w_1,w_2,w_3)$ is determined with a simplex fitting algorithm \citep[e.g.,][]{Numrec} in the interval of values of $a$ covered by the red squares on the figures. For completeness, the values of $\vec{w}$ are listed in Table~\ref{tab:ellipse}.
 \begin{table}
   \begin{tabular}{ll}
     \hline
     Case & \vec{w} \\ \hline
     $\epsilon=(1/6,1/8)$ & $(0.30,8.76,0.20,-4.78)$ \\
     $\epsilon=(3/4,1/2)$ & $(0.12,7.58,0.087,-6.00)$ \\
     $\epsilon=(1,1)$ & $(0.15,8.89,0.11,-5.94)$ \\
     ``CDM'', $L=12.5$ pc$/h$ & $(0.221.31,0.20,-10.9)$ \\
     ``CDM'', $L=25$ pc$/h$ & $(0.21,2.92,0.2,-11.1)$ \\
     \hline
  \end{tabular}
  \caption[]{Parameters used in equation (\ref{eq:ellipsefit}) to draw the portion of ellipse (red squares) on second line of panels of Figs.~\ref{fig:simplices_count} and \ref{fig:simplices_count_cdm}, as well as the corresponding logarithmic slope on third line of panels.}
\label{tab:ellipse}
\end{table}

A non exponential behaviour of the simplices count reflects a quiescent behaviour of the dynamics, or, in other words, the absence of chaos. This is what we find for the early, monolithic, violent relaxation phase of halos growing from three sine waves initial conditions, except possibly in the axisymmetric case, but this is a very degenerate configuration. These results have to be interpreted with caution, because the measurements cover a very limited time range in the highest resolution simulations.  Convergence with respect to spatial resolution $n_{\rm g}$ is not achieved. This is also clearly illustrated by sheet volume measurements discussed furthermore below. This lack of convergence is even more pronounced in the CDM simulations. In fact, the blue curve on right panel of third line of Fig.~\ref{fig:simplices_count_cdm} suggests an exponential behaviour of $N_{\rm s}$ at advanced times in the CDM case.\footnote{more exactly, of the form $\gamma_0\, a^{-\gamma_1} \exp( \gamma_2 a)$, with $\gamma_0, \gamma_1, \gamma_2 > 0$.} This behaviour is seen only in the highest force resolution simulation and only for a short time. This result is inconclusive but indicates that mergers, that actually take place during this small interval of time (see Fig.~\ref{fig:pdens_halo3}), play an important role in introducing some chaotic signature. 

Bottom panels of Figs.~\ref{fig:simplices_count} and \ref{fig:simplices_count_cdm} show the evolution of the phase-space sheet volume as a function of expansion factor. At linear order in the geometric representation, this volume is given by the sum of the elementary volumes of each tetrahedron it is composed of. To compute the three-dimensional volume of a tetrahedron with six-dimensional coordinates, one can for instance first find the 3D submanifold in which this tetrahedron lies, and then compute the volume of the tetrahedron in this submanifold. The phase-space sheet volume is therefore a quantity difficult to interpret because it overlaps between configuration and velocity spaces: it depends on the way velocities are scaled with respect to positions, that is on the choice of metric. Here, the figures assume box size and Hubble constant unity.

Even if it is difficult to interpret it in detail, the volume of the phase-space sheet remains a physical quantity, so it should not depend on any simulation parameter. This is not at all the case as soon as first halos form. Indeed, except in the quasi-linear regime, where predictions of linear Lagrangian perturbation theory successfully reproduce simulation measurements nearly until collapse time (red losanges on bottom panels of Fig.~\ref{fig:simplices_count}),\footnote{To compute the prediction from linear Lagrangian perturbation theory, we create a regular network of vertices with $n_{\rm s}=32$ following equations (\ref{eq:inigrida}) and (\ref{eq:inigridb}), then perturb this network according to Zel'dovich approximation up to the time of interest following equations (\ref{eq:linlag1}) and (\ref{eq:linlag2}). These vertices are tessellated with a set of tetrahedra of which we compute the sum of the volumes as explained just above in the main text.}  the results depend strongly on spatial resolution $n_{\rm g}$ (but not significantly on other control parameters, as expected). 

Halo formation marks the transition between quasi-linear and fast growth of the phase-space volume. When the value of $n_{\rm g}$ is reduced, this growth is slightly delayed and becomes very significantly less prominent.  This reflects the fact that during the violent relaxation phase, the phase-space sheet is subject to many foldings in the centre of the system, where a power-law singularity builds up, as discussed in detail in next section. The number of these foldings and the corresponding augmentation of the volume strongly depend on force resolution. The visual inspections performed in \S~\ref{sec:sparez} and \ref{sec:sparezsli} indeed show that the structure of the halos is pretty insensitive to force resolution in the outer parts of the system, but it becomes more and more intricate in the centre when increasing $n_{\rm g}$. While the phase-space sheet volume does not seem, for this reason, to be a very useful quantity to study, it provides a robust tool to demonstrate rigorous convergence with respect to force resolution. Note finally that the exponential behaviour seen for the simplices count in the high resolution CDM simulation with $L=25$ pc$/h$ is not obvious on bottom right panel of Fig.~\ref{fig:simplices_count_cdm}, but large fluctuations introduced by mergers make the results difficult to interpret. 
 
\section{Profiles}
\label{sec:profiles}
In this section, we perform classic measurements of the radial density profile, $\rho(r)$, as well as the pseudo phase-space density (equation \ref{eq:Qofr}).  Details on how the measurements are performed are provided in Appendix A.2. 

The objective is to revisit phases (ii) and (iii) of the history of dark matter halos depicted in Introduction. After careful tests of force and mass resolution, we examine the early violent relaxation phase of dark matter protohalos and the subsequent convergence to an universal NFW-like profile, with detailed studies of the power-law slopes of the projected and pseudo phase-space densities. One important question is whether mergers represent a {\em sine qua non} condition for the convergence to NFW. This issue will be addressed by comparing the history of CDM halos to that of idealised halos obtained from three sine wave initial conditions and which experience a purely monolithic evolution.

This section is thus organized as follows. Exploring force resolution in \S~\ref{sec:rhosparez} will allow us to build a simple and approximate recipe to select the interval of scales supposedly not affected by softening of the force. Turning to mass resolution in \S~\ref{sec:tirho}, we shall see that particle number in the PM simulations does not affect the results significantly at the coarse level for all the simulations we did, except may be at late time, depending on the desired accuracy. Then, \S~\ref{sec:densprof} examines the different phases of the history of density profiles for all the sine wave initial conditions as well as the five halos extracted from the two ensembles of CDM simulations. Likewise, \S~\ref{sec:Qofr} deals with the pseudo phase-space density. The measurements will be put in perspective with respect to numerous and well know results in the literature. 
\subsection{Force resolution}
\label{sec:rhosparez}
\begin{figure*}[htp!]
  \centerline{\null \hskip 1cm \bf 3 sine waves, force resolution study for $\epsilon=(3/4,1/2)$}
\centerline{\includegraphics[width=17cm]{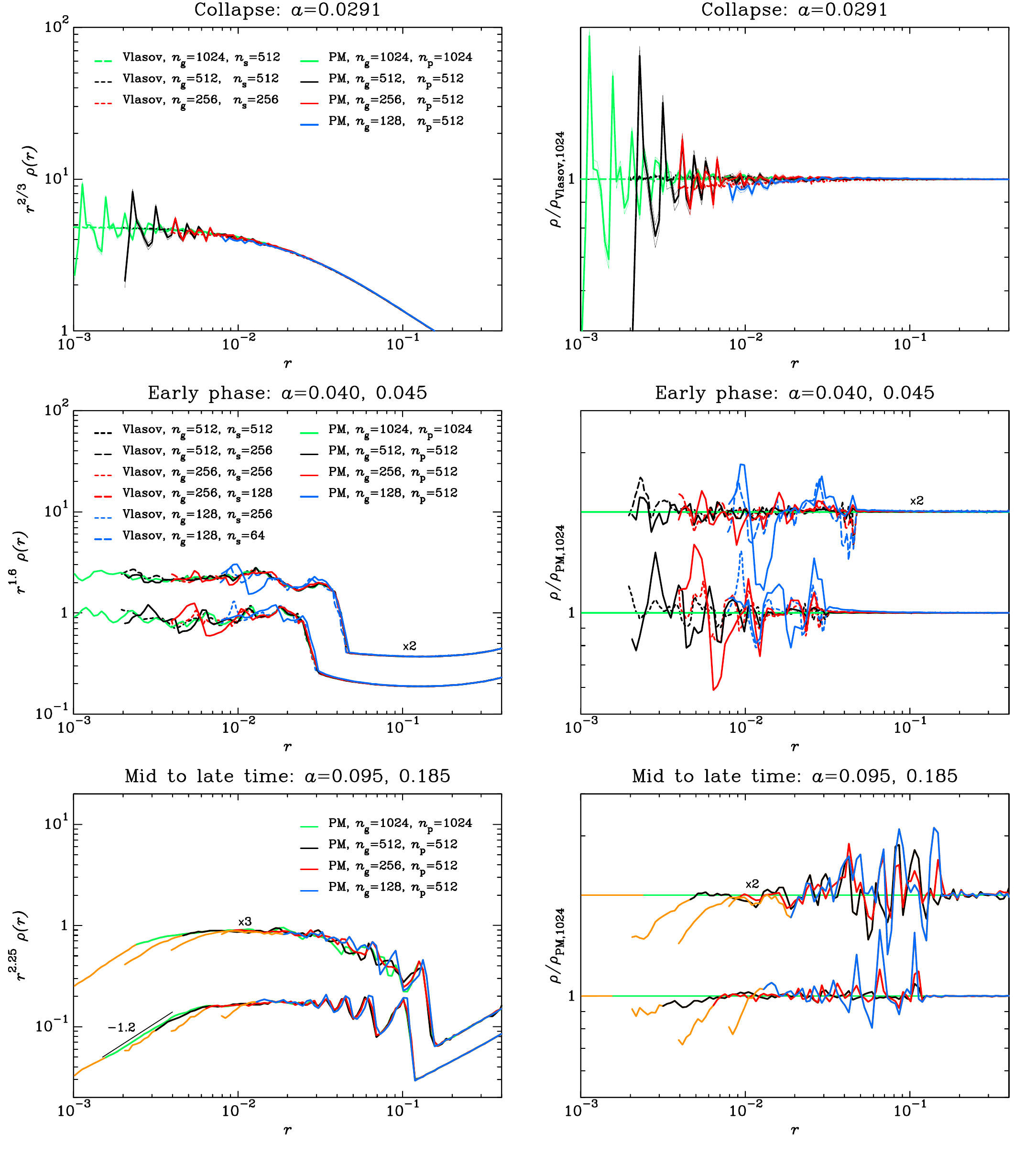}}
\caption[]{Radial density profile: force resolution analysis of the three sine waves simulations with $\vec{\epsilon}=(3/4,1/2)$. Different epochs are considered following conventions of Table~\ref{tab:regimes}, namely collapse time in top panels, early time in middle panels (with a multiplication by a factor 2 of the curves corresponding to $a=0.045$ for clarity) and mid to late time in bottom panels (with a multiplication by a factor 3 and 2 of the curves corresponding to $a=0.185$ on left and right panel, respectively). As discussed in the main text, to have a better view of each regime, the quantity represented on left column is the logarithm of $r^{\alpha} \rho(r)$, with $\alpha=2/3$, $1.6$ and $2.25$, respectively in top, middle and bottom panels. Various simulations are considered, both in the Vlasov and PM cases, as indicated on each panel through the values of $n_{\rm g}$, $n_{\rm s}$ and $n_{\rm p}$ also shown in Table~\ref{tab:tabsim}. Just note that for the long dashed black curve corresponding to a {\tt ColDICE} run with $n_{\rm g}=512$ and $n_{\rm s}=256$, the simulation used is VLA-ANI2-HRS, but VLA-ANI2-HR would provide nearly identical results. On bottom panels, the left part of the curves corresponding to regions supposedly influenced by small scale force softening is displayed in orange, as discussed in the main text. To highlight better the differences between various curves, the right column displays density ratios: on top panel, the quantity displayed is $\rho/\rho_{\rm Vlasov, 1024}$, where $\rho_{\rm Vlasov,1024}$ is the density measured in our highest resolution {\tt ColDICE} run (corresponding to the dashed green curve); on two bottom panels, the quantity displayed is $\rho/\rho_{\rm PM, 1024}$, where $\rho_{\rm PM, 1024}$ is the density measured in our highest resolution PM run (corresponding to the solid green curves).}
\label{fig:resolution_profile_a}
\end{figure*}

Figure \ref{fig:resolution_profile_a} examines in details the effects of changing force resolution for the three sine waves simulations with $\vec{\epsilon}=(3/4,1/2)$. On left panels, the quantity displayed is $r^{\alpha} \rho(r)$, with different values of $\alpha$ to emphasize better various phases of the dynamics, as discussed in details in \S~\ref{sec:tirho}. To highlight the differences between various curves, right panels display ratios between the measured density and the one obtained from the highest available resolution run.

At collapse time considered in top panels, the structure of the local pancake singularity implies $\rho(r) \propto r^{-\alpha}$ with $\alpha=2/3$ in the centre of the halo \citep[e.g.,][]{Arnold82}. Middle panels examine early time, during which violent relaxation takes place. In this case, the radial density is close to a power-law of slope $\alpha \simeq 1.6$, consistent with the literature. Bottom panels show mid and late time, where the system slowly relaxes to a NFW like profile, with a plateau consistent with secondary spherical infall prediction, $\alpha=2.25$ \citep[][]{Bertschinger85}. On can also observe at mid time a regime with $\alpha=1.2$ at small radii. 

As roughly demonstrated by two top lines of panels, the mesh cell size $r_{\rm min}=\varepsilon \equiv L/n_{\rm g}$ seems a good estimate of the lower bound of the trustable dynamical range at collapse time and during the early relaxation phase.\footnote{One can however detect some damping of the profile at small radii in top right panel of Fig.~\ref{fig:resolution_profile_a} for $n_{\rm g} \le 256$. Indeed, one effect of force softening is to delay halo formation, in particular collapse time, as already discussed in \S~\ref{sec:sparezsli} and \S~\ref{sec:comp}.} This means that effective resolution of the codes is close to optimal in this regime, as long as the quantity of interest is the radial density at the coarse level. This includes the PM simulations, despite the additional force softening introduced by Hanning filtering and TSC interpolation.

Softening of the force has however some non trivial consequences at later epochs. Its indeed cumulates with time by contaminating increasingly large scales. While it seems difficult to predict this effect analytically, it can be modelled in a phenomenological way by examining PM runs of various force resolutions in order to isolate the region contaminated by softening, which is represented in orange on bottom panels of Fig.~\ref{fig:resolution_profile_a}. This region is determined with the following phenomenological formula,
\begin{equation}
  r_{\rm min}=\frac{L}{n_{\rm g}} \exp \left\{ \frac{\log(a/a_{\rm early})}{\log(a_{\rm late}/a_{\rm early})} \log 2.4 \right\},
  \label{eq:monrmin}
 \end{equation}
 where $a_{\rm early}=0.045$ and $a_{\rm late}=0.185$ correspond to early and late time, respectively, for $\vec{\epsilon}=(3/4,1/2)$. This equation is such that $r_{\rm min}=L/n_{\rm g}$ for $a=a_{\rm early}$ and $r_{\rm min}=2.4\, L/n_{\rm g}$ for $a=a_{\rm late}$, while $r_{\rm min}$ presents a power-law behavior as a function of expansion factor between these two values. The same formula will be employed for measurements presented in Figs.~\ref{fig:profile_a} and \ref{fig:profile_b}, but with different values of $a_{\rm early}$ and $a_{\rm late}$ for the CDM halos, namely $a_{\rm late}=0.5$ for all halos and  $a_{\rm early}=0.084$, $0.111$, $0.047$, $0.067$ and $0.067$ for halos 1 to 5, respectively.
\begin{figure*}[htp!]
  \centerline{\null \hskip 1cm \bf 3 sine waves, mass resolution study for $\epsilon=(3/4,1/2)$}
 \centerline{\includegraphics[width=17cm]{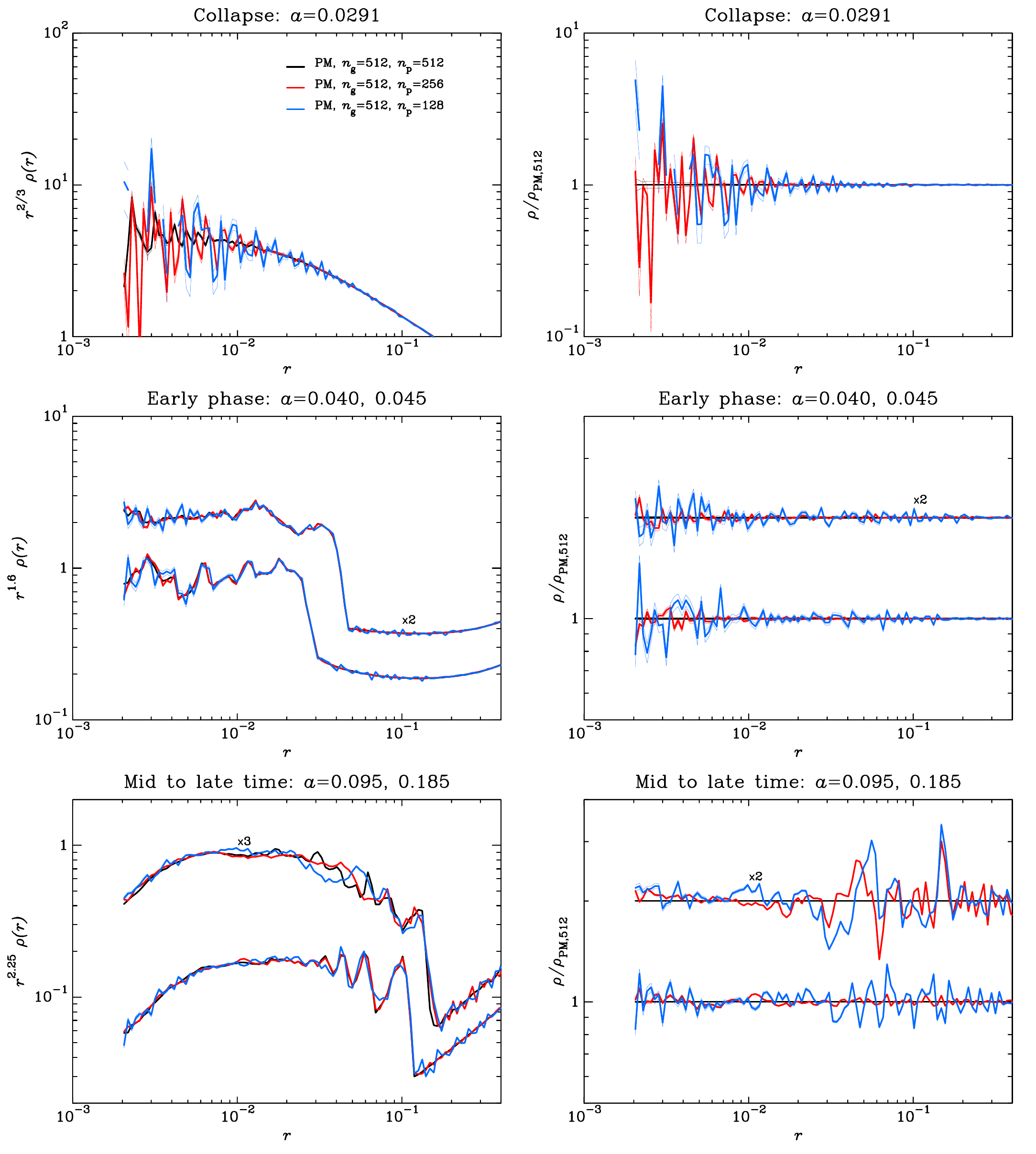}}
\caption[]{Radial density profile: mass resolution analysis of the three sine waves PM simulations with $\vec{\epsilon}=(3/4,1/2)$. This figure is analogous to Fig.~\ref{fig:resolution_profile_a} except that it focuses on the effect of changing the number of particles $n_{\rm p}^3$ in the $N$-body runs.  All the simulations have the same spatial resolution, $n_{\rm g}=512$, but different values of $n_{\rm p}$,  namely $n_{\rm p}=512$ (black, PM-ANI2-HR), 256 (red, PM-ANI2-MR) and 128 (blue, PM-ANI2-LR). On right panels, the ratio considered is  $\rho/\rho_{\rm PM,512}$, where $\rho_{\rm PM,512}$ is the density measured in the PM run with $n_{\rm g}=n_{\rm p}=512$ corresponding to black curves.}
\label{fig:resolution_profile_b}
\end{figure*}

When looking more closely at the density profile, one can observe some fluctuations. On top panels, the density profile should be perfectly smooth, which is almost the case for the Vlasov runs.\footnote{Some small noise can be discerned in the {\tt ColDICE} measurements of top right panel, but this is related to the way the density is calculated, as detailed in Appendix~A.2.} The fluctuations in the PM simulations are simply related to discreteness effects. Their amplitude is controlled by the combination of bin width and number of particles $n_{\rm p}^3$.  Note that these fluctuations are much larger than the one sigma level prediction from Poisson noise shown as very thin lines on top panels. Indeed, at collapse time, the system can still been seen as a distorted regular network of particles, which can introduce significant aliasing effects on the binned radial density.

At later times, fine variations in the density profile are mainly related to properties of the caustic pattern. Obviously, force resolution can affect this pattern, especially in the PM code, for which additional softening of the force cannot be ignored anymore. Keeping in mind the logarithmic scale used in Fig.~\ref{fig:resolution_profile_a}, density fluctuations sufficiently far from the centre of the system still remains pretty insensitive to force resolution, consistently with visual inspections performed in previous sections, except at late time shown in top curves of bottom panels. In the latter case, force resolution seems to influence details of the radial density nearly up to the size of the system, which reflects again the cumulative nature of force softening. Hence, note finally that using fine instead of coarse details of the density profile to determine the available trustable range would impose much more restrictive constraints than equation (\ref{eq:monrmin}) and this even more so for the PM code. 

\subsection{Mass resolution}
\label{sec:tirho}
Figure \ref{fig:resolution_profile_b} is analogous to Fig.~\ref{fig:resolution_profile_a}, but studies mass resolution for the PM code. The main result of this figure is that particle count does not affect significantly the density profile at the coarse level for all the values of $n_{\rm p}$ considered. 

Turning to finer details of the profile, we already mentioned in previous section the aliasing effects on the measured radial density at collapse time due to the memory of the initial set-up of the particles on a regular pattern.  Even with the numerous complex processes already taking place during the early part of the violent relaxation phase, this memory can persist in the multi-stream region, as long as mixing is not sufficiently rich to locally (pseudo-)randomize the particle distribution. Indeed, the intrinsic softening nature of the Green function used to solve Poisson equation can temper, at least for some time, the effects of the discrete nature of the representation of the system with particles. Hence, on second line of panels of Fig.~\ref{fig:resolution_profile_b}, we can suppose that most of the additional fluctuations appearing at various radii when diluting the system are of the same nature as on the top panels, as already discussed in \S~\ref{sec:massrez}. At some point, however, part of these fluctuations are not transient anymore. Along with some more subtle collective effects related to shot noise \citep[e.g.,][]{Colombi15b}, they can grow through gravitational instability and introduce significant deviations from the prediction of the mean field limit. This clearly shows up at late time as noticeable differences between top curves of bottom panels of Fig.~\ref{fig:resolution_profile_b}.
 \subsection{Time evolution: density}
 \label{sec:densprof}
\begin{figure*}[htp!]
\centering
\includegraphics[width=17cm]{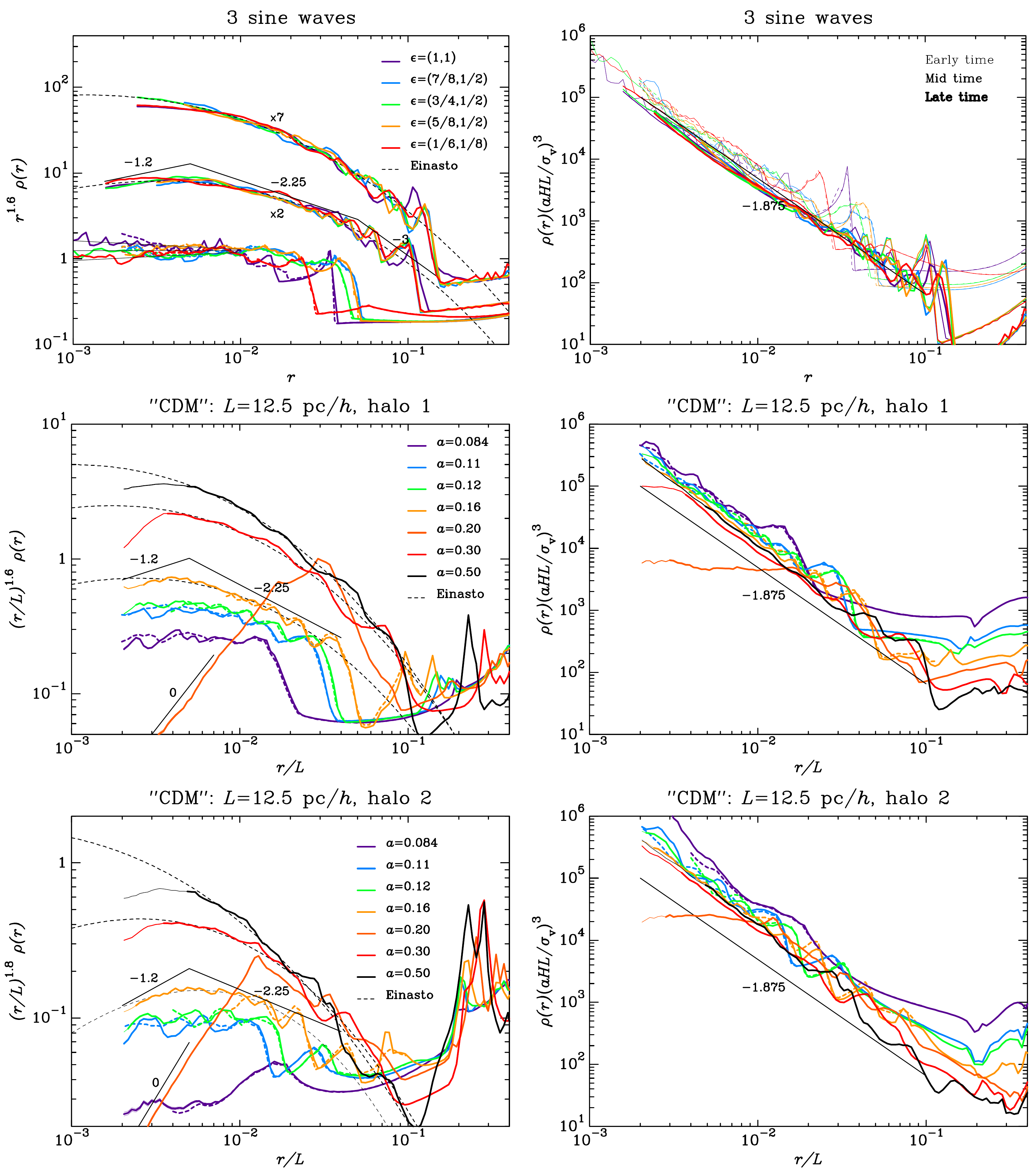}
\caption[]{Time evolution of the radial density profile (left) and of the pseudo phase-space density (right). {\it Top two panels} correspond to the results obtained for the three sine waves simulations for different values of $\vec{\epsilon}$. Three regimes are considered: early time, mid time (multiplied by a factor 2 for clarity on left panel) and late time (multiplied by a factor 7 on left panel), as detailed in Table~\ref{tab:regimes}. The continuous curves of various colors correspond to the highest resolution PM runs, namely PM-Q1D-HR, PM-ANI1-HR, PM-ANI2-UHR, PM-ANI3-HR and PM-SYM-HR in Table~\ref{tab:tabsim}. Only the parts of the curves that are not supposed to be influenced by force softening are shown. In addition, the dashed curves of the same colour provide the measurements obtained at early time in high resolution Vlasov simulations, namely VLA-Q1D-HR, VLA-ANI1-HRS, VLA-ANI2-HR, VLA-ANI3-HRS and VLA-SYM-HR.  To emphasize the very clear power-law behaviour present at early time, the quantity actually displayed on left panel is $r^{\alpha} \rho(r)$, with $\alpha=1.6$. In addition, thin lines indicate different slopes, in particular $\rho(r) \propto r^{-2.25}$ and $Q(r) \propto r^{-1.875}$ as predicted by secondary spherical infall model; dashed curves show Einasto profiles with parameters given in Table~\ref{tab:einasto}; finally, three close very thin lines on top left panel also indicate small variations in the logarithmic slope, $\alpha=1.5$, $1.6$ and $1.7$. {\it Next two lines of panels} are analogous to two top ones, but correspond respectively to halos 1 and 2 extracted from the ``CDM'' runs with $L=12.5$ pc$/h$. Several values of the expansion factor are considered to show various stages of the evolution. Again, the continuous curves of various colors correspond to the PM run PM-CDM12.5-HR. They become thinner at small scales, where force softening is thought to influence the results. The dashed curves of the same colour correspond to Vlasov simulations of the highest possible resolution available, namely VLA-CDM12.5-HR for $a=0.084$ and $0.11$, VLA-CDM12.5-MR for $a=0.12$ and VLA-CDM12.5-LR for $a=0.16$. Mergers, that induce a temporary flattening of the density profile, are emphasized by thin lines with $\alpha=0$.}
\label{fig:profile_a}
\end{figure*}

\begin{figure*}[htp!]
\centering
\includegraphics[width=17cm]{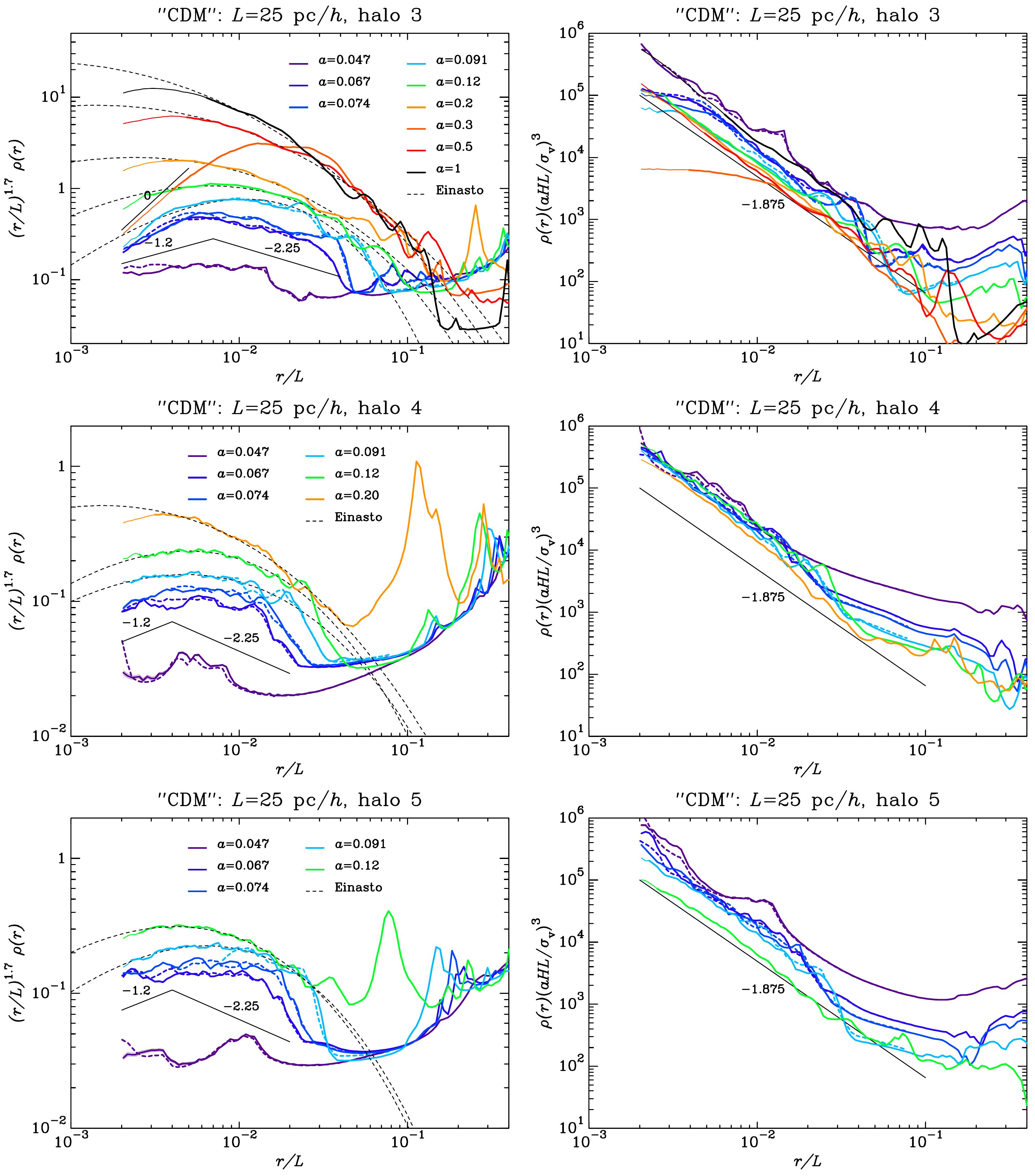}
\caption[]{Time evolution of the radial density profile and the radial pseudo phase-space density: continued. Same as in Fig.~\ref{fig:profile_a} but for ``CDM'' halos 3, 4 and 5, extracted from the ``CDM'' runs with $L=25$ pc$/h$. Note that halos 4 and 5 merge with halo 3. One of these mergers is clearly captured by the orange curve corresponding to $a=0.3$ on top left panel.}
\label{fig:profile_b}
\end{figure*}

Figures~\ref{fig:profile_a} and \ref{fig:profile_b}, show, for all the halos studied in this paper, the radial density profile $\rho(r)$ and the pseudo phase-space density $Q(r)$ (equation~\ref{eq:Qofr}). 

First, we focus on left panels, which display $r^{\alpha} \rho(r)$ as a function of $r$, where $\alpha$ ranges in the interval $[1.5,1.8]$. This value of the  logarithmic slope of the density profile reflects the behaviour seen systematically in the early monolithic phase of the evolution of the protohalos, as found in many previous works, both for the three sine waves case \citep[e.g.,][]{Nakamura85,Moutarde95} and CDM protohalos \cite[][]{Diemand05,Ishiyama10,Anderhalden13,Ishiyama14,Angulo17,Delos18a,Delos18b}. What is mainly new here is that we provide more accurate investigations of the three sine waves initial conditions with a wide variety of values of $\vec{\epsilon}$.

We find for the three sines waves that, after collapse along the three axes,  the early phase of violent relaxation builds always the same kind of power-law profile, with $\alpha \simeq 1.6 \pm 0.1$ (the error is estimated by visual inspection), whatever $\vec{\epsilon}=(\epsilon_a,\epsilon_b)$ with $\epsilon_a >0 $ and $\epsilon_b>0$.\footnote{having one or two of the coordinates of the vector $\vec{\epsilon}$ null reduces the dimensionality of the problem and obviously leads to different slopes.} Note a trend of the slope to increase from $\alpha \simeq 1.5$ to $\alpha \simeq 1.7-1.8$ when going from quasi-1D to axisymmetric configuration, as indicated by the group of three thin lines on top left panel of Fig.~\ref{fig:profile_a}.

While the axisymmetric configuration is locally equivalent, at leading order, to a spherical Gaussian overdensity, $\rho(r) \propto 1- (2\pi r/L)^2/2$, $r \ll L$, the evolution of the system does not lead to the expected slope $\alpha=2.25$ predicted by the secondary spherical infall model \citep[][]{Gott75,Gunn77, Fillmore84,Bertschinger85} and measured approximately in three-dimensional $N$-body simulations of spherical spherical Gaussian overdensities \citep[e.g.,][]{Gosenca17}. One has thus to keep in mind that the axisymmetric three sine waves configuration remains, from the dynamical point of view, strongly anisotropic compared to the purely spherical case.

During the monolithic early violent relaxation phase, the ``CDM'' experiments behave similarly as the three sine waves with a slope ranging approximately in the interval $[1.5,1.8]$, in agreement with the early measurements of  
\citet[][]{Diemand05}, but slightly larger than the value obtained from more recent investigations which rather suggest $\alpha \in [1.3,1.5]$ with a preference for $\alpha \simeq 1.5$ \citep[]{Ishiyama10,Anderhalden13,Ishiyama14,Angulo17,Delos18a,Delos18b}. These slight inconsistencies can be explained at least partly, but in order to understand them, we first need to examine the evolution with time of the various systems under consideration, which we do now.

Whatever the nature of initial conditions, we notice that the power-law behaviour seen at early time disappears at some point and all the systems relax to a ``NFW''-like profile \citep[][]{NFW96,NFW97}, or more precisely, in the analyses performed here, an Einasto-like profile \citep[][]{Einasto65},
\begin{equation}
  \rho(r) \propto \exp\left\{ -\frac{2}{\gamma} \left[\left( \frac{r}{r_0} \right)^\gamma- 1\right]\right\}, \label{eq:einasto}
\end{equation}
that we shall still abusively designate by NFW. The parameters used for the curves displayed in Figs.~\ref{fig:profile_a} and \ref{fig:profile_b} are detailed in Table~\ref{tab:einasto}. The value of $\gamma$ decreases with time\footnote{except for halo 4, but given the crudeness of the measurements and the limited dynamical range at hand, this exception may be ignored.} down to $\gamma=0.15\pm 0.03$,\footnote{except for halo 5 which only reaches mid time prior to merger with halo 3.} in agreement with numerous previous works \citep[e.g.,][]{Navarro04,Merritt06,Duffy08,Gao08,Stadel09,Navarro10,Dutton14,Klypin16}.

This somewhat inevitable evolution towards NFW is already well known in the litterature. In the CDM case, the change of the nature of the profile is generally interpreted as the result of multiple mergers \citep[e.g.,][]{Syer98,Ishiyama14,Ogiya16,Angulo17}.  However, the relaxation to a NFW-like profile is also observed in the three sine waves simulations, that is even in the absence of merger. This result is robust against particle shot noise as illustrated by bottom panels of Fig.~\ref{fig:resolution_profile_b}. The fact that even in the monolithic case, the density profile and its central slope can change and also relax to NFW is not new \citep[see for instance][]{Huss99,MacMillan06,Ogiya18}. Our numerical experiments confirm again the attractor nature of the NFW profile with a spectacular convergence of all the sine wave simulations at late time whatever the value of $\vec{\epsilon}$ we considered, as illustrated by the top curves of upper left panel of Fig.~\ref{fig:profile_a}.

\begin{table}
  \begin{tabular}{llll}
\hline
    Case            & exp. factor $a$ & $\gamma$ &    $r_0$  \\ \hline
    3 sine waves & mid time &  $0.19$ & $0.011$ \\
                          &  late time &  $0.15$ & $0.0049$ \\ \hline
    ``CDM'': halo 1 & $0.16$ & $0.18$ & $0.11$ pc/$h$ \\
                              & $0.30$ & $0.17$ & $0.082$ pc/$h$ \\
                              & $0.50$ & $0.16$  & $0.052$ pc/$h$ \\ \hline
    ``CDM'': halo 2  &  $0.16$ &  $0.27$  & $0.11$ pc/$h$ \\
                              & $0.30$ & $0.19$ & $0.056$ pc/$h$ \\
                              & $0.50$ & $0.12$ & $0.011$ pc/$h$ \\ \hline
    ``CDM'': halo 3 &  $0.091$ & $0.49$ & $0.38$ pc/$h$ \\
                              &  $0.12$ & $0.30$ & $0.29$ pc/$h$ \\
                              & $0.2$ & $0.19$ & $0.14$ pc/$h$ \\
                              & $0.5$ & $0.16$ & $0.094$ pc/$h$ \\
                              & $1.0$ & $0.15$ & $0.041$ pc/$h$ \\ \hline
     ``CDM'': halo 4 & $0.091$ & $0.25$ & $0.23$ pc/$h$ \\
                               & $0.12$ & $0.30$ & $0.21$ pc/$h$ \\
                               & $0.2$ & $0.18$ & $0.096$ pc/$h$ \\ \hline
    ``CDM'': halo 5  & $0.091$ & $0.33$ & $0.26$ pc/$h$ \\
                              & $0.12$ & $0.27$ & $0.19$ pc/$h$ \\ \hline
   \end{tabular}
  \caption[]{Parameters used for the fits with an Einasto profile (eq.~\ref{eq:einasto}) in Figs.~\ref{fig:profile_a} and \ref{fig:profile_b}. They are determined with a simple simplex algorithm \citep[e.g.,][]{Numrec} on some ``trustable'' dynamical range accounting for force softening and a halo extension guessed by visual inspection. Given the rough nature of the measurements due to the limited force resolution of the simulations, no errorbars are provided, so the values of the parameters are only indicative. For the three sine waves case, the fit is performed using the $\vec{\epsilon}=(3/4,1/2)$ highest resolution PM simulation, namely PM-ANI2-UHR.}
\label{tab:einasto}
\end{table}

Note a potentially interesting behaviour that can be observed at mid time and at small radius for $\vec{\epsilon}=(3/4,1/2)$ and $(1,1)$, which seems compatible with the power-law $\rho(r) \propto r^{-1.2}$. When examining close successive snapshots, it can be noticed that this plateau seems to be the result of a progressive change of the slope of the power-law plateau seen at early time along with a reduction of the extension of this region. However, the simulations do not have sufficient force resolution to fully confirm this intermediary regime, especially since this $\alpha \sim 1.2$ slope can also be observed at small scales in the orange part of some of the curves on bottom left panel of Fig.~\ref{fig:resolution_profile_a}. In this case, it results from softening of the force at small scales. Still, keeping this $\alpha \simeq 1.2$ regime in mind, one may then isolate another scaling range with a slope compatible with the prediction $\alpha=2.25$ from secondary spherical infall, evidenced even better for $\vec{\epsilon}=(3/4,1/2)$ in bottom left panel of Fig.~\ref{fig:resolution_profile_a}. Finally, another regime at larger scale compatible with the slope $\alpha=3$ can also be guessed before the cut-off limit corresponding to the halo extension.  Turning to the CDM halos, the same argument may apply as long as their evolution is monolithic. They are not as well resolved as in the three sine waves simulations, so the examination of left panels of Figs.~\ref{fig:profile_a} and \ref{fig:profile_b} is inconclusive to this respect, although one might be tempted to say that halo 1 and halo 2 follow the trend just discussed above for $a=0.16$.

We now come back to the mild discrepancy observed between the logarithmic slope of our CDM microhalos during the early relaxation phase and recent investigations in the literature. To suggest explanations of the differences, one can first notice that the simulations realised in the present work lack dynamical range. The periodic box size $L$ is very small, which makes the interpretation of the results problematic, because the scale corresponding to $L$ should, in reality, become highly nonlinear very quickly, which means that the tidal and merger history of the halos under consideration is unrealistic. In the real CDM scenario, one expects frequent mergers, therefore, a large fraction of protohalos could be composite and pass through the monolithic relaxation phase only shortly if not at all, which implies a smaller value of $\alpha$. Indeed, the picture in which first microhalos form from a single well defined singularity might be oversimplified. It still needs to be refined, both from the theoretical point of view \citep[following footsteps of e.g.,][]{Arnold82,Hidding14,Feldbrugge18}, and the numerical point of view, by studying in detail the topology of the dynamical history of halos in Lagrangian and Eulerian spaces. In the CDM simulations studied in the present work, all the selected halos pass through a monolithic stage, sufficiently long for the establishment of a ``clean'' violent relaxation phase. Therefore, we might expect our CDM halos to have an initial power-law profile with $\alpha$ close to that of the three sine waves case, hence $\alpha \ga 1.5$, while in simulations with larger box sizes such as in \citet[][]{Ishiyama14}, \citet[][]{Angulo17}, the composite yet possibly more realistic nature of the halos can lead to $\alpha \la 1.5$.

Another limit of the simulations in the present work is their force resolution, which implies that only a restricted scale range is available to measure the slope and this might lead to an overestimate of $\alpha$. For instance, the simulation used in \citet[][]{Ishiyama10} has comparable box size (30 pc) to our runs, but much higher force resolution,\footnote{note that there is much less than one particle per softening length in this work, which might be an issue.} which widens considerably the dynamical range for measuring radial density profiles. For the CDM halos analysed, they clearly find $\alpha \simeq 1.5$.  On the other hand, we have seen in the three sine waves simulations that the early power-law plateau seems to become less steep and less extended with time, with $\alpha \simeq 1.2$ at mid time, as a result of natural evolution of the halo. In other words, the slope is time dependent, and measuring it exactly just after early relaxation is non trivial. While resolution issues are probably real for the CDM halos analysed in the present work, this might also explain why the values of $\alpha$ estimated here are slightly larger than in recent analyses in the literature.

In conclusion, for halos going through a truly monolithic violent relaxation phase during their formation, it seems reasonable to think that the logarithmic slope of the power-law density profile building up during this phase can be slightly larger than $1.5$. In the measurements performed for this article, $\alpha$ was indeed found to range in the interval $[1.5,1.8]$. Then, various dynamical processes, such as evolution under slow infall as well as successive mergers decrease the effective value of $\alpha$ and drive progressively the system to the dynamical attractor embodied by the Einasto profile. However, bear in mind again that it has not been proven here that all dark matter halos experience a pure monolithic phase during the early stages of their evolution, nor that a definitive answer to this question exists yet. 
\subsection{Time evolution: pseudo phase-space density}
\label{sec:Qofr}
Right columns of Figs.~\ref{fig:profile_a} and \ref{fig:profile_b} plot the pseudo phase-space density $Q(r)$ as a function of radius $r$.\footnote{more precisely, the dimensionless quantity $\rho(r)\, (aHL/\sigma_v)^3$, where $H\equiv (1/a)\, {\rm d}a/{\rm d}t$ is the Hubble parameter and the density $\rho$ is normalised to unity, $\langle \rho \rangle=1$.} In agreement with previous works \citep[e.g.,][]{Taylor01,Navarro10,Ludlow10},  function $Q(r)$ presents a power-law behaviour compatible with the prediction from secondary spherical infall model, $Q(r) \propto r^{-\alpha_Q}$ with $\alpha_Q=1.875$. However, the fact that this result stands {\it irrespective of the dynamical state of the halos}, even during the early violent relaxation phase and mid time (except of course when a merger perturbs temporarily the profile), is non trivial and somewhat new. To our knowledge, there are indeed only few measurements in the literature of $Q(r)$ for CDM halos during all the stages of the evolution, in particular the early violent relaxation phase. On interesting exception is the recent work of \citet{Ishiyama10}, which suggests deviations from the prediction of secondary spherical infall, with $\alpha_Q \simeq 2.25$ at small $r$. Our measurements are not sufficiently accurate to confirm this. 

A power-law behaviour of function $Q(r)$ is a clear signature of self-similarity. In the pure self-similar framework and assuming spherical symmetry, the projected density $\rho(r)$ and the pseudo phase-space density are both power-laws, with
\begin{equation}
  \alpha_Q=\frac{6-\alpha}{2}, \quad \alpha > 0,
  \label{eq:self}
\end{equation}
\citep[e.g.,][and references therein]{Vogelsberger11}.
When the infall is purely radial, one expects $\alpha \in [2,3]$ \citep[][]{Fillmore84}, hence $\alpha_Q \in [1.5,2]$. The addition of angular momentum, which provides more realistic solutions, adds spread to all the possible values of the density slope, $\alpha \in [0,3]$ \citep[][]{White92, Sikivie1997,Nusser01}, which in turn implies $\alpha_Q \in [1.5,3]$. In the early monolithic phase of their evolution, our halos have a clear power-law behaviour for the density with $\alpha \in [1.5,1.8]$.  Applying naively equation (\ref{eq:self}) implies $\alpha_Q \in [2.1,2.25]$, which is in fact compatible with our measurements at small radii, given their level of accuracy. Along the same line of thought, the value of $\alpha_Q=2.25$ measured at small radii by \citet[][]{Ishiyama10}, whose dark matter halos have $\alpha=1.5$, is strikingly consistent with the self-similar prediction.

Remind however that we are far from spherical symmetry. In the triaxial case, the nature of the self-similar solutions obviously changes, and even if the expected density profile slope at small radius is the same as the one predicted in spherical symmetry with non zero angular momentum, it may be reached only at very small radii \citep[e.g.,][]{Lithwick11}. Additionally, the smooth nature of the initial overdensity deviates from the actual assumptions intrinsic to self-similar solutions, which makes the interpretation of the early monolithic phase of the evolution difficult in this framework. Remind however that in the purely spherical case, as already mentioned in previous section, a smooth overdensity evolves to a state compatible with the secondary infall solution \citep[][]{Gosenca17}.  Subsequent mergers add a tidal torque contribution, that is the generation of angular momentum from the accretion, which complicates further the interpretation of the results \citep[although this can be also approached in spherical symmetry in a self-similar fashion, e.g.,][]{Zukin10a,Zukin10b,Lapi11}.

Even if the evolution of the phase-space density is self-similar, the finite extension of the system can imply quantities such as the projected density $\rho(r)$ or the velocity dispersion $\sigma_v(r)$ not to be pure power-laws due to the effects of the cut-offs. But ratios such as $Q(r)$ might just compensate for this finite size effect and evidence better the self-similar nature of the dynamics \citep[][]{Alard13}. Hence the fact that $Q(r)$ is a pure power-law is not incompatible with $\rho(r)$ not being so, as figures \ref{fig:profile_a} and \ref{fig:profile_b} show. This is also consistent with solutions of Jeans' equation \citep[][]{Dehnen05}. 

It is worth noticing that function $Q(r)$ tends to be more ``universal'' than the density profile. For instance, in the quantitative examples discussed above, its logarithmic slope covers twice a smaller range of values than that of the density. Similarly, according to \citet[][]{Dehnen05}, the resolution of Jeans' equation assuming a pure power-law for $Q_r(r) \equiv \rho(r)/\sigma_{v,r}(r)^3$ with $\sigma_{v,r}(r)^2$ the radial velocity dispersion, provides, in practice, consistent solutions only if $\alpha_{Q,r}\equiv {\rm d} \ln Q_r/{\rm d}\ln r=\alpha_{\rm crit}\equiv 35/18-2\,\beta(r=0)/9$, where
\begin{equation}
  \beta(r) \equiv 1- \sigma_{v,\perp}(r)^2/\sigma_{v,r}(r)^2,
  \label{eq:betadef}
\end{equation}
is the velocity anisotropy parameter assumed to be linearly related to the local logarithmic density slope, with $\sigma_{v,\perp}(r)^2$  the transverse velocity dispersion.  It is known that $Q_r(r)$ and $Q(r)$ present very similar power-law behaviours, with $\alpha_{Q,r}$ very slightly larger than $\alpha_Q$ \citep[see, e.g.,][]{Ludlow10}. In other worlds, $\alpha_Q$ is in practice never expected to be so different from the generic value of secondary spherical infall model, $\alpha_Q=1.875$. 

In conclusion, even if it seems striking that the pseudo phase-space density is roughly compatible with the power-law predicted by standard secondary spherical infall model,  $\alpha_Q=1.875$, it is not surprising. It merely reflects the self-similar nature of the dynamics in phase-space  \citep[][]{Alard13}. Such self-similarity is also clearly suggested  by direct measurements of the history of orbits of particles in dark matter halos \citep[][]{Sugiura20}. While the logarithmic slope of $Q(r)$ is changing little with time, we have seen in previous section that the density profile presents some striking transformations during the course of the dynamics. This is obviously related to changes in the velocity distribution, in particular to evolution of the anisotropy parameter $\beta(r)$. Indeed, it is well known that radial instabilities play an important role in the internal dynamics of halos \citep[e.g.,][]{Huss99,MacMillan06}. 
\section{Conclusion}
\label{sec:conclusion}
In this article, the formation and evolution of dark matter halos have been studied in details with the Vlasov code {\tt ColDICE} and a traditional $N$-body Particle-Mesh (PM) code. Two kinds of initial conditions were considered: on one hand, the highly symmetrical set up with three sine waves, on the other hand, neutralino Cold Dark Matter (CDM)  fluctuations in very small periodic boxes of size $12.5$ and $25$ pc$/h$. In these analyses, that include projected density and phase-space diagrams, radial density profiles and pseudo phase-space density, we paid particular attention to numerical convergence with respect to force resolution traced by the resolution $n_{\rm g}$ of the mesh used to solve Poisson equation and with respect to mass resolution traced by the initial number $n_{\rm s}^3$ of vertices in ${\tt ColDICE}$ and the number  $n_{\rm p}^3$ of particles in the PM simulations. The main results of this paper can be summarised as follows:
\begin{itemize}
\item {\em The $N$-body method is robust when there is typically more than one particle per softening length of the force, that is $n_{\rm p} \ga n_{\rm g}$.} This result is well known \citep[e.g.,][]{Melott97} but was never tested with comparisons of the $N$-body approach to a pure Vlasov code. Of course, because of its high computational cost, {\tt ColDICE} can be used to follow the evolution of halos only in the early relaxation phase. During this phase, discrete noise of particles has little effect on the dynamical evolution of the system  and agreement between PM code and {\tt ColDICE} is excellent. Pushing the PM simulations further, halos profiles are still not affected by $N$-body relaxation at the coarse level, however some instabilities clearly develop at small scales when $n_{\rm p} < n_{\rm g}$.
   \vskip 0.2cm
\item{\em Early violent relaxation phase of protohalos (also called microhalos).} During this phase, the halos are found to display a power-law density profile, $\rho(r) \propto r^{-\alpha}$, with $\alpha \in [1.5,1.8]$, which agrees well with the literature but with a slightly larger values of $\alpha$. Indeed, most previous investigations of CDM microhalos profiles found $\alpha \in [1.3,1.5]$ with a preferred value $\alpha \simeq 1.5$. One obvious explanation of this difference is that the CDM simulations performed in this work lack dynamical range and that the three sine waves simulations are not sufficiently representative of true dark matter halos, given their very high level of symmetry.  Another possible source of the difference might be related to the time at which the slope is measured. Indeed, halo profiles evolve significantly, even in the monolithic stage, which can affect measurements of the logarithmic slope. Finally, the picture in which the halos always first form from a well defined singularity might be oversimplified, even in the context of smooth initial conditions produced by a massive neutralino.  
  \vskip 0.2cm
 \item {\em Complexity.} During the early relaxation phase, it is possible to estimate the level of complexity of the ${\tt ColDICE}$ phase-space sheet by measuring its total volume $V_{\rm s}$ or the total number of simplices $N_{\rm s}$ it is composed of. While both $V_{\rm s}$ and $N_{\rm s}$ increase very quickly with time after collapse, with a growth rate ranging approximately between $a^{7}$ to $a^{30}$ for $N_{\rm s}$, the increase is not exponential in most cases, which suggests the absence of chaos. Only the highest resolution CDM simulation with box size 25 Mpc$/h$, which is the object of several mergers in the period covered by ${\tt ColDICE}$, presents a clear signature of exponential growth for $N_{\rm s}$. However these results are inconclusive in the sense that convergence with force resolution is not demonstrated, especially when examining phase-space sheet volume.
    \vskip 0.2cm
  \item {\em Phase-space structure.} During the early relaxation state, it is possible to examine in detail the structure of the phase-space sheet using phase-space diagrams. The three-sine wave halos display an intricate yet coherent spiral structure which is subject to multiple foldings in phase-space, that can be related to successive collapses along each axis of the dynamics and also show clear signatures of self-similarity. The random nature of CDM initial conditions makes the phase-space structure somewhat fuzzy but still coherent at the coarse level. This structure is of course strongly perturbed by the presence of mergers. Note that the predictions of the Vlasov code seem fairly robust with respect to force resolution when far enough from the centre of the system in terms of softening length of the force field. This is demonstrated as well by the examination of the caustic pattern in the projected density.
     \vskip 0.2cm
   \item {\em  Convergence to NFW-like dynamical attractor.} After careful tests of convergence, the PM code was used to follow furthermore the evolution of the halos. As already well known from many investigations in the literature, the initial power-law behaviour breaks down and the density profile converges to the well known dynamical ``NFW-like'' universal attractor, irrespectively of initial conditions, even in the three-sine wave simulations. This clearly shows again that mergers do not represent a necessary condition for convergence to NFW and that radial instabilities, which change the properties of the velocity distribution, can also play a major role.
      \vskip 0.2cm
\item{\em Pseudo phase-space density.} The pseudo phase-space density $Q(r)=\rho(r)/\sigma_v(r)^3$ measured in all the halos is compatible with the power-law $Q(r) \propto r^{-1.875}$ predicted by secondary infall model at all the times, even during the early relaxation phase. This result is of course well known for relaxed halos, but is non trivial when considering the early and intermediary phases of their evolution, where they display very different forms of the density profile. It represents a clear signature of self-similarity of the dynamics in phase-space. 
\end{itemize}
The analyses performed in this work clearly demonstrate that it is possible to perform $N$-body simulations in a robust way. While the tessellation approach is free of particle shot noise, it is very costly. The extremely quick growth of the phase-space sheet complexity makes this method unaffordable beyond a limited number of dynamical times whatever its level of optimisation. To solve this problem, it has been proposed that a hybrid implementation is employed, relying on the tessellation in regions where relaxation is incomplete and where $N$-body technique can really introduce artificial instabilities, and using particles in dense, dynamically relaxed locations, where the warmness of the system makes the $N$-body approach much more reliable \citep[][]{Stucker20}. However, this hybrid approach, which allows one to use the tessellation method when its cost remains affordable while, at the same time, it corrects for the main defects of the $N$-body approach, might seem unnecessarily complex. Instead, following a rather simple but often ignored ancient numerical strategy \citep[e.g.,][]{Melott97,Splinter98}, a better control of the traditional $N$-body approach could be achieved by making sure that there is everywhere in the computational volume at least one particle per local softening length, the main difficulty being to preserve as much as possible the Hamiltonian nature of the numerical system. This can be achieved with straightforward modifications of current $N$-body codes based on adaptive mesh refinement \citep[][]{Kravtsov97,Teyssier02,Bryan14}, by improving criteria of refinement using constraints based on estimates of local entropy production. 

Another interesting result of the investigations of this article is the relaxation to a universal profile irrespectively of initial conditions, even in the absence of mergers. While this result is not fundamentally new, the detailed analyses of the three-sine wave case in various configurations was never performed at the level of accuracy achieved in this work. However, the simulations, even with $n_{\rm g}=1024$, still lack spatial resolution. It would be clearly worth reinvestigating the halos studied in the present work with high spatial resolution $N$-body simulations, performed in a controlled way just as advocated above, to analyse in detail the evolution of the velocity field structure of the halos. In particular, it would be worth studying how the velocity anisotropy parameter $\beta(r)$ (equation \ref{eq:betadef}) changes with time, to understand the effects of radial instabilities on the evolution of the density profile and compare them to the effects of mergers. 
\begin{acknowledgements}
We thank Thierry Sousbie, who wrote the source code of {\tt ColDICE}, for very fruitful exchanges during several years of collaboration. We also thank Christophe Alard, Mark Neyrinck, Shohei Saga, Andrei Sobolevski and Atsushi Taruya for fruitful discussions. This work was supported by ANR grant ANR-13-MONU-0003. Numerical computation with {\tt ColDICE} was carried out using the HPC resources of CINES (Occigen supercomputer) under the GENCI allocations c2016047568, 2017-A0040407568 and 2018-A0040407568. PM simulations and post-treatment of {\tt ColDICE} data were performed on HORIZON cluster of Institut d'Astrophysique de Paris. 
\end{acknowledgements}
\bibliographystyle{aa}

\section*{Appendix A: Calculation of the projected density and of radial profiles}
\label{app:rhoofr}
While {\tt ColDICE} uses a complex ray-tracing algorithm to compute the exact intersections between the tessellation and the mesh used to solve Poisson equation, post-treatment is performed in a less sophisticated way, by replacing each simplex of the tessellation by a large number of particles. The next two sections explain how the three-dimensional projected density and the radial profiles are calculated from the tessellation structure provided by {\tt ColDICE}. 
\subsection*{A.1: Calculation of projected density}
\label{app:appendixA1}
To compute the projected density used in Figs.~\ref{fig:CDM_allbox}--\ref{fig:pdens_halo3}, a three-dimensional grid of resolution $n_{\rm sub}=512$ is defined on a cube of size $L_{\rm sub}$ that can correspond to the full simulation volume, like in Fig.~\ref{fig:CDM_allbox}, or a portion of it, like in Fig.~\ref{fig:dens_3D_all}. Then each simplex overlapping with this volume is refined recursively in an isotropic fashion $\ell_{\rm max}$ times. At the end of the process, each subsimplex is replaced with a particle lying at the barycentre of the four vertices composing it. Then this particle is affected to the computational grid with a standard cloud-in-cell interpolation \citep[][]{Hockney88}. For each simplex, the final level of refinement $\ell_{\rm max}$ is chosen so that each subsimplex size is small enough compared to the size $\Delta_{\rm sub}=L_{\rm sub}/N_{\rm sub}$ of each voxel of the mesh, namely
\begin{equation}
  \ell_{\rm max}={\rm max}\left\{ {\rm max}_i\lfloor \log_2(s_i/\Delta_{\rm sub}) +1\rfloor,0\right\},
  \label{eq:lmax}
\end{equation}
with
\begin{equation}
  s_i={\rm max}_{j \neq k}(|v_{i,j}-v_{i,k}|),
\end{equation}
where $v_{i,j}$ is the $i^{\rm th}$ coordinate of vertex $j$ ($j=1,\cdots,4$) of the simplex. With the choice of $\ell_{\rm max}$ given by equation (\ref{eq:lmax}), discreteness effects related to the particle representation are sufficiently small to be invisible on the figures.

\subsection*{A.2: Calculation of radial profiles}
\label{app:appendixA2}
To compute the radial profiles, the procedure is slightly different. First, considering $N_{\rm bins}$ logarithmic bins of width
\begin{equation}
  \Delta {\rm log} R=\frac{1}{N_{\rm bins}}\log \left( \frac{R_{\rm max}}{R_{\rm min}} \right)
\end{equation}
in the interval $[R_{\rm min},R_{\rm max}]$,  the simplices are refined isotropically again $\ell_{\rm max}$ times,  but with a value of $\ell_{\rm max}$ defined as follows:
\begin{equation}
  \ell_{\rm max}={\rm max}\left\{ \left\lfloor -\frac{1}{3} \log_2(E^2N_{\rm virtual}) \right\rfloor,0 \right\}
  \label{eq:ellmaxrad}
\end{equation}
with $E=0.005$,
\begin{eqnarray}
  N_{\rm virtual} &= &\frac{4}{3} \pi\,\frac{R_{\rm s}^3}{V_{\rm s}} \exp( 3\, \Delta{\rm log} R-1), \\
R_{\rm s}&=& {\rm max}\left\{ {\rm min}_j \sqrt{\sum_{i=1}^3 v_{i,j}^2},R_{\rm min} \right\},
\end{eqnarray}
and $V_{\rm s}$ is the volume of the smallest possible parallelepiped $P$ containing the simplex with sides aligned with the coordinates axes:
\begin{equation}
  V_{\rm s}=\prod_{i=1}^3({\rm max}_j\,v_{i,j}- {\rm min}_j\, v_{i,j}).
\end{equation}
If this parallelepiped contains the origin of coordinates, $R_{\rm s}$ is also set equal to $R_{\rm min}$. Basically, $N_{\rm virtual}$ represents an estimate of the number of parallelepipeds the smallest radial bin intersecting with the simplex would contain. In equation (\ref{eq:ellmaxrad}), $E$ represents the order of magnitude of the maximum relative error one aims to achieve in the presence of the Poisson noise induced by the replacement of the simplices with particles.

In the last step, instead of replacing the subsimplex with a particle at the barycentre of the vertices composing it, it is replaced with a particle thrown at random inside the volume occupied by the simplex, in order to reduce aliasing effects on the binned radial profile. Then the particle is affected to the bin directly with a weight proportional to the mass of the subsimplex. Additionally, this weight can be multiplied, depending on the quantity of interest, by the radial coordinate of the velocity, its square, or the square of the transverse velocity. The choice of $E=0.005$ is such that fluctuation noise introduced by the procedure remains nearly invisible on the curves shown in Figs.~\ref{fig:resolution_profile_a} to \ref{fig:profile_b}.\footnote{Note however that density ratios shown in top right panel of Fig.~\ref{fig:resolution_profile_a} present small variations related to this noise.} This means that all the irregular features than can be seen on the curves shown in these figures are intrinsic and not induced by sampling of the simplices with particles.
\end{document}